\documentclass[journal]{IEEEtran}
\usepackage{algorithm}
\usepackage{algpseudocode}
\usepackage{amsmath}
\usepackage{amssymb}
\usepackage{amsthm}
\usepackage{bbm}
\usepackage{bm}
\usepackage{braket}
\usepackage{color}
\usepackage{diagbox}
\usepackage{extarrows}
\usepackage{stfloats}
\usepackage{url}
\usepackage{graphicx}
\usepackage{multirow}
\usepackage{subfigure}
\usepackage{times}
\usepackage{ifthen}
\usepackage[letterpaper, left=0.64in,right=0.64in,top=0.76in,bottom=1.1in]{geometry}

\newboolean{includeAppendix}
\setboolean{includeAppendix}{true} 

\newboolean{showcomments}
\setboolean{showcomments}{false}
\newcommand{\meng}[1]{\ifthenelse{\boolean{showcomments}}
{ \textcolor{blue}{(Meng says: #1)}}{}}

\newcommand{\ning}[1]{\ifthenelse{\boolean{showcomments}}
{ \textcolor{magenta}{(Ning says: #1)}}{}}

\newcommand{\yanning}[1]{\ifthenelse{\boolean{showcomments}}
{ \textcolor{red}{(Yanning says: #1)}}{}}

\newcommand{\nikos}[1]{\ifthenelse{\boolean{showcomments}}
{ \textcolor{cyan}{(Nikos says: #1)}}{}}

\newcommand{\guocheng}[1]{\ifthenelse{\boolean{showcomments}}
{ \textcolor{red}{(Guocheng says: #1)}}{}}

\newcommand{\rev}[1]{#1}

\theoremstyle{definition}
\newtheorem{Definition}{Definition}
\newtheorem{Assumption}{Assumption}
\newtheorem{Lemma}{Lemma}
\newtheorem{Theorem}{Theorem}

\newtheorem{Proposition}{Proposition}
\newtheorem{Remark}{Remark}
\newtheorem{Question}{Question}

\makeatletter
\newcommand{\linebreakand}{%
  \end{@IEEEauthorhalign}
  \hfill\mbox{}\par
  \mbox{}\hfill\begin{@IEEEauthorhalign}
}
\makeatother

\begin{document}
\title{Graph Attention Reinforcement Learning for Multicast Routing and Age-Optimal Scheduling}
\author{
Yanning Zhang,~\IEEEmembership{Member,~IEEE,} Guocheng Liao,~\IEEEmembership{Member,~IEEE,} Shengbin Cao,\\
Ning Yang,~\IEEEmembership{Member,~IEEE,} Nikolaos Pappas,~\IEEEmembership{Senior Member,~IEEE}, and Meng Zhang,~\IEEEmembership{Member,~IEEE}
\thanks{
Yanning Zhang and Meng Zhang are with the Zhejiang University - University of Illinois Urbana-Champaign Institute, Zhejiang University (E-mail: yanning.22@intl.zju.edu.cn, mengzhang@intl.zju.edu.cn).

Guocheng Liao is with Sun Yat-Sen University (E-mail: liaogch6@mail.sysu.edu.cn).

Shengbin Cao is with South China Normal University (E-mail: shengbincao@zju.edu.cn).

Ning Yang is with the Institute of Automation, Chinese Academy of Sciences (E-mail: ning.yang@ia.ac.cn).

Nikolaos Pappas is with the Department of Computer and Information Science (IDA), Linköping University (E-mail: nikolaos.pappas@liu.se).
}

}
\maketitle
\pagestyle{plain}
\thispagestyle{plain}
\begin{abstract}
Multicast routing is essential for real-time group applications, such as video streaming, virtual reality, and metaverse platforms, where the \textit{Age of Information (AoI)} acts as a crucial metric to assess information timeliness.
This paper studies dynamic multicast networks with the objective of minimizing the expected average AoI by jointly optimizing multicast routing and scheduling. The main challenges stem from the intricate coupling between routing and scheduling decisions, the inherent complexity of multicast operations, and the graph representation.
We first decompose the original problem into two subtasks amenable to hierarchical reinforcement learning (RL) methods. We propose the first  RL framework to address the multicast routing problem, also known as the Steiner Tree problem, by incorporating graph embedding and the successive addition of nodes and links.
\rev{For graph embedding, we propose the Normalized Graph Attention mechanism (NGAT) framework and establish non-expansiveness of the complete embedding matrix in the Frobenius norm under fixed symmetric row-stochastic attention, while evaluating the learned model's stability and generalization empirically.}
We validate our framework through experiments on four datasets, including the real-world AS-733 dataset. The results demonstrate that our proposed scheme can be up to $9.85\times$ more computationally efficient than traditional multicast routing algorithms, achieving approximation ratios of $1.1$–$1.3$. These results are not only comparable to state-of-the-art (SOTA) methods but also highlight its superior generalization capabilities, performing effectively on unseen and more complex tasks.
Additionally, our age-optimal TGMS algorithm reduces the average weighted AoI by 38.7\% and the weighted peak age by 37.1\% under various scenarios.
\end{abstract}
\section{Introduction}
\subsection{Background and Motivations}
\IEEEPARstart{R}{eal}-time networked applications have garnered significant attention across various domains, including human-computer interaction \cite{liu2023real}, monitoring systems \cite{dini2024real}, and Internet-of-Things (IoT) \cite{abd2019role}. These applications rely on timely updates to ensure accurate and up-to-date information for critical tasks such as decision-making, user interaction, and system control \cite{popovski2022perspective}. While latency has traditionally been a key metric in communication networks \cite{quinn2001ip}, it is now widely acknowledged that minimizing delay alone does not guarantee timely information updates \cite{kaul2012real}. For example, maximizing the frequency of sensor updates may optimize resource utilization but can lead to monitors receiving outdated information due to message backlogs. This highlights the need for a metric that better captures the timeliness of information dissemination. The concept of \textit{Age of Information (AoI)} has emerged as a promising metric in fields such as learning and network protocols \cite{yates2021age}. AoI quantifies the freshness of information available to a monitor about a specific entity or process, making it a suitable performance metric for real-time applications \cite{li2020age,pappas2022age}. \par

Meanwhile, \textit{multicast} has emerged as a critical communication paradigm for the modern real-time group applications. Multicast is widely regarded as a necessity rather than an option for effectively scaling the Internet \cite{quinn2001ip}. Applications such as video streaming \cite{jiang2021survey}, intelligent transportation systems \cite{yin2019improving}, and industrial automation \cite{abd2019role} have experienced significant growth, further emphasizing the demand for efficient multicast solutions. The ability to deliver fresh, synchronized data to multiple endpoints simultaneously is not merely a technical advantage—it is a foundational requirement for enabling next-generation technologies, from autonomous vehicle coordination to massive IoT deployments.

Addressing multicast challenges is of critical importance. A major bottleneck arises in the routing process, which involves attaining the optimal paths from a source to multiple destinations \cite{quinn2001ip}. This problem is closely linked to the design of multicast routing protocols, which are formulated as instances of the \textit{Steiner-Tree Problem (STP)}—a well-known Combinatorial Optimization (CO) problem classified as NP-hard \cite{oliveira2005survey}. The NP-hard nature of these problems renders them computationally intractable for large-scale networks. Although approximation algorithms have been developed (e.g., \cite{byrka2010improved, fischetti2017thinning}), their effectiveness is often constrained in dynamic and resource-limited environments.

To tackle the complexity of multicast problems, researchers have developed various approximation algorithms to optimize the multicast process (see the survey in \cite{paul2002survey}). However, the inherently distributed nature of network systems often limits traditional controllers, forcing them to optimizing routing based on local information. Consequently, existing approaches may struggle to perform effectively in large-scale networks or scenarios requiring age-minimal scheduling.
This motivates us to answer the first key question of this paper:
\begin{Question}
How should one design an efficient and scalable multicast routing framework?
\end{Question}

Additionally, unlike traditional metrics such as delay or cost minimization in static networks, AoI-based scheduling prevents update backlogs and excessive resource depletion, which are critical for real-time applications (e.g., \cite{ramakanth2023monitoring, pu2023aoi}). Although some studies have explored \textit{multicast scheduling} for AoI optimization (e.g., \cite{li2020age}), they often overlook the multicast routing problem. This limitation stems from the black-box nature of traditional TCP/IP networks, which hinders information sharing across layers \cite{fu2013survey}. Cross-layer design methods have been proposed to address these limitations by jointly optimizing routing and scheduling decisions, thus improving network performance (e.g., \cite{chen2006cross, kuran2010cross}). Consequently, a cross-layer multicast framework is well-suited for optimizing AoI through joint routing and scheduling decisions. \par

Furthermore, real-world networks often operate under energy constraints, where the overall energy consumption must be carefully managed \cite{cui2005energy}. This introduces further complexity, as a trade-off between energy consumption and AoI \cite{xie2021age} renders existing multicast algorithms inadequate. Notably, energy consumption in multicast networks is closely tied to scheduling decisions. For instance, a multicast tree with more destinations may reduce AoI but increase energy consumption. Thus, there is a pressing need for a novel multicast framework that provides a holistic solution for optimizing AoI in energy-constrained networks. \par

In this paper, we aim to further answer the second key question:
\begin{Question}
    How can multicast scheduling and routing algorithms be designed to minimize the age of information?
\end{Question}
\subsection{Key Challenges and Solution Approach}

We now summarize the key challenges of answering the above question as follows:
\begin{enumerate}
    \item \textbf{Cross-Layer Design}. 
	Multicast scheduling and routing are inherently interdependent, significantly influencing the AoI at destination nodes. However, existing approaches often address these aspects in isolation, resulting in suboptimal solutions. Furthermore, the high-dimensional nature of the joint optimization problem makes it computationally intractable to solve directly. Thus, it is crucial to design a cross-layer framework that can jointly optimize multicast scheduling and routing decisions.
    \item \textbf{Computationaal Complexity}.
    Multicast routing problems typically fall under the category of combinatorial optimization, which is known to be NP-hard. The complexity escalates when accounting for the time-varying nature of networks, particularly in large-scale deployments. While approximation algorithms have been proposed, they often fail to meet the real-time requirements of multicast applications, necessitating more efficient solutions.
    \item \textbf{High-dimensional Graph Information}. 
    Traditional methods are inefficient when extracting relevant graph features due to the non-Euclidean nature of graphs, even for machine learning methods. However, graph information is crucial for making routing decisions in multicast networks. Therefore, an effective graph embedding method is required to extract hidden graph information while reducing the dimensionality.
\end{enumerate}

First, to jointly solve the age-optimal multicast scheduling and routing problem, we decompose the original problem into two subproblems: (1) the scheduling subproblem and (2) the tree-generating subproblem. This decomposition reduces the complexity of the original problem and facilitates a cross-layer \textit{reinforcement learning (RL)} architecture design. \par

Second, the NP-hard nature of multicast routing problems makes obtaining optimal solutions infeasible, particularly under the stringent time constraints of real-time applications. Inspired by \cite{khalil2017learning}, we propose a tree-generating heuristic algorithm, incrementally constructing multicast trees by adding nodes and edges while adhering to tree constraints. Prior work \cite{khalil2017learning} has demonstrated that integrating a greedy meta-algorithm within an RL framework yields near-optimal solutions for various combinatorial optimization problems. 
 Unlike traditional multicast routing algorithms, which are computationally expensive for large and dynamic networks, our RL framework learns approximately optimal routing decisions through interaction with the environment and generalizes effectively to unseen network topologies, offering a scalable solution for real-time applications.
\par

Third, to address the high-dimensional nature of graph data, we propose a graph-based framework to extract latent features while reducing dimensionality. Building on prior work \cite{khalil2017learning}, which combines graph embeddings with RL for combinatorial optimization, we extend this approach to multicast networks using Graph Attention Networks (GATs) \cite{velickovic2017graph}, known for their effectiveness in modeling graph-structured data. We design a quality function to evaluate multicast tree performance, integrating it into a hierarchical RL framework for routing. \rev{Additionally, we introduce a normalized GAT and prove conditional non-expansiveness in the Frobenius norm for fixed symmetric row-stochastic attention.}

We summarize our key contributions in the following:
\begin{itemize}
    \item \textbf{Joint Multicast Scheduling and Routing Problem.}
	We formulate the first problem of joint multicast routing and scheduling, accounting for energy constraints and potential network dynamics. We decompose the original problem into two subproblems: (1) a scheduling subproblem and (2) a tree-generating subproblem, and prove the equivalence between the subproblems and the original problem. \textit{To the best of our knowledge, this is the first work to consider the problem of joint multicast scheduling and routing for minimizing AoI.}
    
    \item \textbf{Hierarchical RL Framework with Graph Representation.}
    To address Challenges 1 and 3, we propose a fully unsupervised hierarchical RL framework that learns the multicast decisions end-to-end with a convergence analysis. To address Challenge 3, we leverage the graph embedding method to extract hidden graph information, enabling the RL agent to make informed decisions. To increase the generalization capability of our model, we design 
    \rev{a normalized GAT variant with a Frobenius-norm non-expansiveness guarantee under fixed symmetric row-stochastic attention and empirical evaluation of the learned model's stability.}

    \item \textbf{Practical Implementation Design.}
    We propose a practical implementation of our designed algorithm with a cross-layer architecture. We design a hierarchical two-timescale operation mechanism, where the multicast tree is generated at a slower timescale than the scheduling policy to reduce signaling overhead. Also, we deploy our framework based on routing tables, core mechanism of most modern multicast protocols (e.g., \cite{deering1995protocol}) to demonstrate its practicality.

    \item \textbf{Experimental Results.}
	We validate our approach using four datasets, including the AS-733 real-world dataset. We compare our model with various baselines independently on the two subproblems. For the Steiner-Tree Problem (STP), our proposed tree generator outperforms state-of-the-art (SOTA) algorithms by achieving up to $9.85\times$ faster computation while maintaining comparable approximation ratios. 
    For the joint multicast scheduling and routing problem, the results show that our proposed framework achieves an average reduction of $38.7\%$ in AoI and $37.1\%$ in peak AoI compared to baselines including a SOTA age-optimal scheduling algorithm \cite{xie2021age}. Moreover, it demonstrates superior generalization capabilities, effectively handling unseen graphs. We also conduct extensive experiments to validate the robustness of our algorithm in dynamic networks with different levels of dynamics.
\end{itemize}

\section{Related Work}

\subsection{Multicast Routing}

\textit{Multicast routing} poses a significant challenge in various networked systems (e.g., ad hoc networks \cite{singal2021qos} and Internet of Things (IoTs) \cite{kumar2022game}) and game-theoretic scenarios (e.g., cooperative games \cite{hatano2018computational}). 
Extensive research has been dedicated to optimizing multicast routing across various scenarios. Given the distributed nature of network entities, spanning trees are widely regarded as one of the most efficient and practical mechanisms for multicast routing \cite{paul2002survey}.
Considering various requirements, different types of trees have been proposed, with the Steiner tree being one of the most popular types \cite{ljubic2021solving}. Given a set of terminal nodes, the STP finds a tree that connects all terminal nodes at the minimum cost. The STP falls within the domain of CO problems, which are proven to be NP-hard \cite{oliveira2005survey}. Researchers have proposed various algorithms to solve the STP. For instance, Byrka \textit{et al.} \cite{byrka2010improved} developed an LP-based approximation algorithm to solve the STP with a $1.55$ approximation ratio. Fischetti \textit{et al.} \cite{fischetti2017thinning} reformulates the STP as a mixed-integer linear programming problem and proposes a linear-time algorithm to solve it. At the same time, the edge costs are assumed to be the same. \textit{However, the same optimization problem needs to be re-optimized at each time slot for the above methods due to the time-varying nature of networks, which is computationally infeasible.} \par
\subsection{Age of Information}
The concept of \textit{AoI} was initially introduced by Kaul \textit{et al.} \cite{kaul2012real} and garnered significant attention in wireless scenarios (e.g., \cite{kadota2018scheduling,sun2021age,yang2020optimizing,chen2022age}) and queue theory (e.g., \cite{yates2018age,bedewy2019minimizing,huang2015optimizing}). In multicast networks, a substantial body of research has primarily concentrated on analyzing stopping schemes to optimize AoI. For instance, Li \textit{et al.} \cite{li2020age} considered fixed and randomly distributed deadlines for multicast status updates in a real-time IoT system. Buyukates \textit{et al.} \cite{buyukates2019age} considered a multi-hop multicast network and determined an optimal stopping rule for each hop for exponential link delays. \textit{Despite the success of these works in reducing AoI, the considered problem settings are relatively simple for real scenarios.} Some studies focus on serving policies to optimize multicast AoI. For example, Bedewy \textit{et al.} \cite{bedewy2019age} prove that a preemptive last-generated, first-served policy results in smaller age processes than other causal policies. \textit{However, their work does not consider energy consumption.} \par
On the other hand, to deal with the complexities of real-world scenarios, researchers have delved into leveraging \textit{Machine Learning Methods} to optimize AoI in diverse fields, including UAV-assisted systems (e.g., \cite{sun2021aoi, oubbati2022synchronizing, eldeeb2023traffic, fu2023age}), IoTs systems (e.g., \cite{wu2020deep}), and intelligent transportation systems (e.g., \cite{bai2023aoi}). This line of work mainly focuses on optimal resource allocation and trajectory design to achieve AoI optimization. 
A few studies along this direction considered scheduling policies for optimizing AoI. 
For instance, \cite{elgabli2019reinforcement, yin2020application, zhang2024distributed} focused on scheduling algorithms for real-time applications in AoI-aware scenarios.
\textit{However, these works did not consider the graph information in multicast networks, making them inapplicable to multicast scenarios.} \par
Furthermore, AoI has also gained research interest in energy harvesting networks, especially in IoT scenarios (e.g., \cite{stamatakis2019control, hatami2023age, xie2023age}). Deep RL methods are also applied to optimize the trade-off between AoI and energy consumption (e.g., \cite{li2022deep}). Some studies have considered this trade-off for multicast networks. For instance, Xie \textit{et al.} \cite{xie2021age} considered different stopping schemes for multicast networks to maximize energy efficiency while minimizing AoI. Nath \textit{et al.} \cite{nath2018optimum} considered a similar problem and proposed an optimal scheduling strategy.
\textit{However, these approaches did not account for the relation between multicast routing and scheduling.} \par
\section{System Model}
\label{Sec: System Model}
\subsection{System Overview}
\begin{figure}[!t]
    \centering
    \includegraphics[width = 0.35\textwidth]{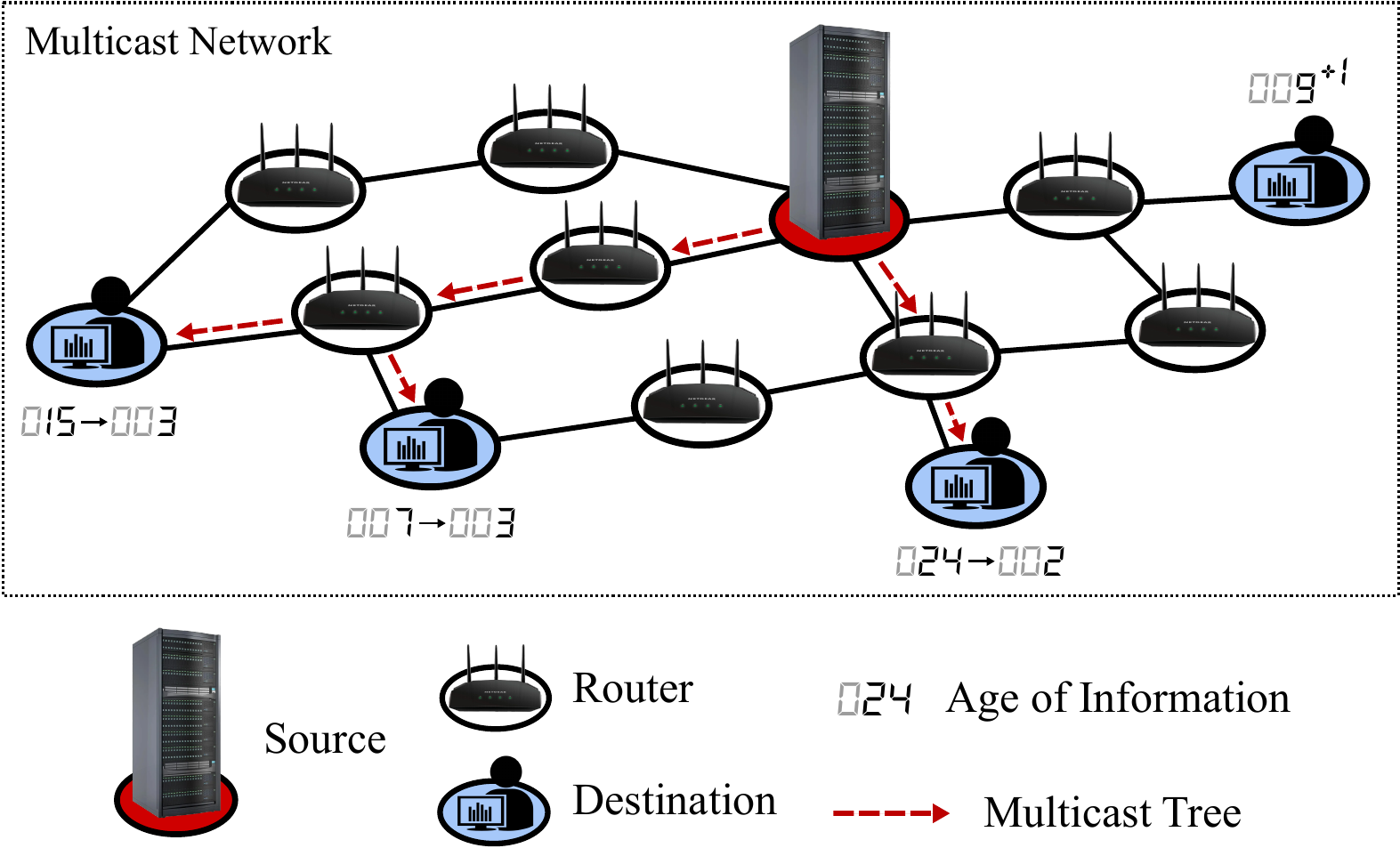}
    \caption{An example of a multicast network. The nodes are connected by links with different costs. At the beginning of each time slot, the source generates update packets, which are then forwarded to a part of the destinations by routers. The packets traveling through different paths may not arrive at the destinations simultaneously, making the AoI of destinations different.}
    \label{Fig: Multicast Network}
\end{figure}
We consider a real-time multicast network, as illustrated in Fig. \ref{Fig: Multicast Network}. Such a network is applicable to a wide range of real-time applications. For instance, in virtual reality scenarios, users may participate in live performances, immersive online gaming sessions, or interactive experiences within the metaverse, all requiring seamless real-time data transmission. Similarly, a traffic management center can multicast real-time traffic updates to multiple vehicles or roadside units in intelligent transportation systems.

The system operates for an infinite horizon of discrete slotted time $t\in\mathbb{N}$, where $\mathbb{N}$ is the set of all natural numbers. We characterize the network topology by a dynamic (undirected) graph. The graph at time $t$ is denoted as $\mathcal{G}_t= \{\mathcal{V}_t, \mathcal{E}_t\}$, where $\mathcal{V}_t$ represents the set of nodes and $\mathcal{E}_t$ represents the set of undirected links. There are three distinct categories of nodes in each $\mathcal{V}_t$:
\begin{itemize}
    \item \textbf{The Source Node.} 
    The network includes a designated source node that generates updates for a multicast group, e.g., a video streaming server. We assume that the source node can generate at least one update packet for each destination at each time slot.
    \item \textbf{Router Nodes.}
    Router nodes are responsible for forwarding update packets to destinations. The forwarding process is based on multicast routing decisions.
    \item \textbf{Destination Nodes.}
    Destination nodes are expected to receive status updates continuously from the source node, e.g., subscribers in a video streaming service. The set of destinations at time $t$ is denoted as $\mathcal{U}_t \subset \mathcal{V}_t$. 
\end{itemize}
\begin{Remark}
    Multicast communication can be classified into two types: source-specific multicast and group-shared multicast \cite{sahasrabuddhe2000multicast}. The above network setting can also be applied to group-shared multicast, where the source node is the node that sends data to other destinations at time $t$. When multiple sources transmit simultaneously, we consider the transmitting process for each of them independently.
\end{Remark}

\textit{Network Dynamics.} Real-world networks often exhibit \textit{dynamic behaviors}. For instance, in wireless networks, topology changes occur due to node mobility or link instability. We assume the network topology $\mathcal{G}_t$ follows a stochastic process. This assumption has been widely adopted (e.g., \cite {sinha2016throughput}). For the network topology $\mathcal{G}_t$ and the users $\mathcal{U}_t$, we have:
\begin{equation}
    \{\mathcal{G}_{t+1}, \mathcal{U}_{t+1}\} = o(\mathcal{G}_t, \mathcal{U}_t, \xi_t),
\end{equation}
where $o(\cdot)$ represents the dynamic transition function and $\xi_t$ is the random variable that characterizes the network dynamics. The statistical properties of $o(\cdot)$ and $\xi_t$ rely on specific network scenarios, which are assumed unknown a priori. We assume at least one path exists from the source to each destination $u\in\mathcal{U}_t$ at each time slot $t$. \par
\subsection{Multicast Process}
We adopt a \textit{generate-at-will} model (e.g., \cite{akar2025age, champati2021minimum}) for multicasting status updates. Specifically, the source node can generate multiple update packets at the start of each time slot. These packets are forwarded through router nodes and ultimately delivered to the destination nodes.

\textit{Delay:}
We make the following assumptions on packet delay:
\begin{Assumption}\label{Assum1}
    The delay between the source and the destination is linearly proportional to the hops between them.
\end{Assumption}
Therefore, we normalize the \textit{one-hop delay for each packet to a single time slot}. The above assumption has been justified in many multi-hop scenarios (e.g., \cite{haenggi2005routing, bui2009novel, ogbe2023optimal}). When Assumption \ref{Assum1} is not satisfied, extending to a more general case is conceptually straightforward by using the sum of delays between nodes instead of hops. This does not alter our results.

We consider an error-free transmission (e.g., \cite{vuran2009error}) with an adaptive power scheme to compensate for potential deep fading scenarios (e.g., \cite{asghari2010adaptive}). Hence, a packet can be transmitted from node $i$ to node $j$ if the link $(i,j)$ exists in $\mathcal{E}_t$, which takes one time slot with an average energy cost of $C_{i,j}$. If multiple packets arrive at a destination node simultaneously, only the packet with the smallest AoI is considered. \par
Next, we describe the \textit{multicast scheduling decisions}. At the start of time slot $t$, a controller can select a subset of destination nodes $\mathcal{U}'_t\subseteq\mathcal{U}_t$ to generate updates for (the updates may not be received immediately), which we refer to as multicast scheduling in this paper. It is crucial to trade off energy consumption and AoI reduction. \par
Now we define the \textit{multicast routing decisions}. Due to the one-source-multiple-destination nature of multicast problems, multicast trees are widely used to represent multicast routing decisions \cite{oliveira2005survey}. Let $\mathcal{T}_t$ denotes a multicast tree at time slot $t$ of $\mathcal{G}_t= \{\mathcal{V}_t, \mathcal{E}_t\}$, i.e.:
\begin{equation}
    \mathcal{T}_t=\{\mathcal{V}^\mathcal{T}_t, \mathcal{E}^\mathcal{T}_t\}, \mathcal{V}^{\mathcal{T}}_t \subseteq {\mathcal{V}}_t, \mathcal{E}^{\mathcal{T}}_t \subseteq {\mathcal{E}}_t.
\end{equation}
The multicast tree $\mathcal{T}_t$ has two properties: (i) It is a connected acyclic directed graph. (ii) The source node and the selected destinations $\mathcal{U}'_t$ are included. Let $\Omega(\mathcal{U}'_t)$ denote the set of all possible multicast trees at time slot $t$. The multicast routing process can be described as follows: at time slot $t$, the source generates $|\mathcal{U}'_t|$ packets, which will be transmitted to their destinations along $\mathcal{T}_t\in\Omega(\mathcal{U}'_t)$ in the following time slots. \par
Due to possibly different hops from the source to different destinations in $\mathcal{T}_t$, the arrival times of packets generated at time $t$ may be different (see Fig. \ref{Fig: Multicast Network}). When a new multicast tree is generated, we let the previously generated packets adhere to the plan, i.e., following the multicast tree at its generation. In a dynamic network, a link may be deactivated before packets arrive. In this situation, the packet will be dropped when it reaches the inactive link. We assume that the network topology change is slower than a packet transmission.
\par
\subsection{Age of Information}
\begin{figure}[!t]
    \centering
    \includegraphics[width = 0.35\textwidth]{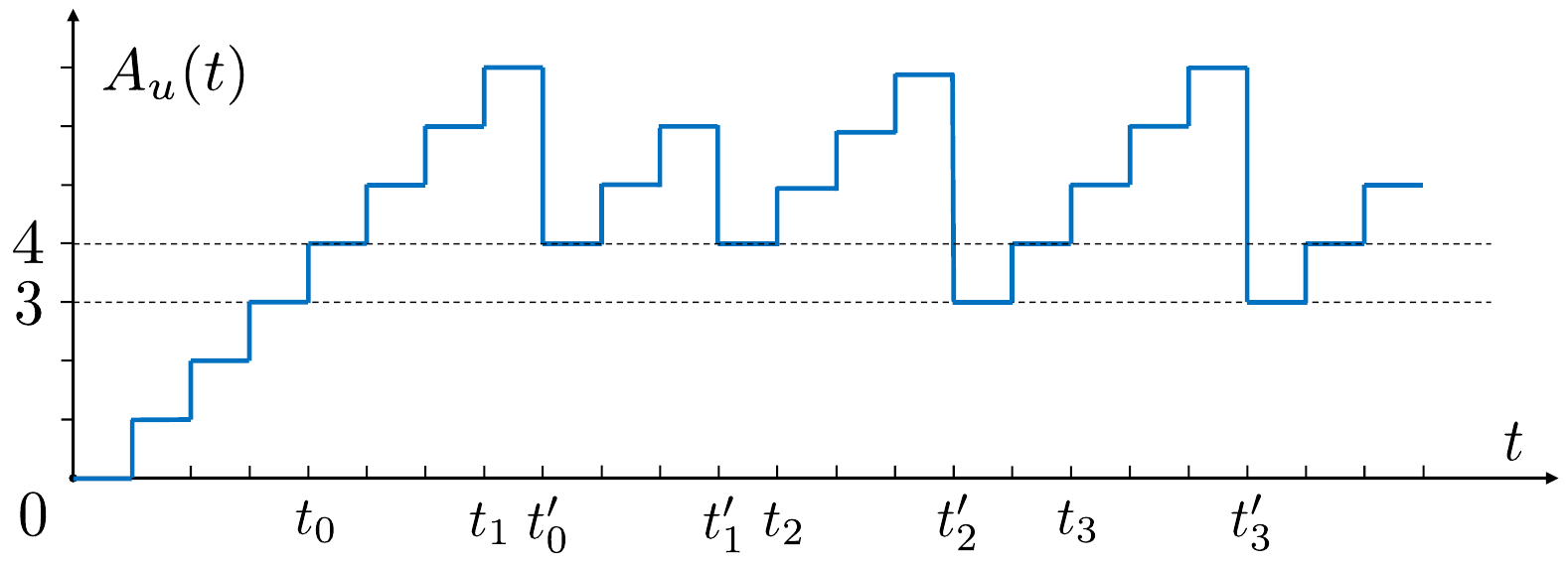}
    \caption{An example of AoI evolution. The $k$-th packet is generated at time $t_k$ and delivered at time $t'_k$. The AoI of destination $u$ is updated upon the reception of a packet update and based on the age of the packet at time $t'_k$.}
    \label{Fig: AoI Curve}
\end{figure}

To analyze the AoI of each destination, we will first need the definition of the \textit{age of packet}. For an arbitrary packet $p\in\mathbb{N}$, the age of packet $p$ is defined as:
\begin{equation}
    \label{AoI of a packet}
    \hat A_p(t) \triangleq t-t_p,\quad \forall t \ge t_p,
\end{equation}
where $t_p\in \mathbb{N}$ represents the time at which packet $p$ is generated and $\hat A_p(t)$ increases linearly in time.
The AoI of destination $u\in\mathcal{U}_t$ is thus defined as:
\begin{equation}
    \label{Def: AoI of destination 1}
    A_u(t) \triangleq
    \begin{cases}
        \ \hat A_p(t), & \text{if } d_{u, p}(t) = 1,\\
        \ A_u(t-1) + 1, & \text{otherwise},
    \end{cases}
\end{equation}
where $d_{u, p}(t)$ is an indicator representing whether packet $p$ arrives at destination $u$ at time $t$. If packet $p$ arrives at destination $u$ at time $t$, $d_{u, p}(t) = 1$; otherwise, $d_{u, p}(t) = 0$. As shown in Fig. \ref{Fig: AoI Curve}, if a destination $u$ receives a packet at time $t$, $A_u(t)$ is updated to be the AoI of the received packet at that time. Otherwise, $A_u(t)$ grows linearly with time.
We note that the AoI process in \eqref{Def: AoI of destination 1} is not inherently bounded if a destination is not updated for an arbitrarily long time. For the theoretical convergence analysis, we consider a finite practical AoI upper bound $A_{\max}$, equivalently using the truncated state $A_u(t)\leftarrow \min\{A_u(t),A_{\max}\}$. This treatment is consistent with practical systems with finite counters, buffers, deadlines, or packet expiration mechanisms. In the experiments, $A_{\max}$ can be chosen sufficiently large so that the truncation is never activated and therefore does not affect the reported numerical results.
The average weighted AoI of the network is defined as:
\begin{equation}
    \label{Def: Average Weighted AoI}
    \overline{A} \triangleq \limsup_{T\to\infty} \frac{1}{T} \sum_{t=0}^{T} \sum_{u\in\mathcal{U}_t} \omega_u A_u(t),
\end{equation}
where $\omega_u\in(0, 1)$ represents the weighted importance of a destination $u\in\mathcal{U}_t$.

Considering the multicast process described above, we derive that the time required for a packet to reach destination $u$ is equal to the number of hops between the source and destination $u$ in a multicast tree $\mathcal{T}_t$, denoted as $h_{\mathcal{T}_t}(u)$. To establish the relation between the AoI and multicast trees, we refer to the following lemma:
\begin{Lemma}[\cite{west2001introduction}]
    \label{Lemma: Unique Path}
    Any two vertices in a tree can be connected by a unique simple path.
\end{Lemma}
From Lemma \ref{Lemma: Unique Path}, we know that $h_{\mathcal{T}_t}(u)$ is unique. Then \eqref{Def: AoI of destination 1} can be rewritten as:
\begin{equation}
    \label{Def: AoI of destination 2}
    A_u(t+h_{\mathcal{T}_t}(u))=
    \begin{cases}
        h_{\mathcal{T}_t}(u), &\text{if } u\in \mathcal{V}^{\mathcal{T}}_t, \\
        A_u(t+h_{\mathcal{T}_t}(u)-1)+1, &\text{otherwise}.
    \end{cases}
\end{equation}
That is, $A_u(t+h_{\mathcal{T}_t}(u))$ will drop to $h_{\mathcal{T}_t}(u)$ when a packet generated at $t$ arrives at destination $u$ at time $t+h_{\mathcal{T}_t}(u)$. Note that $A_u(t)$ depends on two factors: (i) the generation intervals of packets for destination $u$ and (ii) the age of packets upon arrival $h_{\mathcal{T}_t}(u)$, both of which depend on multicast trees. \par

\subsection{Energy Consumption}
We consider the \textit{energy consumption} in multicast processes. The cost (per multicast) $\mathcal{T}_t=\{\mathcal{V}^\mathcal{T}_t, \mathcal{E}^\mathcal{T}_t\}$ is defined as: 
\begin{equation}
    \label{Def: Energy Consumption}
    C(\mathcal{T}_t) \triangleq \sum_{e\in \mathcal{E}(\mathcal{T}_t)}c_{e}.
\end{equation}
This means that if we multicast packets from the source to the selected destinations $\mathcal{U}'_t$ through the multicast tree $\mathcal{T}_t$, the total energy consumption is the sum of the energy costs of all links in $\mathcal{T}_t$. The energy consumption of a network is constrained by a long-term power budget $\overline{C}$, which is represented as follows:
\begin{equation}
    \label{Def: Energy Consumption Constraint}
    \limsup_{T\to\infty} \frac{1}{T}\sum_{t=0}^{T}C(\mathcal{T}_t)\le \overline{C}.
\end{equation}
The presence of an energy consumption constraint will lead to a trade-off between energy consumption and AoI, which is a crucial consideration in multicast networks \cite{xie2021age}. A multicast tree with more destinations may lead to lower AoI but higher energy consumption. Therefore, it is necessary to carefully select the destinations included in each multicast tree, underscoring the significance of multicast scheduling.

\subsection{Age-minimal Multicast Scheduling and Routing}
We now formally define the problem of age-minimal multicast scheduling and routing. Let $\pi \triangleq \{\mathcal{T}_0, \mathcal{T}_1, \dots, \mathcal{T}_T\}$ denotes an update policy. We consider \textit{causal policies}, in which control decisions are made based on the network's history and current information. Specifically, $\mathcal{T}_t$ is determined based on $\{\mathcal{G}_k, \mathcal{T}_k, A_u(k)|0\le k \le t-1, u\in\mathcal{U}_k \}$. Here we focus on stationary policies, where $\pi$ does not change over time. Therefore, the limit superior can be dropped in dynamic networks \cite{sun2017update}. Our goal is to find a policy $\pi$ to minimize the expected time-average AoI while satisfying the energy constraint. This problem can be formulated as follows:
\begin{subequations}
    \label{Original Problem}
    \begin{align}
        \textbf{OP:}\quad \min_{\pi} &\  \lim_{T\to\infty} \frac{1}{T} \mathbb{E}_{\pi}
        \left[\sum_{t=0}^{T} \sum_{u\in\mathcal{U}_t} \omega_u A_u(t)\right],\\
        \label{Long-Time Energy Constraint}
        \text{s.t.} &\  \lim_{T\to\infty}\frac{1}{T}\mathbb{E}_{\pi}
        \left[\sum_{t=0}^{T} C(\mathcal{T}_t) \right]\le \overline{C}.
    \end{align}
\end{subequations}
The dual problem can be formulated as:
\begin{equation}
    \label{Dual Problem} 
    \begin{aligned}
        \textbf{DP:} \quad \max_{\lambda\geq 0} & \ \min_{\pi} \lim_{T\to\infty} \frac{1}{T} \mathbb{E}_{\pi} \left[
         \sum_{t=0}^{T} \sum_{u\in\mathcal{U}_t} \omega_u A_u(t) + \right. \\
        & \quad\quad\quad\quad\quad\quad \left.
         \lambda \left(\sum_{t=0}^{T}C(\mathcal{T}_t) - T \overline{C}\right)\right],
    \end{aligned}
\end{equation}
where $\lambda$ is the Lagrangian multiplier corresponding to \eqref{Long-Time Energy Constraint}. Inspired by Theorem 3.6 of \cite{altman2021constrained}, we establish the following lemma to formalize the equivalence between \textbf{OP} and \textbf{DP} as follows:
\begin{Lemma}
    The optimal solution of the dual problem (Eq. \ref{Dual Problem}) is indeed the optimal solution of the primal problem (Eq. \ref{Original Problem}) under the following Slater condition:
    \begin{equation}
        \lim_{T\to\infty} \frac{1}{T} \mathbb{E}_\pi \left[ \sum_{t=0}^{T} C(\mathcal{T}_t) \right] < \overline{C}, \exists \pi \in \Pi,
    \end{equation}
    where $\Pi$ is the set of all stationary policies.
\end{Lemma}
The above condition can be satisfied by choosing a policy that always selects the empty action. 
The problem above encounters several challenges: (i) Policy $\pi$ couples the multicast scheduling and routing decisions; (ii) the solution space of multicast trees is exponential in the number of destinations, making it computationally infeasible to solve the problem directly; (iii) the network dynamics are unknown, making it challenging to obtain the optimal solutions. Therefore, we reformulate the problem into subproblems and solve them sequentially, corresponding to the cross-layer design.
\section{Problem Reformulation}
Our main approach is to decompose the problem \textbf{DP} in \eqref{Dual Problem} into a scheduling subproblem and a tree-generating subproblem. The scheduling subproblem is a primary problem that focuses on selecting the destinations that need to be updated in each time slot, typically at a higher layer, such as the transport layer. The tree-generating subproblem aims to obtain an optimal multicast tree for the chosen destinations, typically operating at a lower layer, such as the network layer. The Lagrangian multiplier $\lambda$ acts as a signal between two layers. 
\subsection{Problem Decomposition}
We start with considering  two subproblems of \textbf{DP} in \eqref{Dual Problem} as follows:
\begin{Definition}[\textbf{Scheduling Subproblem}]
    Given a network $\mathcal{G}_t$, find a set of destination sets $\mathcal{U}'_t\subseteq\mathcal{U}_t$ to solve:
    \begin{equation}
        \label{Scheduling Problem}
        \begin{aligned}
            \textbf{P1:}\quad
            \max_{\lambda\geq 0} \ 
            \min_{\mathcal{U}_t'} \lim_{T\to\infty} \frac{1}{T} \mathbb{E}_{\mathcal{U}_t'}\left[\sum_{t=0}^T g(\lambda, \mathcal{U}_t')\right],
        \end{aligned}
    \end{equation}
    where 
    $g(\lambda, \mathcal{U}'_t)$ is defined by the following tree-generating subproblem:
\end{Definition}
\begin{Definition}[\textbf{Tree-generating Subproblem}]
    Given a network $\mathcal{G}_t$ and a set of destinations $\mathcal{U}'_t \subseteq \mathcal{U}_t$, obtain a multicast tree ${\mathcal{T}=\{\mathcal{V}^{\mathcal{T}}, \mathcal{E}^{\mathcal{T}}\}}$ such that:
    \begin{equation}
        \label{Tree-generating Problem}
        \begin{aligned}
            \textbf{P2:} \ 
            g(\lambda, \mathcal{U}'_t)=\!\! \min_{\mathcal{T} \in \Omega(\mathcal{U}'_t)} 
         \mathbb{E}_{\mathcal{T}}
            \left[ \sum_{u\in\mathcal{U}'_t} \omega_u \sum_{t'=0}^{T}  A_u(t') \right.
            \\
             \left. + \lambda \left(\sum_{t'=0}^{T}C(\mathcal{T}_{t'}) - T \overline{C} \right) \right].
        \end{aligned}
    \end{equation}
\end{Definition}
Here, $\sum_{t'=0}^{T} A_u(t')$ measures the long-term AoI incurred by multicast tree $\mathcal{T}$. This metric will be further analyzed later, as the AoI is not updated immediately—a point discussed in Section \ref{Sec: System Model}. 
\begin{Remark}
    We observe that \textbf{P2} reduces to the standard STP: For an undirected graph $\mathcal{G}=(\mathcal{V}, \mathcal{E})$, the STP is formulated as
    \begin{align}
        \label{Problem: STP}
        \textbf{STP:} \quad \min_{\mathcal{T}\subseteq \mathcal{G}}\quad C(\mathcal{T})\triangleq \sum_{e\in\mathcal{E}(\mathcal{T})}c_e, 
    \end{align}
    which is among Karp's 21 most fundamental NP-hard problems  \cite{karp2010reducibility}. For the classical \textbf{STP}, an approximation ratio of $\alpha = \ln(4) + \epsilon \approx 1.39$ can be achieved in polynomial time \cite{byrka2013steiner}, which is the currently best-guaranteed performance known for polynomial complexity.
\end{Remark}

We can observe the equivalence between the two subproblems and the problem \textbf{DP}. The scheduling subproblem \textbf{P1} aims to select a subset of destinations to update at each time slot. In contrast, the tree-generating subproblem \textbf{P2} aims to generate multicast trees for the selected destinations. The two subproblems are coupled through the Lagrangian multiplier $\lambda$. To deal with the complexity of both subproblems, we model them as \textit{Markov Decision Processes (MDPs)}, which RL algorithms can solve.

\subsection{Scheduling MDP}
The scheduling problem \textbf{P1} is a primary problem that aims to optimally select a subset of destinations to update at each time slot. Therefore, we naturally reformulate it as an MDP, denoted as $\mathcal{M}_1=\{\mathcal{S}_1, \mathcal{A}_1, \mathbb{P}_1, r_1\}$ and described as follows:
\begin{itemize}
    \item \textbf{State Space:} Let $\hat{\mathcal{V}}=\max_{t\in\mathbb{N}_0}|\mathcal{V}_t|$ denote the max number of nodes over $t$, the state $s_t$ is defined as:
    \begin{equation}
        s_t =\{\mathcal{G}_t, \mathbf{x}_t\}, s_t \in \mathbb{R}^{\hat{\mathcal{V}}^2+6\hat{\mathcal{V}}},
    \end{equation}
    where $\mathbf{x}_t\in\mathbb{R}^{6\hat{\mathcal{V}}}$ denotes node features at time $t$, including one-hot encoding of node type (3 dims), weighted node importance ($\omega_u$) (1 dim), AoI of the node ($A_u(t)$) (1 dim), and number of transmitting packets (1 dim).
    The topology $\mathcal{G}_t$ is represented by an adjacency matrix with $\hat{\mathcal{V}}^2$ elements. 
    \item \textbf{Action Space:} The action space is the power set of $\mathcal{U}_t$, i.e., the set of
    all subsets of $\mathcal{U}_t$, denoted by:
    \begin{equation}
        \label{S: Actions}
        \mathcal{A}_1 = {\rm Pow}(\mathcal{U}_t).
    \end{equation}
    where ${\rm Pow}(\mathcal{X})$ denotes the power set of $\mathcal{X}$, i.e., the set of all subsets of $\mathcal{X}$.
    The action of $\mathcal{M}_1$ is denoted as $a_t\in\mathcal{A}_1$. It follows that the cardinality $|\mathcal{A}_1|$ is $2^{|\mathcal{U}_t|}$, which leads to the curse of dimensionality \cite{barto2003recent}.
    \item The transition probability is denoted as $\mathbb{P}_1$, which is unknown a priori in practice.
    \item \textbf{Rewards:} The reward function $r_1$ is defined as:
    \begin{equation}
        \label{S: Reward Function}
        r_1(s_t, a_t) = g(\lambda, \mathcal{U}'_t),
    \end{equation}
    where $g(\lambda, \mathcal{U}'_t)$ is the objective function of problem \textbf{P2}.
\end{itemize}
Note that $g(\lambda, \mathcal{U}'_t)$ is unknown at time $t$ (see \eqref{Tree-generating Problem}). Therefore, we need to find an equivalent transformation of \textbf{P2} to make $r_1(s_t, a_t)$ computable, as shown in the following lemma:
\begin{Lemma}\label{L2}
    \label{Lemma: Another form of P2}
    Given network $\mathcal{G}_t$ and a set of destinations $\mathcal{U}'_t\subseteq\mathcal{U}_t$, the Problem \textbf{P2} is equivalent to the following problem:
    \begin{equation}
        \begin{aligned}
        \textbf{P2-B:} \ g(\lambda, \mathcal{U}'_t)= & \!\! \max_{\mathcal{T} \in \Omega(\mathcal{U}'_t)} \sum_{u\in\mathcal{U}'_t} \omega_u \left(1-\frac{h_{\mathcal{T}}(u)}{\hat h_{\mathcal{G}_t}}\right) A_u(t)
        \\
        & \quad\quad\quad\quad\quad
        - \lambda ( C(\mathcal{T}) - \overline{C} ),
        \end{aligned}
    \end{equation}
    where $h_{\mathcal{T}}(u)$ is the number of hops of the path
    between the source and destination $u$ in $\mathcal{T}$, which is unique because of Lemma \ref{Lemma: Unique Path}; $\hat{h}_{\mathcal{G}_t}$ is the maximum number of hops for all node pairs of $\mathcal{G}_t$, which is also called the length of $\mathcal{G}_t$.
\end{Lemma}
The proof of Lemma \ref{L2} is given in the Appendix. 
We observe that \textbf{P2-B} in Lemma \ref{L2} remains a CO problem and is, therefore, NP-hard, necessitating further approximation techniques to achieve efficient solutions with high optimality.

\section{Tree Generator-based Multicast Scheduling} \label{TGMS}
As mentioned above, we face two challenges: (i) the curse of dimensionality of MDP $\mathcal{M}_1$; (ii) the NP-hardness of the tree-generating subproblem \textbf{P2-B}. We first reformulate the tree-generating subproblem \textbf{P2-B} as an MDP to address these challenges. Then, we propose a Tree Generator-based Multicast Scheduling (TGMS) algorithm. Our proposed approach consists of a tree generator and a scheduler, which utilize graph embedding methods and DRL techniques. The proposed graph embedding method lowers the dimensions from $\mathcal{O}(\hat{\mathcal{V}}^2)$ to $\mathcal{O}(\hat{\mathcal{V}})$ with key information extracted.

\subsection{Reformulating STP as an MDP}
Our approach to tackling the subproblem \textbf{P2-B} is mainly motivated by the greedy meta-algorithm design in \cite{khalil2017learning}. 

\textit{Procedure Outline:} We present the outline of approximately solving the STP below.
Since a tree with $n$ nodes has exactly $n-1$ links, we can generate a multicast tree by incrementally selecting nodes and links. Specifically, we initiate with an empty subgraph $\mathcal{P}$, which is the partial solution of a multicast tree. Then at each step, we greedily select a node $v$ and an edge to $\mathcal{P}$.
\begin{figure}[!t]
    \centering
    \includegraphics[width = 0.35\textwidth]{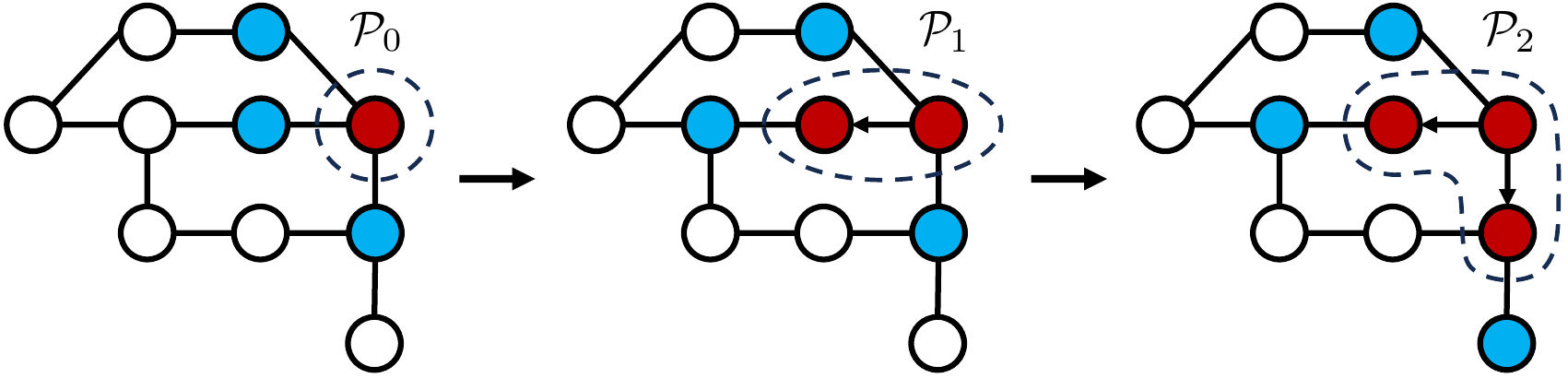}
    \caption{An example of the tree-generating process. The red nodes denote the selected nodes of a partial solution, and the blue nodes denote the selectable nodes. At each step, a node $v$ is selected and added to the partial solution. The destinations are not shown for simplicity.}
    \label{Fig: Partial Solution}
\end{figure}
The node $v$ is selected from a heuristic quality function, denoted as $Q(\mathcal{P}, v)$. It represents the quality of $\mathcal{P}$ after adding $v$ to $\mathcal{P}$. An example of the tree-generating process is shown in Fig. \ref{Fig: Partial Solution}. The above process can be described as follows:
\begin{equation}
    \mathcal{P} \gets (\mathcal{P}, v^*), \text{where } v^*=\mathop{\arg\max}\limits_{v\in\bar{\mathcal{P}}} Q(\mathcal{P},v).
\end{equation}
As outlined above, the tree-generation process can be interpreted as a sequential decision-making problem, where nodes and links are included step by step, which can be formulated as an MDP.
Hence, we further reformulate \textbf{P2-B} as follows. \par

\textit{STP-MDP.}
To measure the quality of a partial solution $\mathcal{P}$, we can transform the objective of \textbf{P2-B} into a quality function by substituting $\mathcal{P}$ for $\mathcal{T}$. Next, we formally define the tree-generating MDP. We introduce a virtual timescale $\tau$ and a partial solution $\mathcal{P}_\tau=\{\mathcal{V}^{\mathcal{P}}_\tau, \mathcal{E}^{\mathcal{P}}_\tau\}$ with a source node at $\tau=0$. Recall that problem \textbf{P2-B} is influenced by the selected destinations $a_t \subseteq \mathcal{U}_t$ (see \eqref{S: Actions}). Given $a_t$, an induced MDP $\mathcal{M}_2(a_t)=\{\mathcal{S}_2, \mathcal{A}_2, \mathbb{P}_2, r_2\}$ is defined as follows:
\begin{itemize}
    \item \textbf{State Space:} The state $s_{\tau}$ is the partial solution $\mathcal{P}_\tau$, i.e.:
    \begin{equation}
        \label{TG: State}
        s_{\tau}=\{\mathcal{P}_\tau\}, s_{\tau}\in \mathbb{R}^{\hat{\mathcal{V}}^2 + \hat{\mathcal{V}}}.
    \end{equation}
    The state $s_\tau$ includes two indicators for $\mathcal{V}^{\mathcal{P}}_\tau$ and $\mathcal{E}^{\mathcal{P}}_\tau$, respectively. When $a_t$ is covered by $\mathcal{P}_\tau$, $s_{\tau}$ will be a terminal state\footnote{For $\gamma<1$, an MDP with a terminal state can be embedded into a standard discounted infinite-horizon MDP by treating the terminal state as an absorbing state with zero (or fixed) reward thereafter \cite{puterman2014markov}. For the undiscounted case $\gamma=1$, this interpretation requires finite/proper termination; otherwise, improper policies may cause non-terminating trajectories.} and $\mathcal{P}_\tau$ is the generated multicast tree.
    \item \textbf{Action Space:} The action space is defined as:
    \begin{equation}
        \label{TG: Action Space}
        \mathcal{A}_2 = \{v|v\in \mathcal{V}_t, v\notin \mathcal{V}^{\mathcal{P}}_\tau, \exists u\in \mathcal{V}^{\mathcal{P}}_\tau, (u, v)\in \mathcal{E}_t\}.
    \end{equation}
    That is, the action $a_\tau\in\mathcal{A}_2$ is to choose a node $v$ that is not included in $\mathcal{P}_\tau$ and the node $v$ should be connected to at least one node $u$ in $\mathcal{P}_\tau$. In other words, each action $a_{\tau}\in\mathcal{A}_2$ is a neighbor of the current partial solution $\mathcal{P}_\tau$.
    \item \textbf{Transition:} The transition is deterministic here and corresponds to tagging the selected node and an edge, satisfying:
    \begin{align}
        s_{\tau+1}= s_{\tau}\cup \{a_\tau, (v^*_\tau, a_\tau)\}.
    \end{align}
    \item \textbf{Rewards:}
    The quality function $q_2(s_\tau)$ can be defined as:
    \begin{equation}
        \label{TG: Quality Function}
        \begin{aligned}
            q_2(s_\tau) =\! \sum_{u\in \mathcal{U}'_t \cap \mathcal{V}^{\mathcal{P}}_\tau}
            \omega_u \left(1-\frac{h_{\mathcal{P}_\tau}(u)}{\hat h_{\mathcal{G}_t}}\right) A_u(t) \\
            -\lambda(C(\mathcal{P}_\tau)-\overline{C}),
        \end{aligned}
    \end{equation}
    where $h_{\mathcal{P}_\tau}(u)$ is the number of hops between the source and destination $u$ in ${\mathcal{P}}_\tau$, and $\hat{h}_{\mathcal{G}_t}$ is the diameter of $\mathcal{G}_t$. Subsequently, the reward function $r_2$ is defined as:
    \begin{equation}
        \label{TG: Reward Function}
        r_2(s_\tau, a_\tau) = q_2(s_{\tau+1}) - q_2(s_{\tau}).
    \end{equation}
\end{itemize}
When considering a candidate $a_\tau$ for inclusion in the partial solution and choose the link with the lowest cost from the set of candidate links that connect the nodes $\mathcal{V}_{\tau}^\mathcal{P}$ and node $a_\tau$:
    \begin{equation}
        \label{Eq: TG-Link Selection}
        (v^*_\tau, a_\tau) = \mathop{\text{arg min}}_{v\in\mathcal{V}^{\mathcal{P}}_\tau , (v, a_\tau)\in\mathcal{E}^{\mathcal{P}}_\tau} C_{v, a_\tau}.
    \end{equation}
This approach effectively reduces the complexity of $\mathcal{A}_2$ while constraining the cost of $\mathcal{P}_\tau$. The following proposition ensures that the generated subgraph is a multicast tree of $\mathcal{U}'_t$ and the generated multicast tree is a feasible solution:
\begin{Proposition}
    \label{Proposition: Multicast Tree}
    At a terminal state $s_{\tau'}$ of $\mathcal{M}_2(a_t)$, $\mathcal{P}_{\tau'}$ is ensured to be a multicast tree including $a_t$.
    \begin{proof}
        The proof sketch is to show that $\mathcal{P}_{\tau'}$ is connected, acyclic, and includes $a_t$. The detailed proof is provided in the appendix.
    \end{proof}
\end{Proposition}
\begin{Remark}

    While we build upon the general idea of \cite{khalil2017learning}, the original method is not directly applicable to solve the variant of Steiner tree problem in our scenario. The main challenges are: (i) training instability from dynamic action spaces. In the iterative process of constructing the multicast tree ($\mathcal{M}_2$), the set of valid actions (i.e., nodes that can be added to the partial solution $\mathcal{P}_\tau$) changes at every step, leading to unstable training processes; (ii) limited generalization to unseen graphs. The training instability described above makes it difficult for the model to learn the underlying structural properties of the graph, which is essential for generalization.
    Moreover, another arising challenge in our formulated problem is the interdependence between the routing and scheduling, in contrast to static combinatorial optimization problems.

\end{Remark}

\begin{figure}[!t]
    \centering
    \includegraphics[width = 0.45\textwidth]{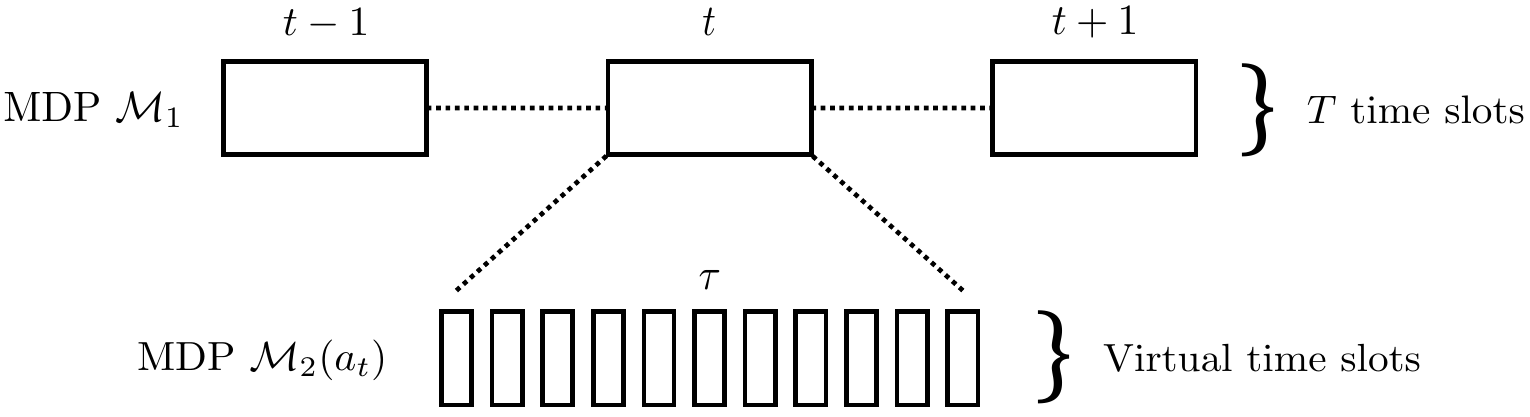}
    \caption{Temporal relation between two MDPs. $\mathcal{M}_1$ selects a subset of destinations $a_t\subseteq\mathcal{U}_t$ at each time slot $t$, which is utilized by $\mathcal{M}_2(a_t)$ to generate multicast trees. 
    $\tau$ is a virtual timescale that discretizes the generation process of a multicast tree. }
    \label{Fig: MDP Decomposition}
\end{figure}

\textbf{Relation between Scheduling and Tree-Generating MDPs.} Here, we analyze the relation between $\mathcal{M}_1$ and $\mathcal{M}_2(a_t)$. The temporal relation is illustrated in Fig. \ref{Fig: MDP Decomposition}. This formulation coincides with the core idea of hierarchical RL (e.g., the MAXQ decomposition \cite{dietterich2000hierarchical}), which decomposes a complex MDP into a set of simpler MDPs, including a root MDP. In our case, $\mathcal{M}_1$ is the root MDP, and $\mathcal{M}_2$ is a sub-MDP. In other words, the scheduling subproblem is a primitive subtask, and the tree-generating subproblem is an invoked subtask. The hierarchical structure can effectively address the challenges of coupled decision variables.
The equivalence between $\mathcal{M}_2(a_t)$ and \textbf{P2} can be described as follows:
\begin{Lemma}
    \label{Lemma: equivalence between M2 and P2}
    Given a subset of destinations $\mathcal{U}'_t\subseteq\mathcal{U}_t$, assume the selected destinations are reachable from the source and the tree-generation trajectory is proper, i.e., it terminates after finitely many node additions. Then, with discount factor $\gamma=1$, solving $\mathcal{M}_2(\mathcal{U}'_t)$ is equivalent to solving problem \textbf{P2}.
    \begin{proof}
        Under the proper-termination condition, each step adds a previously unselected node to the partial solution and the finite graph admits only finitely many such additions before termination. The cumulative reward of $\mathcal{M}_2(\mathcal{U}'_t)$ can be denoted as:
        \begin{equation}
            \label{TG: Cumulative Reward}
            \begin{aligned}
                & R_2 = \sum_{\tau} \gamma^{\tau} r_2(s_\tau, a_\tau) \xlongequal{\gamma=1} q_2(s_{\tau'}) - q_2(s_0) \\
                & = \sum_{u\in \mathcal{U}'_t}
                \omega_u \left(1-\frac{h_{\mathcal{P}_{\tau'}}(u)}{\hat h_{\mathcal{G}_t}}\right) A_u(t) -\lambda(C(\mathcal{P}_{\tau'})-\overline{C}),
            \end{aligned}
        \end{equation}
        where $s_{\tau'}$ is a terminal state. The equality for $\gamma=1$ follows from the telescoping sum induced by the shaped reward $r_2(s_\tau,a_\tau)=q_2(s_{\tau+1})-q_2(s_\tau)$. Note that $\mathcal{P}_{\tau'}$ for $\mathcal{M}_2(\mathcal{U}'_t)$ can be written as $\mathcal{T}_t$ and the objective of problem \textbf{P2-B} is the same as $R_2$. In addition, the two problems have the same solution space. Hence, we conclude that solving $\mathcal{M}_2(\mathcal{U}'_t)$ is equivalent to solving \textbf{P2-B}. From Lemma \ref{Lemma: Another form of P2}, the proof is completed. In the implementation, we use $\gamma=0.99$ as a discounted training surrogate for stable value estimation; this discounted training choice should be interpreted as an approximation to the finite undiscounted equivalence above, rather than as the exact telescoping identity.
    \end{proof}
\end{Lemma}

\subsection{Representation: Graph Embedding Methods}
To overcome the curse of dimensionality $\mathcal{M}_1$ and $\mathcal{M}_2(a_t)$, it is crucial to represent the graph embedding. Graph embedding methods (e.g., GNNs) are valuable for extracting graph information and have demonstrated their efficacy in various CO problems \cite{khalil2017learning}. Some studies have focused on the attention mechanism in GNNs, which has the advantage of selectively aggregating information from neighbors (e.g., \cite{velivckovic2017graph}). Those methods that utilize the attention mechanism are called Graph Attention Networks (GATs). Among them, the GATv2 \cite{brody2021attentive} is a representative model that can effectively capture the importance of nodes in a graph.
\rev{We introduce the following Normalized GAT (NGAT) to control the scale of the embedding update and analyze its non-expansiveness under fixed-attention conditions. Stability of the embedding-dependent attention model is evaluated empirically.}\footnote{$\text{LeakyReLU}(x) = \max(\alpha x, x)$ where $\alpha$ is a hyperparameter.}:
\begin{subequations}
    \label{Eq: NGAT}
    \begin{equation}
        \label{Eq: NGAT node importance}
        \phi(\mathbf{h}_i, \mathbf{h}_j) = \mathbf{a}^\text{T} \text{LeakyReLU}(\mathbf{W}_1 \mathbf{h}_i + \mathbf{W}_2 \mathbf{h}_j + \mathbf{W}_e\mathbf{e}_{i, j}),
    \end{equation}
    \begin{equation}
        \label{Eq: NGAT attention coefficients}
        \alpha_{ij} = \frac{\exp(\phi(\mathbf{h}_i, \mathbf{h}_j))}{\sum_{k\in\mathcal{N}_i \cup \{i\}}\exp(\phi(\mathbf{h}_i, \mathbf{h}_k))},
    \end{equation}
    \begin{equation}
        \label{Eq: NGAT attention sum}
        \begin{aligned}
            f_{\text{NGAT}}(\mathbf{h}_i, \mathbf{x}) &= \frac{1}{\rev{\Vert\mathbf{W}_1\Vert_2}}(\alpha_{ii}(\mathbf{W}_1 \mathbf{h}_i+\mathbf{W}_3\mathbf{x}_i) \\ 
            & \quad\quad\quad\quad + \sum_{j\in\mathcal{N}_i}\alpha_{ij}(\mathbf{W}_1\mathbf{h}_j+\mathbf{W}_3\mathbf{x}_j)).
        \end{aligned}
    \end{equation}
\end{subequations}
where $\mathbf{W}_1$, $\mathbf{W}_2$, and $\mathbf{W}_e$ are learnable parameters, $\mathbf{x}=\{\mathbf{x}_i\}$ denotes the node features, $\mathbf{e}_{i, j}$ denotes the edge features of edge $(i, j)$, and $\mathbf{h}_i\in\mathbb{R}^d$ denotes the embedding vector of node $i$. When updating the node embedding $\mathbf{h}_i$, NGAT aggregates information from neighbors and itself, which is weighted by the attention coefficients $\alpha_{ij}$. By utilizing NGAT, we can reduce the representation of the state space from $\mathcal{O}(\hat{\mathcal{V}}^2)$ to $\mathcal{O}(\hat{\mathcal{V}})$, effectively mitigating the curse of dimensionality. \rev{Here and throughout the NGAT analysis, vector norms are Euclidean, $\|\cdot\|_2$ for a matrix denotes its spectral norm, and $\|\mathbf{W}_1\|_2>0$ is required; $\mathbf{W}_3$ is also a learnable parameter. The score function and row-wise softmax do not generally produce symmetric attention, and attention depends on the current embeddings. The theorem below concerns a common fixed symmetric row-stochastic attention matrix, rather than every attention matrix generated by the learned model.} \par
\rev{For a fixed finite graph with $n$ nodes, arrange the embeddings as the rows of $\mathbf{H}\in\mathbb{R}^{n\times d}$. We measure deviations of the complete embedding matrix by the Frobenius metric}
\begin{equation}
    \rev{d_F(\mathbf{H},\mathbf{H}')=\|\mathbf{H}-\mathbf{H}'\|_F
    =\left(\sum_{i\in\mathcal{V}}\|\mathbf{h}_i-\mathbf{h}'_i\|_2^2\right)^{1/2}.}
\end{equation}
\rev{Unlike the norm of the sum of node deviations, this metric is zero only when all corresponding node embeddings agree.}
\begin{Theorem}[Conditional non-expansiveness of NGAT]
    \label{Theorem: Contraction Mapping of NGAT}
    \rev{Fix the graph, node features $\mathbf{X}$, and weight matrices, with $\|\mathbf{W}_1\|_2>0$. Let $\mathbf{A}=[\alpha_{ij}]\in\mathbb{R}^{n\times n}$ have nonnegative entries, zero entries outside the neighborhood including self-loops, $\mathbf{A}=\mathbf{A}^{\top}$, and $\mathbf{A}\mathbf{1}=\mathbf{1}$. Treat $\mathbf{A}$ as fixed and use the same $\mathbf{A}$ for both inputs. Then the matrix update $F_{\mathbf{A}}$ induced by \eqref{Eq: NGAT attention sum} satisfies}
    \begin{equation}
        \rev{\|F_{\mathbf{A}}(\mathbf{H};\mathbf{X})
        -F_{\mathbf{A}}(\mathbf{H}';\mathbf{X})\|_F
        \leq \|\mathbf{H}-\mathbf{H}'\|_F}
    \end{equation}
    \rev{for every pair of embedding matrices of the same dimensions.}
\end{Theorem}
The proof of the above theorem can be found in the appendix.
\begin{Remark}
    \rev{This result controls perturbations of the complete embedding matrix for a common fixed attention matrix. For a sequence of layers, the bound composes provided that, at every layer, the two trajectories use the same graph, features, weight matrices, and attention matrix satisfying the theorem's assumptions; these quantities may vary across layers. Under these conditions, their Frobenius distance does not increase with depth. Non-expansiveness alone does not imply convergence of either trajectory or uniqueness of a fixed point. In learned NGAT, $\mathbf{A}(\mathbf{H})$ and $\mathbf{A}(\mathbf{H}')$ can differ and need not be symmetric. Thus the theorem does not establish non-expansiveness of that embedding-dependent map or convergence of RL training. These behaviors are assessed empirically.}
\end{Remark}

\subsection{System Architecture}
\begin{figure}[!t]
    \centering
    \includegraphics[width=0.44\textwidth]{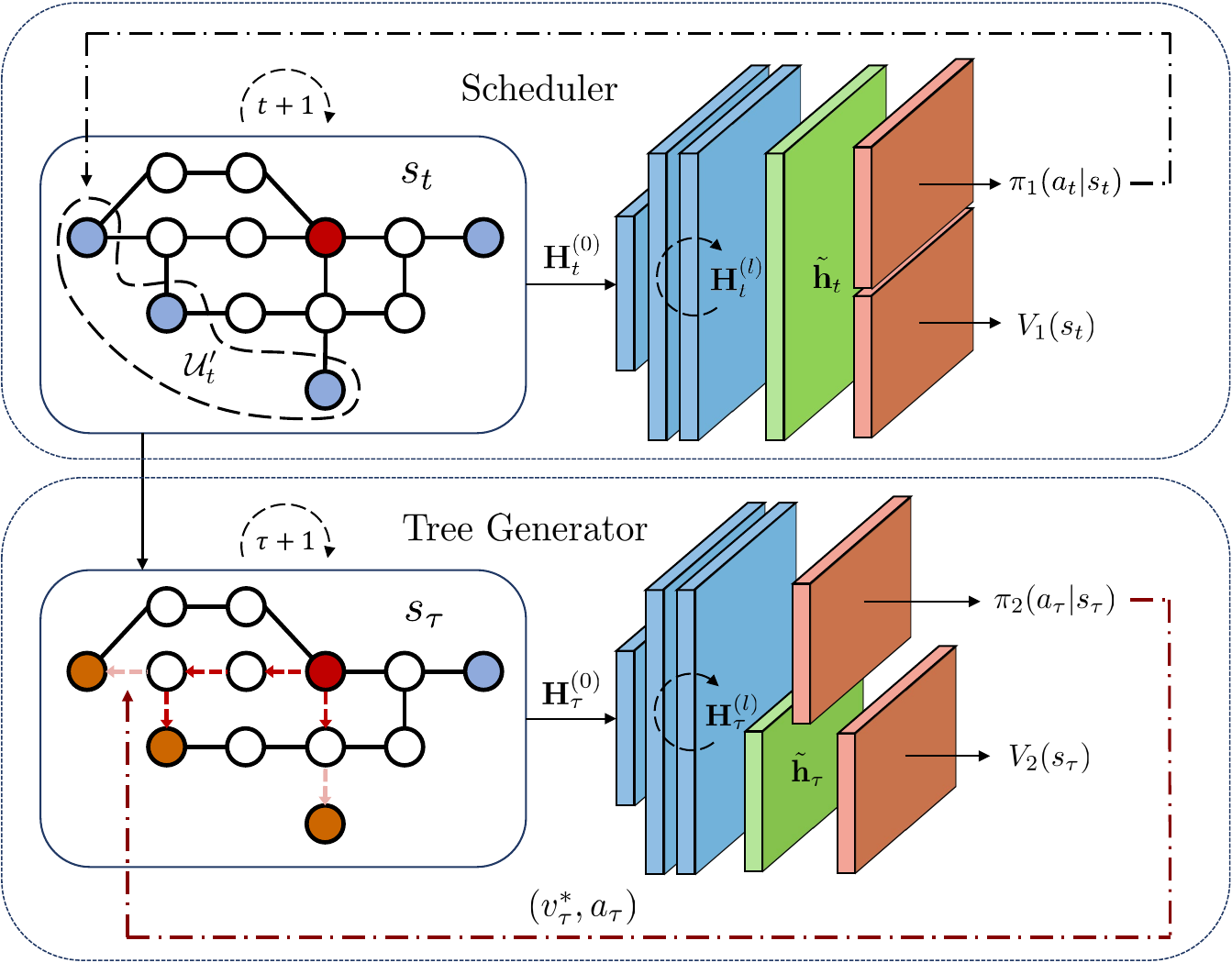}
    \caption{The system architecture of TGMS. A scheduler is performed to select a set of destinations, which are utilized by the tree generator to generate a multicast tree. The initial graph embedding is fed into NGAT layers (blue blocks) to be iteratively updated. The outputs are passed through a pooling layer (green block) to aggregate global information and then forwarded to two distinct heads (red blocks) for policy and value predictions.}
    \label{Fig: TGMS}
\end{figure}

The \textit{system architecture} of TGMS is illustrated in Fig. \ref{Fig: TGMS}.
At each time slot $t$, the scheduler selects a subset of destinations $\mathcal{U}'_t$ from $\mathcal{U}_t$. The tree generator is then invoked to generate a multicast tree $\mathcal{T}_t$ for $\mathcal{U}'_t$. The graph is first embedded with NGAT layers (blue blocks) to capture the spatial structure of the network. The outputs are then passed through a pooling layer (green block) to aggregate global information. The resulting graph embeddings are fed into an actor and a critic (red blocks) for policy and value predictions.

Here, we formally describe the \textit{forwarding processes}. Recall that the action space of $\mathcal{M}_1$ is excessively large (see Eq. \eqref{S: Actions}), making it impractical for the efficient exploration of classical MDP algorithms. In addition, the size of $\mathcal{U}_t$ varies between different networks, further complicating the problem. To address the above challenges, we consider a continuous action output. Specifically, we predict a Bernoulli distribution for each node to overcome the ``Curse of Dimensionality". This allows the scheduler to effectively select an action from an exponentially large space in a linearly computationally tractable manner. These predictions are then used to sample the destinations. The forwarding process of the scheduler can be summarized as:
\begin{subequations}
    \label{Eq: Scheduler}
    \begin{align}
        & \label{Eq: S-Combine}
        \mathbf{H}_t^{(0)} = s_t,
        \mathbf{H}_t^{(l+1)} = f_{\text{NGAT}}(\{\mathbf{h}_{t, i}^{(l)}\}_{i\in\mathcal{V}_t}, \mathbf{x}_t), \\
        & \label{Eq: S-Pooling}
        \tilde{\mathbf{h}}_t = \frac{1}{|\mathcal{V}|} \sum_{i=1}^{|\mathcal{V}|} \mathbf{h}^{(L)}_t, \\
        & \label{Eq: S-Policy head}
        \pi_1(a_t|s_t) =  \text{Sigmoid}(\mathbf{W}_1 \sigma(\mathbf{W}_2 \tilde{\mathbf{h}}_t)), \\
        & \label{Eq: S-Critic head}
        V_1(s_t) = \mathbf{W}_3 \sigma(\mathbf{W}_4\tilde{\mathbf{h}}_t),
    \end{align}
\end{subequations}
where $\sigma$ denotes an activation function. The graph embedding is updated by the NGAT defined by \eqref{Eq: NGAT node importance}-\eqref{Eq: NGAT attention sum}. The outputs are passed through a pooling layer aggregating global information (see \eqref{Eq: S-Pooling}). The resulting graph embeddings are fed into: (i) a policy head responsible for selecting destinations (see \eqref{Eq: S-Policy head}); and (ii) a critic head predicting the value of the state $s_t$ (see \eqref{Eq: S-Critic head}). The graph embedding layers and the heads together constitute the parameterized RL agent.
\par
For the tree generator, we employ graph embedding methods similar to those used in the scheduler. The forwarding process of the tree generator is described as follows:

\begin{subequations}
    \label{Eq: Tree Generator}
    \begin{align}
        & \label{Eq: T-Combine}
        \mathbf{H}_\tau^{(0)} = s_t,
        \mathbf{H}_\tau^{(l+1)} = f_{\text{NGAT}}(\{\mathbf{h}_{\tau, i}^{(l)}\}_{i\in\mathcal{V}_t}, \mathbf{x}_t), \\
        & \label{Eq: T-Pooling}
        \tilde{\mathbf{h}}_\tau = \frac{1}{|\mathcal{V}|} \sum_{i=1}^{|\mathcal{V}|} \mathbf{h}^{(L)}_\tau, \\
        & \label{Eq: TG-Policy head}
        \pi_2(a_\tau|s_\tau) = \log\text{softmax}(\mathbf{W}'_1\sigma(\mathbf{W}'_2\mathbf{H}_\tau^{(L)})), \\
        & \label{Eq: TG-Value head}
        V_2(s_\tau) = \mathbf{W}'_3 \sigma(\mathbf{W}'_4\tilde{\mathbf{h}}_\tau),
    \end{align}
\end{subequations}
where the policy $\pi_2(a_\tau|s_\tau)$ is masked to ensure valid actions. \par
The \textit{inference process} of the TGMS is described in Algorithm \ref{PC: TGMS}. 
\begin{algorithm}[!t]
    \caption{Tree Generator-based Multicast Scheduling (TGMS)}
    \label{PC: TGMS}
    \begin{algorithmic}[1]
    \Require Trained scheduler $\bm{\theta}^*_1$, trained tree generator $\bm{\theta}^*_2$, network $\mathcal{G}_0$ with node features $\mathbf{x}_0$ and edge features $\mathbf{e}_0$.
    \Ensure Scheduling decisions $\bm{\mathcal{U}}'=\{\mathcal{U}'_0, \mathcal{U}'_1, \dots\}$ and multicast trees $\bm{\mathcal{T}}=\{\mathcal{T}_0, \mathcal{T}_1, \dots\}$.
        \State $t \leftarrow 0$.
        \While{$t < T$}
            \State Get state $s_t=\{\mathcal{G}_t, \mathbf{x}_t\}$.
            \State Calculate graph embedding $\tilde{\textbf{h}}_t$ and compute $\pi_1(a_t, s_t)$ from \eqref{Eq: Scheduler} based on $\bm{\theta}^*_1$. \Comment{$O(|\mathcal{V}|+|\mathcal{E}|)$}
            \State $\bm{\mathcal{U}'} \leftarrow$ Sample $\mathcal{U}'_t\in\mathcal{U}_t$ from $\pi_1(a_t, s_t)$.
            \State Initialize $\tau \leftarrow 0$ and partial solution $\mathcal{P}_\tau$.
            \While{$s_\tau$ is not a terminal state} 
                \State Get state $s_\tau=\{\mathcal{P}_\tau\}$.
                \State Calculate graph embedding $\tilde{\textbf{h}}_\tau$ and compute $\pi_2(a_\tau, s_\tau)$ from \eqref{Eq: Tree Generator} based on $\bm{\theta}^*_2$.
                \Comment{$O(|\mathcal{V}|+|\mathcal{E}|)$}
                \State Sample $a_\tau\in\mathcal{N}(\mathcal{P}_\tau)$ from $\pi_2(a_\tau, s_\tau)$. 
                \State $\mathcal{P}_\tau \leftarrow$ select a link $(v^*_\tau, a_\tau)$.
                \State $\tau \leftarrow \tau+1$.
            \EndWhile
            \State $\bm{\mathcal{T}} \leftarrow$ Multicast with $\mathcal{P}_\tau$.
            \State $t \leftarrow t+1$.
        \EndWhile
    \end{algorithmic}
\end{algorithm}
Given the initial network graph topology and features, the TGMS algorithm iteratively selects a subset of destinations $\mathcal{U}'_t$ and generates a multicast tree $\mathcal{T}_t$ for each time slot $t$. The scheduling process is performed by the scheduler, which is based on $\mathcal{M}_1$. The tree-generating process is invoked to generate a multicast tree for the selected destinations based on $\mathcal{M}_2(a_t)$. The Lagrangian multiplier $\lambda$ will be updated in the training process. The Graph embedding methods are acting as feature extractors in both the scheduler and the tree generator.
Specifically, graph embedding methods are used to extract graph information being utilized by the actor and critic heads in the forwarding process. When we train RL agents, we are actually backwarding from the actor and critic heads to the trainable Graph embedding layers. The complexity analysis of TGMS is provided in the appendix. We also give a pseudocode list of the whole algorithm architecture in the appendix.

\subsection{Training: Jointly Train Two Agents}

\subsubsection{Advantage Actor-Critic Algorithm}
Our proposed TGMS algorithm consists of two agents: a scheduler and a tree generator. Each agent is learned from the Advantage Actor-Critic (A2C) algorithm \cite{mnih2016asynchronous}. Let $\bm{\theta}$ and $\bm{\omega}$ denote the parameters of the actor and the critic. We consider the following state-value function:
\begin{equation}
    v_{\pi_{\bm{\theta}}}(s) =\mathbb{E}_{\substack{a_t\sim\pi_{\bm{\theta}}(\cdot|s_t) \\ s_{t+1}\sim \mathbb{P}_{\pi_{\bm{\theta}}}(\cdot|s_t, a_t)}}
    \left[\sum_{t=0}^{\infty}\gamma^t r(s_t,a_t)|s_0=s\right],
\end{equation}
where $\pi_{\bm\theta}$ is the parameterized policy under $\bm{\theta}$ and $\mathbb{P}_{\pi_{\bm{\theta}}}(\cdot|s_t, a_t)$ is the transition probability under policy $\pi_\theta$. 
We aim to maximize the expected state value of the initial state $s_0$. Let $J(\bm{\theta}):=v_{\pi_{\bm\theta}}(s_0)$ denote the objective function. We first analyze how action $a_t$ outperforms other actions under state $s_t$. It is measured by the advantage function, denoted as:
\begin{equation}
    A_{\pi_{\bm{\theta}}}(s_t,a_t)=r(s_t,a_t)-v_{\pi_{\bm\theta}}(s_t).
\end{equation}
According to the policy gradient theorem \cite{sutton1999policy}, the gradient of $J(\bm{\theta})$ is given by:
\begin{equation}
    \label{Eq: Policy Gradient of the Actor}
    \nabla_{\bm{\theta}} J(\bm{\theta})=\mathbb{E}_{s_t\sim \mu_{\bm{\theta}}, a_t\sim\pi_{\bm{\theta}}}
    [A_{\pi_{\bm{\theta}}}(s_t,a_t) \nabla_{\bm{\theta}}\log\pi_{\bm{\theta}}(a_t|s_t) ].
\end{equation}
However, due to high-dimensional nature of $\mathcal{M}_1$ and $\mathcal{M}_2(a_t)$, it is infeasible to directly access $v_{\pi_{\bm{\theta}}}$. Hence, the critic is used to approximate the value of state $s_t$, denoted as $V_{\bm{\omega}}(s_t)$. The pseudo-code of A2C is described in Algorithm \ref{PC: A2C}. 
\begin{algorithm}[!t]
    \caption{Advantage Actor-Critic (A2C)}
    \label{PC: A2C}
    \begin{algorithmic}[1]
        \Require An episode $\{s_0, a_0, r_0, s_1, a_1, r_1, \dots, s_T, a_T, r_T\}$.
        \Ensure $d\bm{\theta}$ - gradient of actor, $d\bm{\omega}$ - gradient of critic.
        \State $t \leftarrow 0$.
        \While{$t < T$}
            \State $A_t \leftarrow \sum_{k=t}^{T} \gamma^k r_k-V_{\bm{\omega}}(s_t)$.
            \State $d\bm{\theta} \leftarrow d\bm{\theta} + \nabla_{\bm{\theta}}(\log\pi_{\bm\theta}(a_t|s_t)A_t + H_{\bm{\theta}})$.
            \State $d\bm{\omega} \leftarrow d\bm{\omega} + \partial(R-V_{\bm{\omega}}(s_t))^2 / \partial \bm{\omega}$.
            \State $t\leftarrow t+1$.
        \EndWhile
    \end{algorithmic}
\end{algorithm}
The actor is updated by $\eqref{Eq: Policy Gradient of the Actor}$ and the critic is updated from the error between $v_{\pi_{\bm{\theta}}}$ and $V_{\bm{\omega}}(s_t)$. The convergence analysis below follows the standard actor-critic template (e.g., \cite{zhang2020global}), which is stated for abstract MDPs satisfying four input conditions: (C1) bounded reward $|r(s,a)|\leq R$; (C2) Lipschitz score function with bounded norm $\|\nabla\log\pi_{\bm\theta}(a|s)\|\leq B_{\bm\theta}$; (C3) bounded critic features $\|\phi(s)\|\leq 1$; and (C4) bounded critic output $|V_{\bm\omega}(s)|\leq B_{\bm\omega}$. \rev{The analysis is conditional on (C1)--(C4). Below, we relate these conditions to the graph-AoI MDP and distinguish the finite-depth NGAT representation bound from the separate bounded-feature assumption used in the actor-critic analysis:}
\begin{itemize}
\item (C1) follows from the AoI truncation $A_{\max}$ introduced in Section III-C and a finite Lagrangian multiplier $\lambda$, as stated in Assumption \ref{Assumptions of Reward and Actor}(a).
\item (C2) is stated as Assumption \ref{Assumptions of Reward and Actor}(b); the bounded-norm part is enforced in implementation by the gradient-clipping technique noted in the footnote of that assumption.
\item \rev{For (C3), NGAT provides a finite-depth representation bound, rather than an unconditional unit-norm guarantee. Let $M_\ell=\max_i\|\mathbf{h}_i^{(\ell)}\|_2$ and $X_{\max}=\max_i\|\mathbf{x}_i\|_2$. For finite weights, nonzero $\|\mathbf{W}_1^{(\ell)}\|_2$, and nonnegative row-stochastic attention, \eqref{Eq: NGAT attention sum} gives}
\begin{equation}
    \rev{\begin{aligned}
    M_{\ell+1}&\leq M_\ell+C_\ell,\\
    C_\ell&=\frac{\|\mathbf{W}_3^{(\ell)}\|_2}{\|\mathbf{W}_1^{(\ell)}\|_2}X_{\max},\\
    M_L&\leq M_0+\sum_{\ell=0}^{L-1}C_\ell.
    \end{aligned}}
\end{equation}
\rev{Bounded inputs and finite parameters therefore give a finite bound at fixed finite depth $L$. This does not establish a depth-uniform bound, a bound uniform over unconstrained training parameters, or $\|\phi(s)\|_2\leq1$ by itself. The unit-norm feature condition (C3) remains an explicit assumption of the actor-critic theorem; it can be imposed by a separate feature normalization. Theorem \ref{Theorem: Contraction Mapping of NGAT} instead bounds the Frobenius distance between two embedding trajectories only when they share the same attention and parameters at every layer under its assumptions. It is not used to infer boundedness of an individual trajectory or convergence of training.}
\item (C4) is stated as Assumption \ref{Assumptions of Critic}(c); under (C3) with $\|\phi(s)\|\leq 1$, this is equivalent to bounding $\|\bm\omega\|$, a standard stability assumption maintained in actor-critic implementations.
\end{itemize}
\rev{These conditions make the roles of $A_{\max}$, $\lambda$, graph size, and the finite-depth NGAT representation bound explicit. Subject to the separately stated actor-critic assumptions, the following theorem gives a conditional stationarity bound; it is not a consequence of embedding non-expansiveness alone:}
\begin{Theorem}[Convergence of Algorithm \ref{PC: A2C}]
    \label{Theorem: Convergence Theorem}
    Let the averaged critic error be
    \begin{equation}
        \bar{\mathcal E}_c(n,k)
        = \frac{1}{k-n+1}\sum_{i=n}^{k}\mathbb{E}[\|\bm\omega_i-\bm\omega^*\|^2].
    \end{equation}
    Under Assumptions \ref{Assumptions of Reward and Actor} and \ref{Assumptions of Critic}, choose $\alpha_t=c_\alpha/(1+t)^\sigma$ with $0<\sigma\leq 1/2$. For $k\geq 2n-1$, we have
    \begin{equation}
        \min_{n \leq i \leq k} \mathbb{E}[\|\nabla J( \bm\theta_i)\|^2]
        \leq C_1\bar{\mathcal E}_c(n,k) + \frac{C_2}{n^\sigma}
        + C_3\epsilon_v^2 + C_4\epsilon_m,
    \end{equation}
    where $C_1=\mathcal{O}(B_{\bm\theta}^2)$, $C_2$ depends on $R$, $L$, $B_j=(R+B_{\bm\omega})B_{\bm\theta}$, $c_\alpha$, $\gamma$, and $\sigma$, $C_3=\mathcal{O}(B_{\bm\theta}^2)$, and $C_4$ is an absolute constant (see Appendix~\ref{App-Proof-T2} for explicit forms). The quantities $\bar{\mathcal E}_c(n,k)$, $\epsilon_v$, and $\epsilon_m$ denote the averaged critic error, the critic function-approximation residual, and the Markovian-noise residual, respectively.
\end{Theorem}
Theorem \ref{Theorem: Convergence Theorem} should therefore be interpreted as an NGAT-aware specialization of standard A2C convergence analysis. If the critic error $\bar{\mathcal E}_c(n,k)$ and the residual terms $\epsilon_v,\epsilon_m$ vanish, the stationarity bound reduces to the standard diminishing-step-size order. Otherwise, the algorithm converges to a neighborhood whose size depends on the critic error, function-approximation error, and Markovian noise. Larger values of $A_{\max}$ or $\lambda$, and looser critic approximation, enlarge the constants and may slow convergence. The specific proof can be found in the Appendix.

\subsubsection{Training Pipeline}
The training pipeline of TGMS is described in Algorithm \ref{PC: Training Pipeline}. 
\begin{algorithm}[!t]
    \caption{Training Pipeline of TGMS}
    \label{PC: Training Pipeline}
    \begin{algorithmic}[1]
        \Require Training dataset $\mathcal{D}$, initialized scheduler $\{\bm{\theta}_1, \bm{\omega_1}\}$ and tree generator $\{\bm{\theta}_2, \bm{\omega_2}\}$.
        \Ensure Trained scheduler $\{\bm{\theta}^*_1, \bm{\omega}^*_1\}$ and tree generator $\{\bm{\theta}^*_2, \bm{\omega}^*_2\}$.
        \While{not converged}
            \State Sample a graph $\mathcal{G}\sim \mathcal{D}$.
            \State $t\gets 0$ and initialize $\mathcal{G}_t$.
            \While{$t < T$}
            \State Get state $s_t=\{\mathcal{G}_t, \mathbf{x}_t\}$.
            \State Run the scheduler based on $\{\bm{\theta}_1, \bm{\omega_1}\}$ and get a transition $\delta_t = \{\pi_1(a_t, s_t), V_1(s_t), r_1(s_t, a_t)\}$.
            \State Store transition in the buffer $B_s \gets \delta_t$.
            \State $\tau \leftarrow 0$.
            \While{$s_\tau$ is not a terminal state}
                \State Get state $s_\tau=\{\mathcal{P}_\tau\}$.
                \State Run the tree generator based on $\{\bm{\theta}_2, \bm{\omega_2}\}$ and get a transition $\delta_\tau = \{\pi_2(a_\tau, s_\tau), V_2(s_\tau), r_2(s_\tau, a_\tau)\}$.
                \State Store transition in the buffer $B_t \gets \delta_\tau$.
                \State $\tau \leftarrow \tau+1$.
            \EndWhile
            \State Update the tree generator $\{\bm{\theta}_2, \bm{\omega_2}\}$ with \textbf{Algorithm \ref{PC: A2C}} based on $B_t$.
            \State Store the extra energy cost $B_\lambda \gets C(\mathcal{T}_t)-\overline{C}$.
            \State $t\gets t+1$.
            \EndWhile
            \State Update the scheduler $\{\bm{\theta}_1, \bm{\omega_1}\}$ with \textbf{Algorithm \ref{PC: A2C}} based on $B_s$.
            \State Update the Lagrangian multiplier $\lambda$ from $B_\lambda$ as $\lambda\gets\max(0,\lambda+\alpha B_\lambda)$.
        \EndWhile
    \end{algorithmic}
\end{algorithm}
The training process is performed iteratively. The scheduler and tree generator are learned from the A2C algorithm. The training process is repeated until the end of the episode. The Lagrangian multiplier $\lambda$ is updated using a subgradient descent method \cite{bertsekas2014constrained} at each time slot to ensure that the energy cost is within the budget. The training process is performed end-to-end, where the scheduler and tree generator are jointly optimized. Due to the complexity of the tree-generating process, there are several challenges in training the tree generator. First, due to the action masking of the tree generator, the tree generator may converge to a local minimum, and the gradient may explode. However, a small learning rate may slow down the convergence. Second, the tree generator may suffer from the exploration-exploitation dilemma, leading to suboptimal solutions. Therefore, we propose the following strategies:

\textbf{Dynamic $\epsilon$-Greedy Exploration.} We introduce a dynamic $\epsilon$-greedy exploration strategy for the tree generator to balance the exploration and exploitation. Specifically, a common $\epsilon$-greedy strategy is to select the action with the highest probability with probability $1-\epsilon$ and randomly select an action with probability $\epsilon$ as follows:
\begin{equation}
    a_\tau = \begin{cases}
        \arg\max_{a_\tau} \pi_2(a_\tau|s_\tau) & \text{with probability } 1-\epsilon, \\
        \text{randomly select } a_\tau & \text{with probability } \epsilon.
    \end{cases}
\end{equation}
However, a small $\epsilon$ may lead to suboptimal solutions, while a large $\epsilon$ may slow down the convergence. To address this issue, we dynamically adjust the value of $\epsilon$ during the training process according to the entropy of the policy. The entropy of the policy is defined as:
\begin{equation}
    H(\pi_2(\cdot|s_\tau)) = -\sum_{a_\tau\in\mathcal{A}_2} \pi_2(a_\tau|s_\tau) \log \pi_2(a_\tau|s_\tau).
\end{equation}
Obviously, for each action $a_\tau$, the maximum entropy is $1/e$. Then the normalized entropy can be defined as:
\begin{equation}
    H_{\max}(\pi_2(\cdot|s_\tau)) = \frac{e H(\pi_2(\cdot|s_\tau))}{|\mathcal{A}_2|}.
\end{equation}
The value of $\epsilon$ is then updated as:
\begin{equation}
    \epsilon = \max\left(\epsilon_{\text{min}}, \epsilon_{\text{max}}\left(1-\frac{H(\pi_2(\cdot|s_\tau))}{H_{\max}(\pi_2(\cdot|s_\tau))}\right)\right),
\end{equation}
where $\epsilon_{\text{min}}$ and $\epsilon_{\text{max}}$ are hyperparameters. The above strategy ensures that the tree generator focuses on optimizing the policy at the beginning of the training process and gradually explores the action space as the training progresses. This method is beneficial for avoiding the local minimum while accelerating the convergence of the tree generator.

\textbf{Selective Backpropagation.} The tree generator may suffer from the action masking problem, which may lead to the gradient explosion. To address this issue, we propose a selective backpropagation strategy. Specifically, the following situations will not trigger the backpropagation process:
\begin{itemize}
	\item \textbf{Greedy Exploration:} When the action is randomly selected from the action space with the $\epsilon$-greedy method, the backpropagation process will not be triggered. Note that the corresponding transition is still valid when the reward is accumulated.
    \item \textbf{Action with High Probability:} When the action is selected with high probability, it is likely that the tree generator has already converged to a local minimum. Therefore, we set a threshold to determine whether the action should be considered for backpropagation. When the probability of the action is higher than the threshold, the backpropagation process will not be triggered.
\end{itemize}
The above strategies are beneficial for exploring the action space while avoiding the gradient explosion problem.

\subsection{Implementation Details}

In the final part of this section, we discuss additional implementation details, including cross-layer integration,  distributed deployment, and hierarchical two-timescale operation.

\textbf{Cross-layer Implementation}. Here, we clarify how to implement a cross-layer design and how AoI information is available for layer operation. The foundation of our cross-layer design is the problem reformulation that decomposes the original task into a higher-level scheduling subproblem and a lower-level tree-generation subproblem, linked by AoI as a clear cross-layer signal. This aligns with classic cross-layer design principles (see Fu et al. \cite{fu2013survey}). Concretely, the process begins at the application layer: each destination node computes its AoI from packet timestamps. This freshness metric is periodically sent to the central controller hosting our TGMS agent. Based on the received AoI information, the agent makes a hierarchical, two-stage decision. First, the scheduler (higher-level logic) observes AoI across destinations and selects the subset that most urgently requires updates to maintain freshness. Second, the tree generator (network-layer logic) takes this subset and computes a resource-efficient multicast tree to reach the chosen destinations. 

\textbf{Distributed Implementation.} 
This paper adopts a centralized approach for two main reasons. First, jointly optimizing multicast scheduling and routing is highly complex due to the tight coupling between these decisions. A centralized agent, with global knowledge of the network (e.g., all nodes' AoI and the full graph topology), is best suited to produce a high-quality, coordinated solution that properly balances trade-offs across the entire system. Second, our design aligns naturally with modern SDN architectures, in which a central controller handles network intelligence and orchestration.

In practice, some scenarios require a decentralized implementation, either to ensure robustness against single points of failure or to operate in networks without a central controller (e.g., ad-hoc networks). The decentralized implementation of our proposed method is feasible by leveraging distributed graph-embedding methods (e.g., \cite{zhou2025protecting}) and cooperative multi-agent reinforcement-learning algorithms (e.g., \cite{lowe2017multi}). However, achieving an effective decentralized policy remains challenging, as each node must act using only local information while still contributing to a globally coherent objective. We view this as an important direction for future research.

\textbf{Hierarchical Two-timescale Operation.} One may be concerned about the overhead of running both the scheduler and the tree generator in each time slot. We chose the per-slot formulation because it captures the fine-grained dynamics of AoI and energy consumption theoretically. In real implementations, our complex hierarchical RL can be naturally extended to a two-timescale design, with routing operating on a longer timescale and scheduling on a shorter one.
The rationale behind the two-timescale design is as follows. Let $t$ denote the shorter timescale of scheduling, and $e$ denote the longer timescale of routing, where each epoch $e$ includes $T$ time slots $t$. For the scheduler, only a single forwarding operation is required in each time slot $t$, which is lightweight. 
\begin{figure}[!t]
    \centering
    \includegraphics[width=0.35\textwidth]{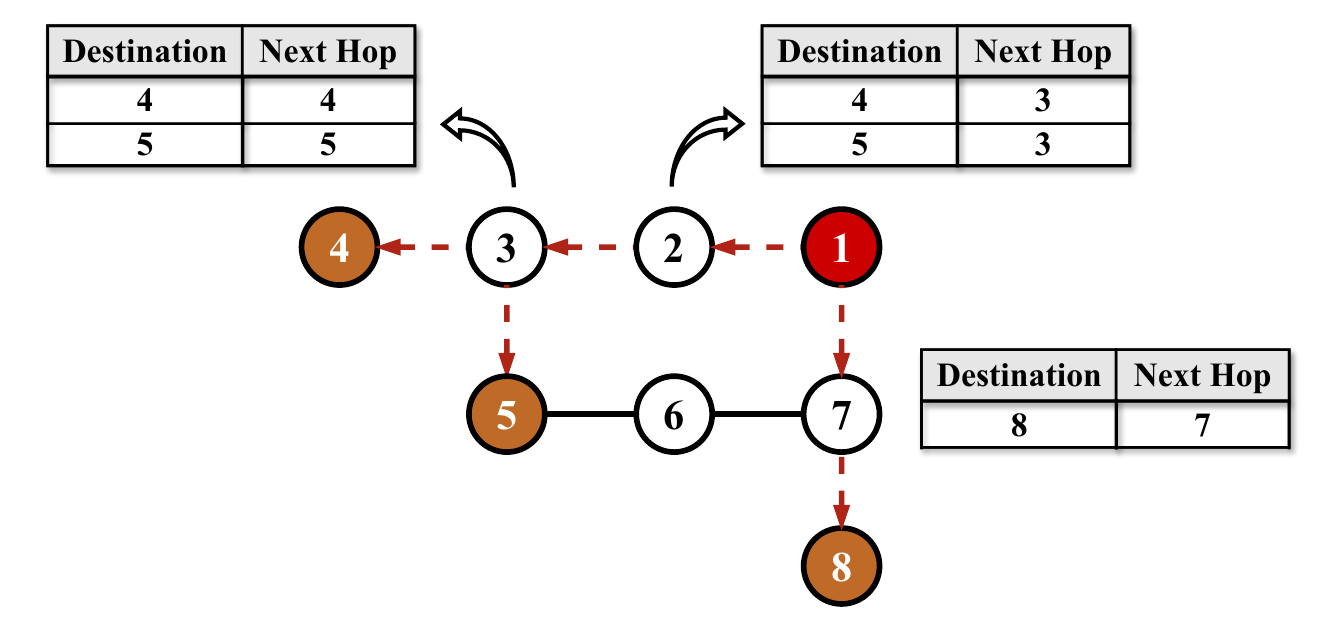}
    \caption{Routing Tables Updates based on Multicast Tree. The red dashed lines represent the multicast tree. When a multicast tree is generated by the tree generator, only the routing tables of nodes in the tree (2, 3, 7 here) are updated. Other routers (6 here) keep their previous routing tables unchanged.}
    \label{fig: routing tables}
\end{figure}
The extended tree-generating scheme is based on the most modern multicast routing protocols (e.g., Protocol Independent Multicast (PIM) \cite{deering1995protocol}), described as follows. When a multicast tree is generated, only the routing tables of nodes in the tree are updated. Other nodes keep their previous routing tables unchanged, which is illustrated in \textbf{Figure \ref{fig: routing tables}}. That means the tree generator only needs to run once in each timescale $e$ (e.g., every $T$ time slots) to update the routing tables of nodes in the multicast tree. The parameter $T$ can be adjusted based on the network size and dynamics, which significantly reduces the overhead of the tree generator. Appendix~\ref{App: Update Interval} reports average AoI for different update intervals.

\section{Experimental Evaluation}
In this section, we conduct extensive experiments to evaluate the performance of our approach. Our experiment mainly includes two parts: (i) the performance evaluation on the standard STP problem (i.e., problem \textbf{P2}); (ii) the performance evaluation on the multicast scheduling problem (i.e., problem \textbf{DP}). We evaluate all baselines on the same devices. The details not mentioned in the main text are in the Appendix.
\subsection{Performance Evaluation on the STP Problem}
In this part, we evaluate the performance of our proposed Tree Generator (TG) and other baseline algorithms on the STP problem. Our MDP formulation of the STP problem is described in the Appendix. The considered STP Datasets are presented in Table \ref{Table: STP Datasets} from the SteinLib Testdata Library \cite{KMV00}.
\begin{table}
    \centering
    \caption{STP Datasets}
    \label{Table: STP Datasets}
    \begin{tabular}{c|c|c|c}
        \hline
        \textbf{Dataset} & \textbf{Num. of Nodes} & \textbf{Num. of Edges} & \textbf{Num. of Terminals} \\
        \hline
        I080 & 80 & $[120, 3160]$ & $[6, 20]$ \\
        I160 & 160 & $[240, 12720]$ & $[7, 40]$ \\
        I320 & 320 & $[480, 51040]$ & $[8, 80]$ \\
        I640 & 640 & $[960, 204480]$ & $[9, 160]$ \\
        \hline
    \end{tabular}
\end{table}
Each dataset has 100 instances. The distributions of the number of edges and the number of terminals are independent. For example, a graph with 80 nodes may have 120 edges and 6 terminals or 120 edges and 20 terminals. The known optimal costs are provided in the datasets. For the algorithms that need to be trained, we divide the dataset into training and testing sets. We randomly choose 80 instances for training, while the remaining 20 instances are used for testing. 

\textit{Baselines:} We consider the following baselines:
\begin{itemize}
	\item \textbf{Random:} This algorithm is based on the partial-solution generation method described in Eq. \eqref{Eq: TG-Link Selection}. The algorithm randomly selects a node from $\mathcal{A}_2$ (see Eq. \eqref{TG: Action Space}) and connects it to $\mathcal{P}_\tau$ with its minimum-cost link.
    \item \textbf{MST:} This algorithm generates a minimum spanning tree for a given graph. All nodes are included.
	\item \textbf{IRR \cite{byrka2013steiner}:} This algorithm is a state-of-the-art algorithm (which is on the Iterated Roundings and Rounding (IRR) algorithm) for the STP problem. It has an approximation ratio of $\rho=\ln(4)+\epsilon \approx 1.39$. 
    \item \textbf{TG-MLP:} This algorithm is a variant of our proposed tree generator. It only consists of the MLP without any graph embedding methods.
\end{itemize}
The parameters of the trainable algorithms are presented in the appendix. For fairness, we test all algorithms on the same test set. We consider the following metrics:
\begin{itemize}
    \item \textbf{Approximation Ratio ($\rho$):} The approximation ratio is defined as the ratio of the cost of the solution generated by the algorithm to the known optimal cost. The lower the ratio, the better the algorithm.
	\item \textbf{Compute Time per Graph:} The calculation time per graph is the average time to generate a solution for a single instance.
\end{itemize}
When testing the TG algorithm, we use a model trained on the I160 dataset to test the I320 and I640 datasets due to memory limitations. This method exhibits the ability to generalize our model. 
\begin{figure}[!t]
    \centering
    \subfigure[Approximation Ratio.]{
        \includegraphics[width=0.47\linewidth]{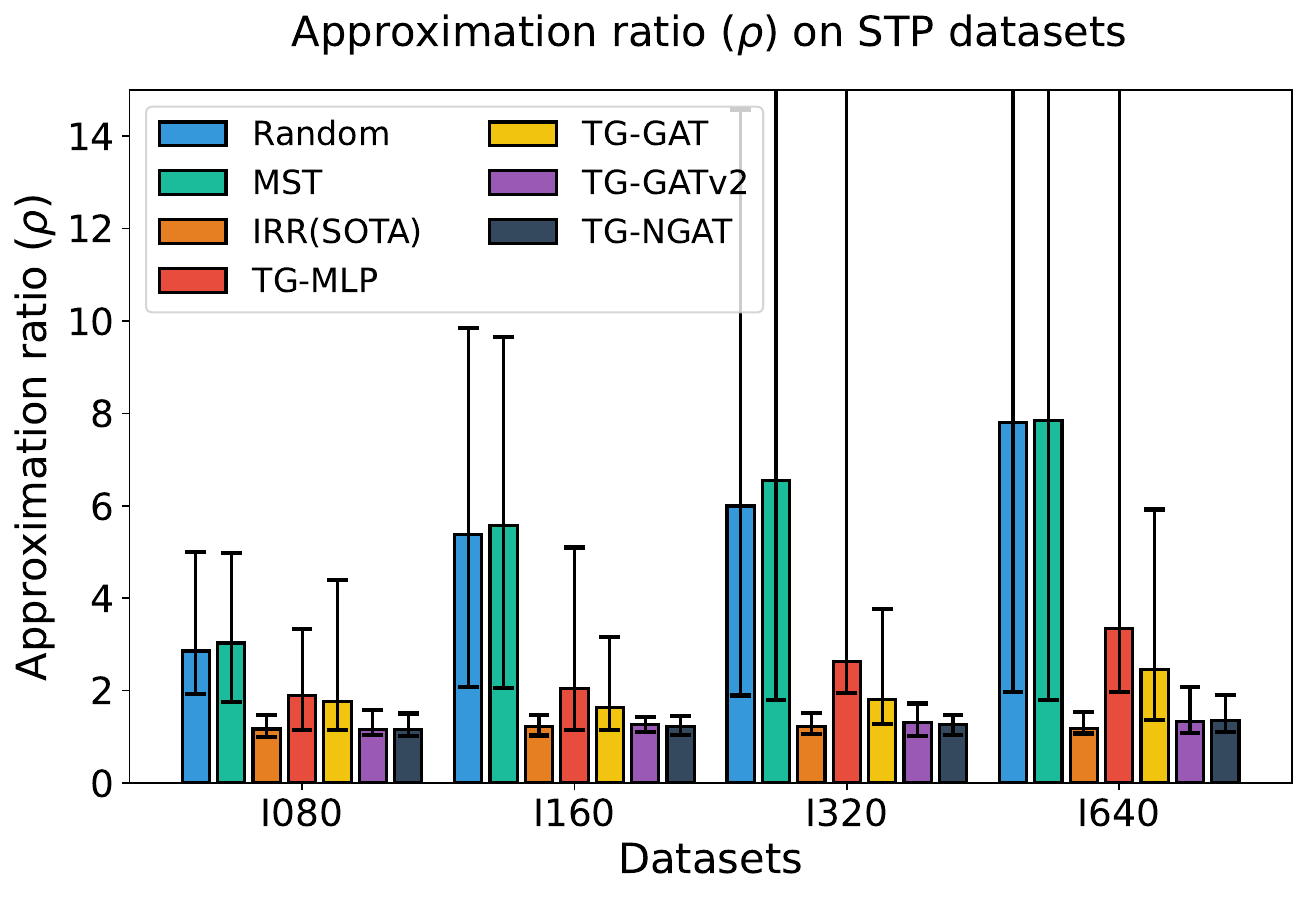}
        \label{Fig: STP_accuracy}
    }%
    \subfigure[Calculation Time per Graph.]{
        \includegraphics[width=0.47\linewidth]{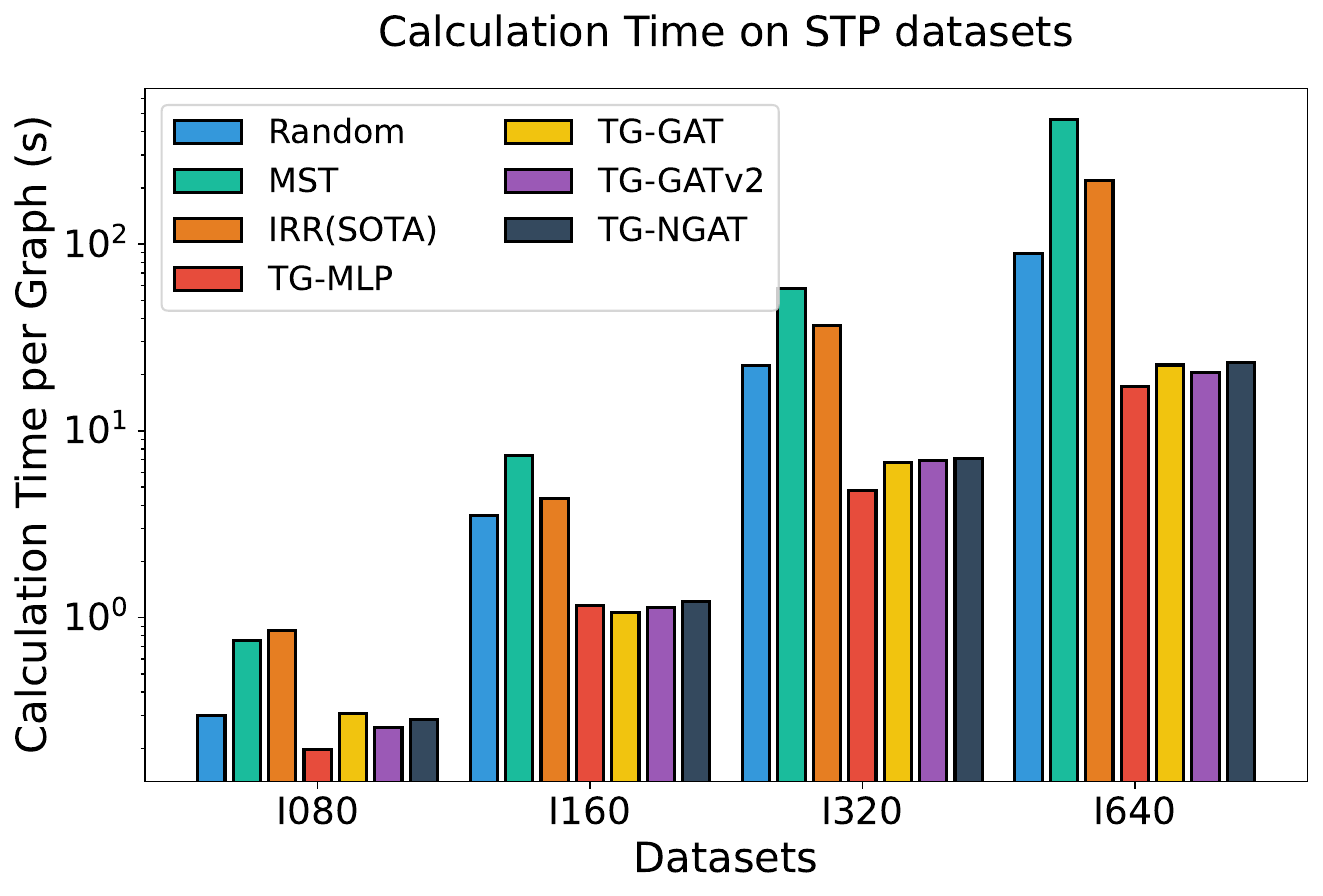}
        \label{Fig: STP_calculation_time}
    }
    \caption{Performance evaluation on the STP datasets.}
\end{figure}
\textit{Performance.} In Fig. \ref{Fig: STP_accuracy}, we observe that the Random, Greedy, and MST algorithms perform poorly regarding the approximation ratio. The poor performance of the Random and Greedy algorithms is due to their failure to consider the graph's global information. 
The MST algorithm also performs poorly because it must include all nodes in the resulting tree, including non-terminal ones. Note that TG-MLP performs better than the above baselines. However, it is not robust in larger graphs since it cannot capture the spatial structure of the graph. The TG and IRR algorithms outperform other baselines in terms of approximation ratio and robustness. 

\textit{Generalization.}  Fig. \ref{Fig: STP_accuracy} demonstrates that, despite not being trained on the I320 and I640 datasets, the TG model performs effectively on these datasets. This highlights the TG algorithm's strong generalization ability to handle unseen tasks that are 4$\times$ more complex.

\textit{Compute Time.}
We evaluate the compute (inference) time in Fig. \ref{Fig: STP_calculation_time}; the TG and TG-MLP algorithms have competitive calculation times compared to other baselines. Since the calculation time of a graph is closely related to the density of the graph, we do not draw the error bars. We observe that the MST algorithm has the longest calculation times since it needs to calculate the cost of all valid edges. The IRR algorithm also has a long calculation time due to its iterative nature. Our proposed TG algorithm has a competitive calculation time compared to other baselines. 
Specifically, the TG algorithm accelerates computation time by $6.35\times$ compared to the non-TG baselines and $5.85\times$ compared to the IRR algorithm.
This advantage is more significant in larger graphs. The specific results are presented in Table \ref{Table: STP Results}, where the comparison is made between the TG-NGAT algorithm and other baselines. The results show that the TG-NGAT algorithm achieves a performance ratio between $1.1$ and $1.3$, close to the IRR (SOTA) algorithm.

\begin{table*}
    \setlength\tabcolsep{1.5pt}
    \centering
    \caption{Experiment Results of STP}
    \begin{tabular}{c|cccc|cccc|cccc|cccc|}
    \hline
        \multirow{2}{*}{\textbf{Algorithm}} & \multicolumn{4}{c|}{\textbf{I080}} & \multicolumn{4}{c|}{\textbf{I160}} & \multicolumn{4}{c|}{\textbf{I320}} & \multicolumn{4}{c|}{\textbf{I640}} \\
        & $\rho$ & $\rho_{worst}$& $\rho_{best}$ & $t(s)$ & $\rho$ & $\rho_{worst}$& $\rho_{best}$ & $t(s)$ & $\rho$ & $\rho_{worst}$& $\rho_{best}$ & $t(s)$ & $\rho$ & $\rho_{worst}$& $\rho_{best}$ & $t(s)$ \\
        \hline
        \textbf{Random} & 3.092	& 4.996 & 1.927 & 0.298 & 5.384 & 9.85 & 2.076 & 3.508 & 6.295 & 14.578 & 1.894 & 22.50 & 7.612 & 30.01 & 1.962 & 89.37
        \\
        \textbf{MST} & 3.03 & 4.971 & 1.752 & 0.751 & 5.582 & 9.648 & 2.063 & 7.393 & 6.544 & 17.94 & 1.795 & 58.08 & 7.847 & 29.837 & 1.801 & 463.0
        \\ 
        \textbf{IRR (SOTA)} & 1.168 & 1.471 & 1.003 & 0.851 & 1.223 & 1.469 & 1.027 & 4.354 & 1.226 & 1.511 & 1.067 & 36.86 & 1.187 & 1.53 & 1.072 & 219.2
        \\
        \textbf{TG-MLP} & 1.885 & 3.338 & 1.144 & 0.196 & 2.047 & 5.097 & 1.151 & 1.163 & 2.638 & 16.34 & 1.067 & 4.794 & 3.351 & 28.13 & 1.969 & 17.27
        \\
        \textbf{TG-GATv1} & 1.77 & 4.398 & 1.147 & 0.307 & 1.642 & 3.169 & 1.151 & 1.06 & 1.809 & 3.764 & 1.281 & 6.778 & 2.45 & 5.92 & 1.369 & 22.53
        \\
        \textbf{TG-GATv2} & 1.167 & 1.577 & 1.042 & 0.259 & 1.262 & 1.427 & 1.095 & 1.137 & 1.303 & 1.72 & 1.024 & 6.912 & 1.326 & 2.078 & 1.079 & 20.65
        \\
        \textbf{TG-NGAT} & 1.149 & 1.504 & 1.018 & 0.285 & 1.231 & 1.448 & 1.041 & 1.227 & 1.257 & 1.464 & 1.036 & 7.082 & 1.345 & 1.905 & 1.106 & 22.245
        \\
        Comp. (Ave.) & $\uparrow$ 52.7\% & $\uparrow$ 60.6\% & $\uparrow$ 34.8\% & 2.22$\times$ & $\uparrow$ 69.7\% & $\uparrow$ 79.3\% & $\uparrow$ 39.5\% & 4.14$\times$ & $\uparrow$ 73.0\% & $\uparrow$ 73.2\% & $\uparrow$ 87.1\% & 5.53$\times$ & $\uparrow$ 75.8\% & $\uparrow$ 90.7\% & $\uparrow$ 31.4\% & 11.56$\times$ \\
        Comp. (SOTA) & $\uparrow$ 24.8\% & $\downarrow$ 3.3\% & $\uparrow$ 1.6\% & 2.99$\times$ & $\downarrow$ 0.7\% & $\uparrow$ 1.4\% & $\downarrow$ 1.4\% & 3.55$\times$ & $\downarrow$ 2.5\% & $\uparrow$ 3.1\% & $\uparrow$ 2.9\% & 5.20$\times$ & $\downarrow$ 13.3\% & $\downarrow$ 24.5\% & $\downarrow$ 3.2\% & 9.85$\times$ \\
    \hline
    \end{tabular}
    \label{Table: STP Results}
\end{table*}

\begin{figure*}[!t]
    \centering
    \subfigure[I080 Dataset.]{
        \includegraphics[width = 0.22\textwidth]{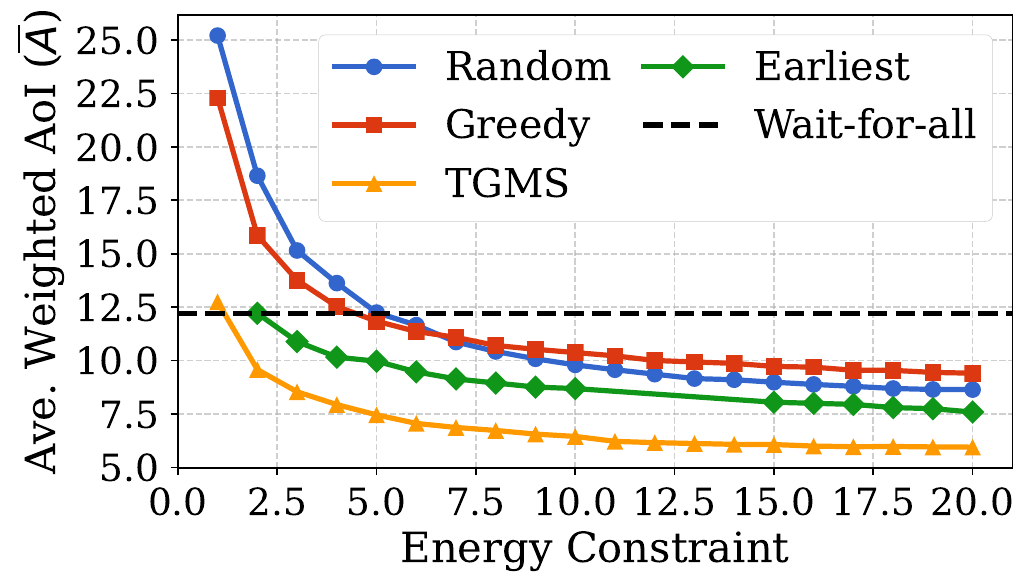}
    }
    \subfigure[AS-733 Dataset.]{
        \includegraphics[width = 0.22\textwidth]{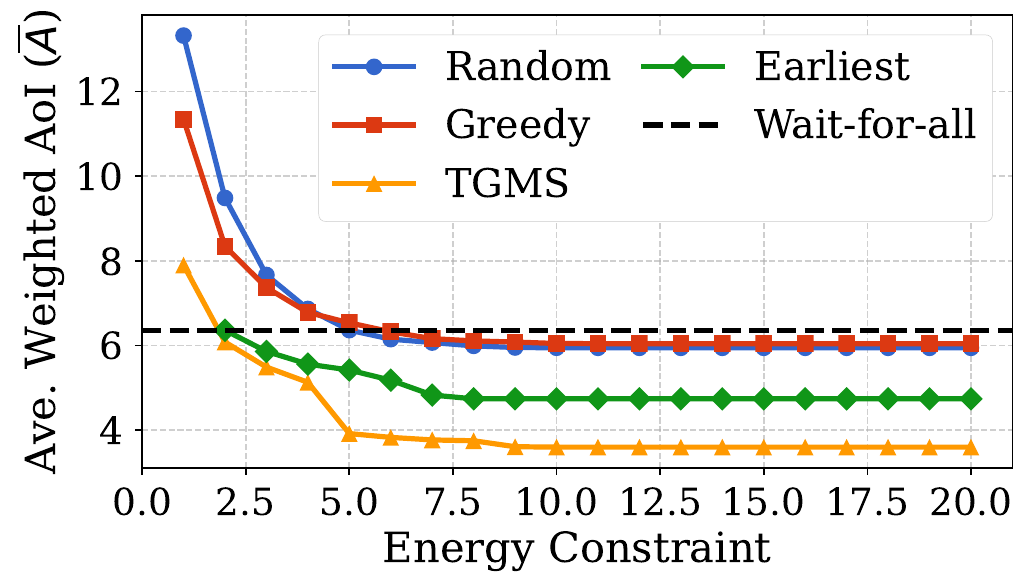}
    }
    \subfigure[BA Graphs.]{
        \includegraphics[width = 0.22\textwidth]{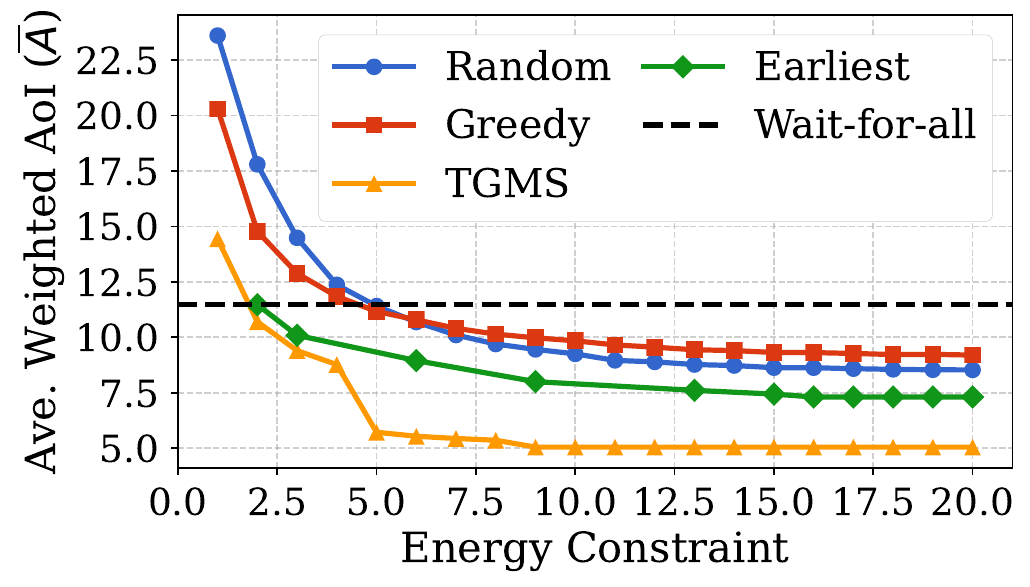}
    }
    \subfigure[ER Graphs.]{
        \includegraphics[width = 0.22\textwidth]{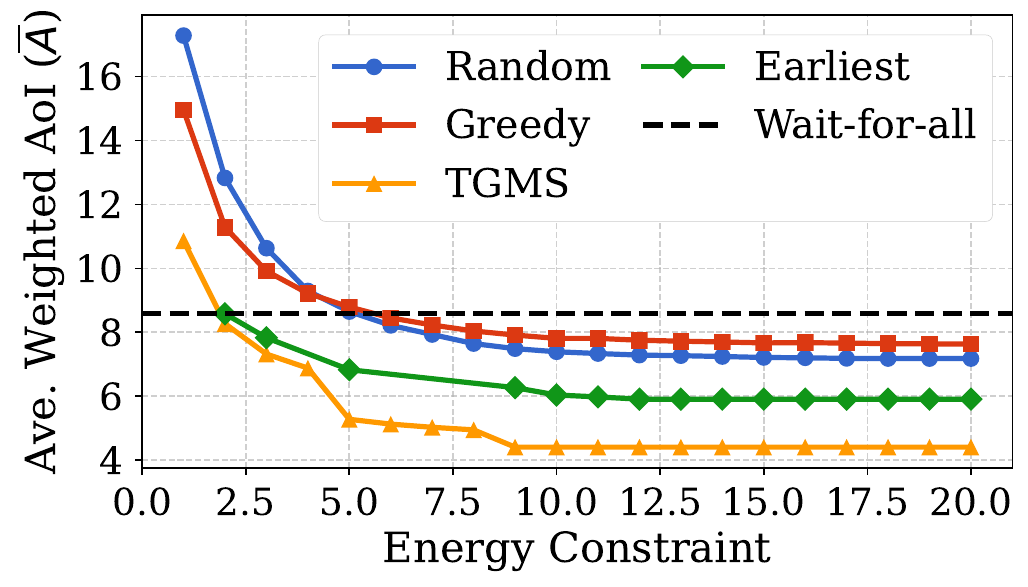}
    }
    \caption{Average weighted AoI over datasets with $|\mathcal{V}|=80$.}
    \label{Fig: Average Weighted AoI}
\end{figure*}
\begin{figure*}[!t]
    \centering
    \subfigure[I080 Dataset.]{
        \includegraphics[width = 0.22\textwidth]{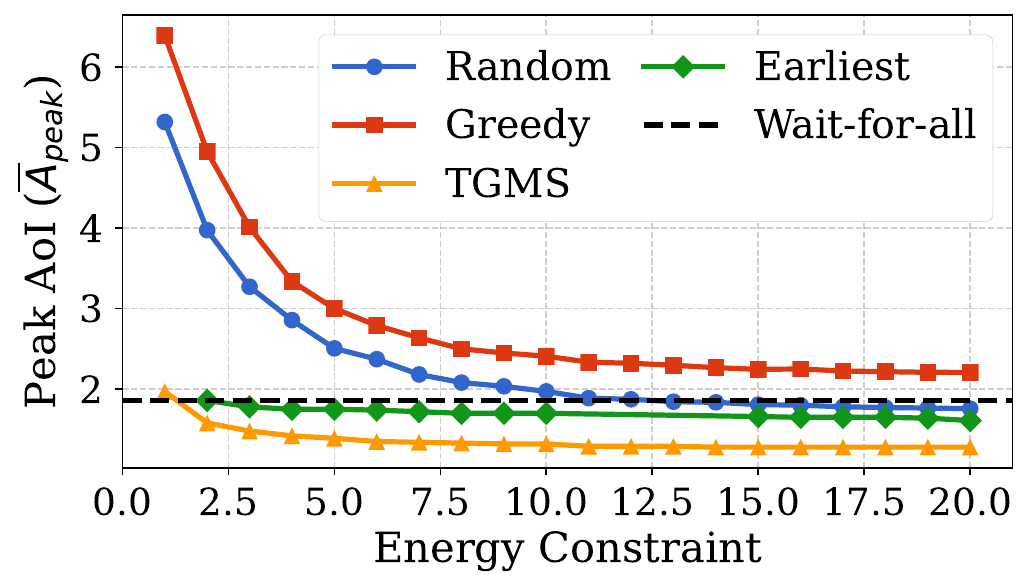}
    }
    \subfigure[AS-733 Dataset.]{
        \includegraphics[width = 0.22\textwidth]{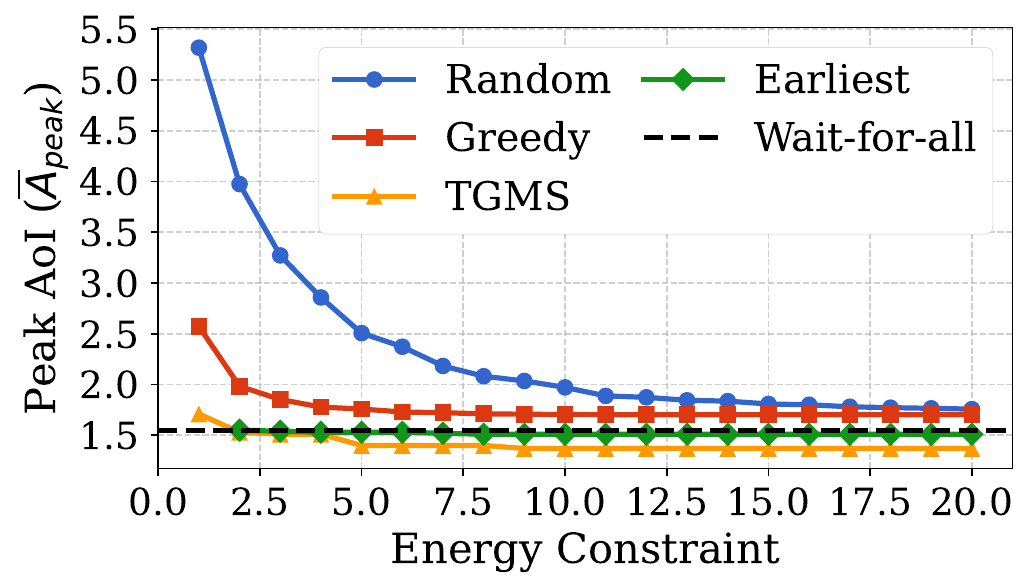}
    }
    \subfigure[BA Graphs.]{
        \includegraphics[width = 0.22\textwidth]{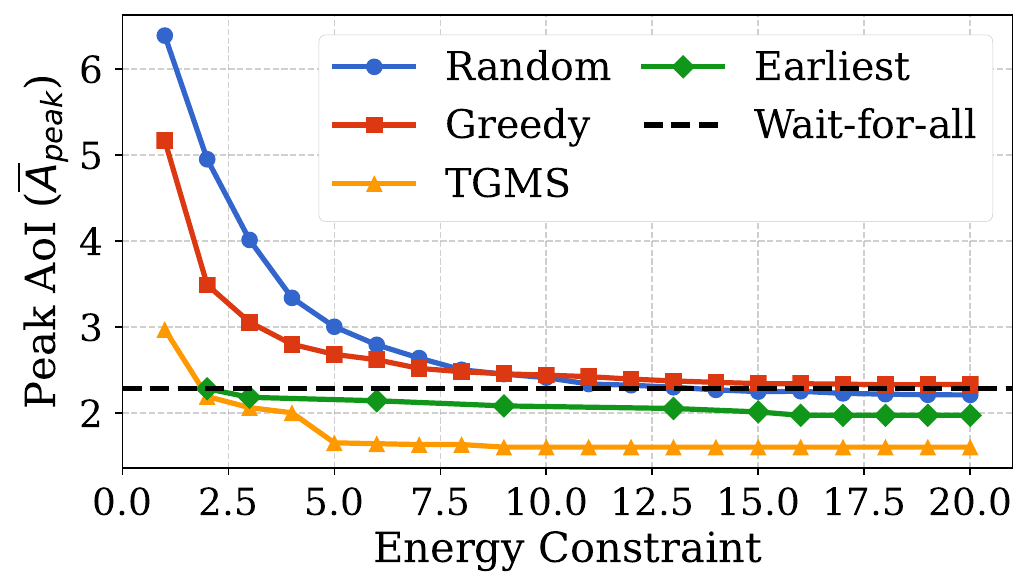}
    }
    \subfigure[ER Graphs.]{
        \includegraphics[width = 0.22\textwidth]{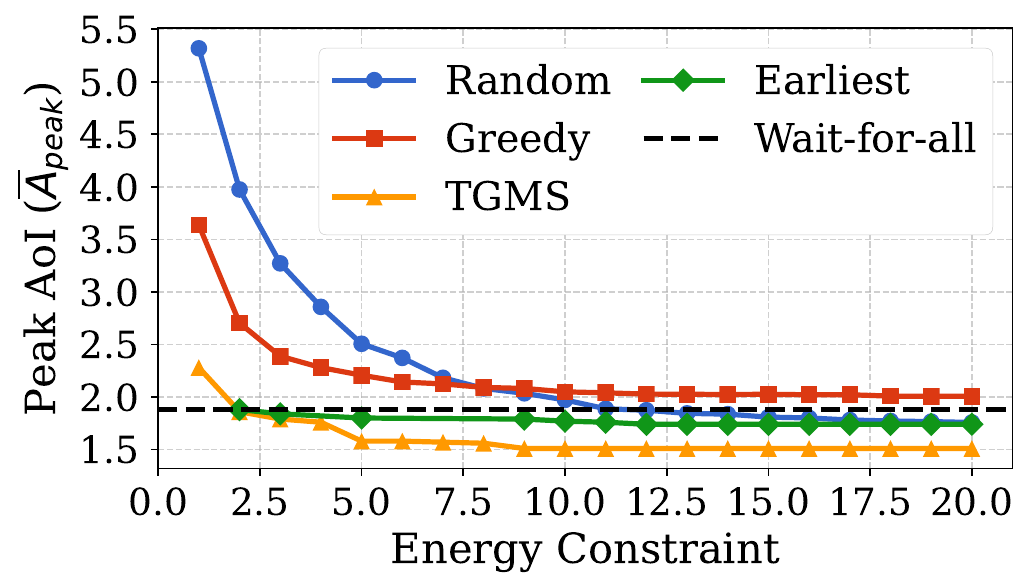}
    }
    \caption{Peak weighted age over datasets with $|\mathcal{V}|=80$.}
    \label{Fig: Peak Weighted Age}
\end{figure*}

\subsection{Performance Evaluation on the Multicast Scheduling Problem}
In this subsection, we mainly evaluate the performance of scheduling algorithms on the multicast scheduling problem.
\subsubsection{Experiment Setting}
We consider four graph topologies to simulate real-world multicast scheduling problems, including a real-world network dataset. The following datasets are considered:
\begin{itemize}
    \item \textbf{I080:} A dataset for evaluating the STP algorithms
    described as above.
    \item \textbf{AS-733:} The Autonomous Systems (AS)-733 dataset is a real-world dataset collected from the University of Oregon Route Views Project \cite{leskovec2005graphs}. It contains 733 abstracted graphs of Autonomous Systems.
    \item \textbf{ER Graphs:} The Erdös-Renyi (ER) random graphs are a type of graph in which each pair of nodes is connected with a fixed probability $p$. They are widely used in multicast networks (e.g., in \cite{mukherjee2016achieving}).
    \item \textbf{BA Graphs:} In Barabasi-Albert (BA) graphs, each newly introduced node connects to $m$ pre-existing nodes. The probability is proportional to the number of links existing nodes already possess. These graphs find significant applications within multicast systems (e.g., in \cite{yu2019online}).
\end{itemize}
Some examples of the above datasets are shown in the Appendix.
We add some necessary features to the above datasets to simulate a multicast network. We set $|\mathcal{V}|=80$ for all datasets to compare the performance of different algorithms fairly. For those datasets that do not contain the terminal nodes, we randomly select $10\%$ of nodes as destinations. The source node is randomly selected from the remaining nodes. The remaining nodes are considered router nodes for those datasets that do not contain the costs of edges. We randomly assign the costs of edges from $\{1, 2, \dots, 10\}$. Some examples of the above datasets are displayed in the Appendix.
We only train the model on the I080 dataset for the algorithms that need to be trained. The remaining datasets are only used for testing. 

\subsubsection{Performance Evaluation}
We first compare several scheduling algorithms to evaluate the performance of our proposed TGMS algorithm. The TGMS model is trained with the training dataset. When testing, we only change the scheduler and keep the tree generator unchanged for fairness. Each graph will be tested for 50 time slots. We consider the following algorithms as baselines:
\begin{itemize}
    \item \textbf{Random:} We first randomly select a sample ratio from a uniform distribution $[0, 1]$ and then randomly select destinations from $\mathcal{U}_t$ according to the sample ratio. 
    \item \textbf{Greedy:} We greedily select half destinations with the most weighted AoI from $\mathcal{U}_t$.
    
    \item \textbf{Earliest-$l$ Scheme \cite{xie2021age}:} The scheduler waits until $l$ ratio of the last destinations receive the updates. We change the ratio $l$ to satisfy different energy constraints. However, there are still some constraints unable to be achieved by changing $l$. 
    \item \textbf{Wait-for-all Scheme \cite{xie2021age}:} The scheduler waits until all last destinations receive the updates. It is a special case of the earliest-$l$ scheme where $l=1$.
    
\end{itemize}
The Random and Greedy algorithms can not guarantee the energy consumption constraint. Therefore, when testing, we force the scheduler to select an empty set if the history energy consumption exceeds the given energy constraint. We consider the following AoI metrics:
\begin{itemize}
    \item \textbf{Average Weighted AoI ($\overline{A}$):}  The main metric of our considered problem (see Eq. \eqref{Def: Average Weighted AoI}).
    \item \textbf{Weighted Peak Age ($\overline{A}_{peak}$):} The weighted peak age is defined as:
    \begin{equation}
        \overline{A}_{peak} = \lim_{N\to\infty} \frac{1}{N} \sum_{p=1}^N \sum_{u\in\mathcal{U}_t} \omega_u A_u(T_u(p)),
    \end{equation}
    where $T_u(p)$ denotes the time at which $u$ receives the $p$th update. The peak age is the average of age peaks, which happen just before a packet arrives.
\end{itemize}

For each dataset, we consider a set of different energy constraints $\overline{C}\in\{1, 2, \dots, 20\}$ and compare the AoI metrics. The results are shown in Figs. \ref{Fig: Average Weighted AoI} and \ref{Fig: Peak Weighted Age}.

    From Fig. \ref{Fig: Average Weighted AoI}, we observe that our scheduler performs superior to the baselines for all datasets. Compared with other baselines, our scheduler reduces the average weighted AoI by $38.65\%$ over all energy constraints. From Fig. \ref{Fig: Peak Weighted Age}, our scheduler reduces the peak weighted age by $36.06\%$ over all energy constraints. In low-energy scenarios ($\overline{C}<5$), our scheduler reduces the peak weighted age by $47.51\%$, which exhibits the robustness of the scheduler over various graph topologies.

\section{Conclusions}
In this paper, we have formulated the first AoI-optimal multicast problem via joint scheduling and routing. We have applied an RL framework to learn the heuristics of multicast routing, based on which we have performed problem
    decomposition amenable to hierarchical RL methods.
\rev{We have further proposed a normalized GAT for extracting graph information, with conditional non-expansiveness of the complete embedding matrix in the Frobenius norm under fixed symmetric row-stochastic attention.}
Experimental results demonstrate that our proposed scheme is up to $9.85\times$ more computationally efficient than traditional multicast routing algorithms, while achieving performance comparable to SOTA methods and showcasing superior generalization capabilities.
Additionally,
TGMS outperforms over other benchmarks, including non-cross-layer design and methods without graph embedding, maintaining a high level of performance and reduced the average weighted AoI by $38.65\%$ and the weighted peak age by $36.06\%$ under low-energy scenarios in joint multicast routing and scheduling problems. \par
We emphasize the main assumptions of our work as follows:
\begin{itemize}
    \item \textbf{Packet Delay:} The packet delay is linearly proportional to the number of hops.
    \item \textbf{Network Dynamics:} The network topology changes are slower than the packet transmission process.
    \item \textbf{Transmission Model:} The use of a "generate-at-will" model with error-free transmissions.
\end{itemize}
\par
A limitation of our algorithm is that the scheduler needs to be retrained if the energy constraint changes. An important future research direction is to design a generalizable scheduler that can adapt to different energy constraints without retraining. Also, the decentralized control is a crucial goal for systems requiring massive scalability or resilience to single points of failure. 

\bibliographystyle{IEEEtran}
\bibliography{ref}

\begin{IEEEbiography}[{\includegraphics[width=1in,height=1.25in,clip,keepaspectratio]{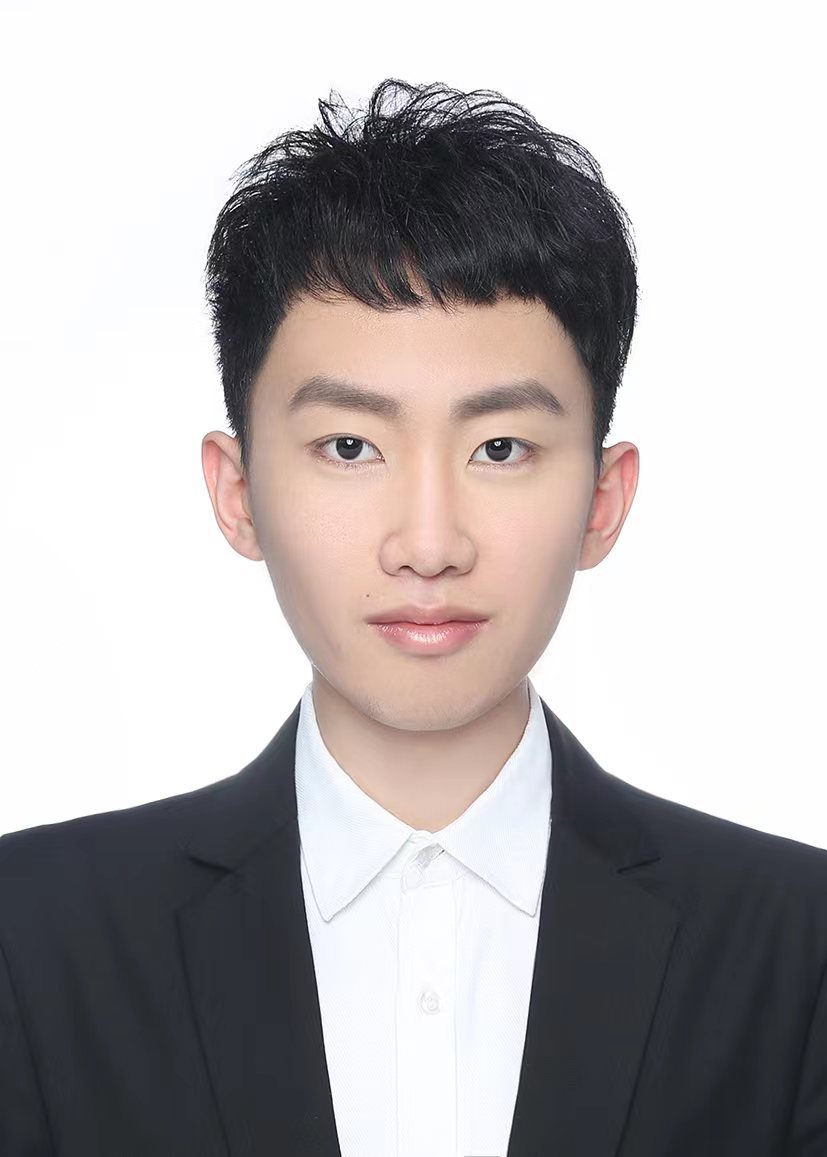}}]{Yanning Zhang}
received the B.Eng. degree in communication engineering from Northeastern University, Shenyang, China, in 2022. He is currently pursuing the Ph.D. degree in information and communication engineering with Zhejiang University, Hangzhou, China. His research interests include large language models, graph neural networks, reinforcement learning, and edge intelligence, with a focus on learning-based network optimization and efficient collaborative inference.
\end{IEEEbiography}

\begin{IEEEbiography}[{\includegraphics[width=1in,height=1.25in,clip,keepaspectratio]{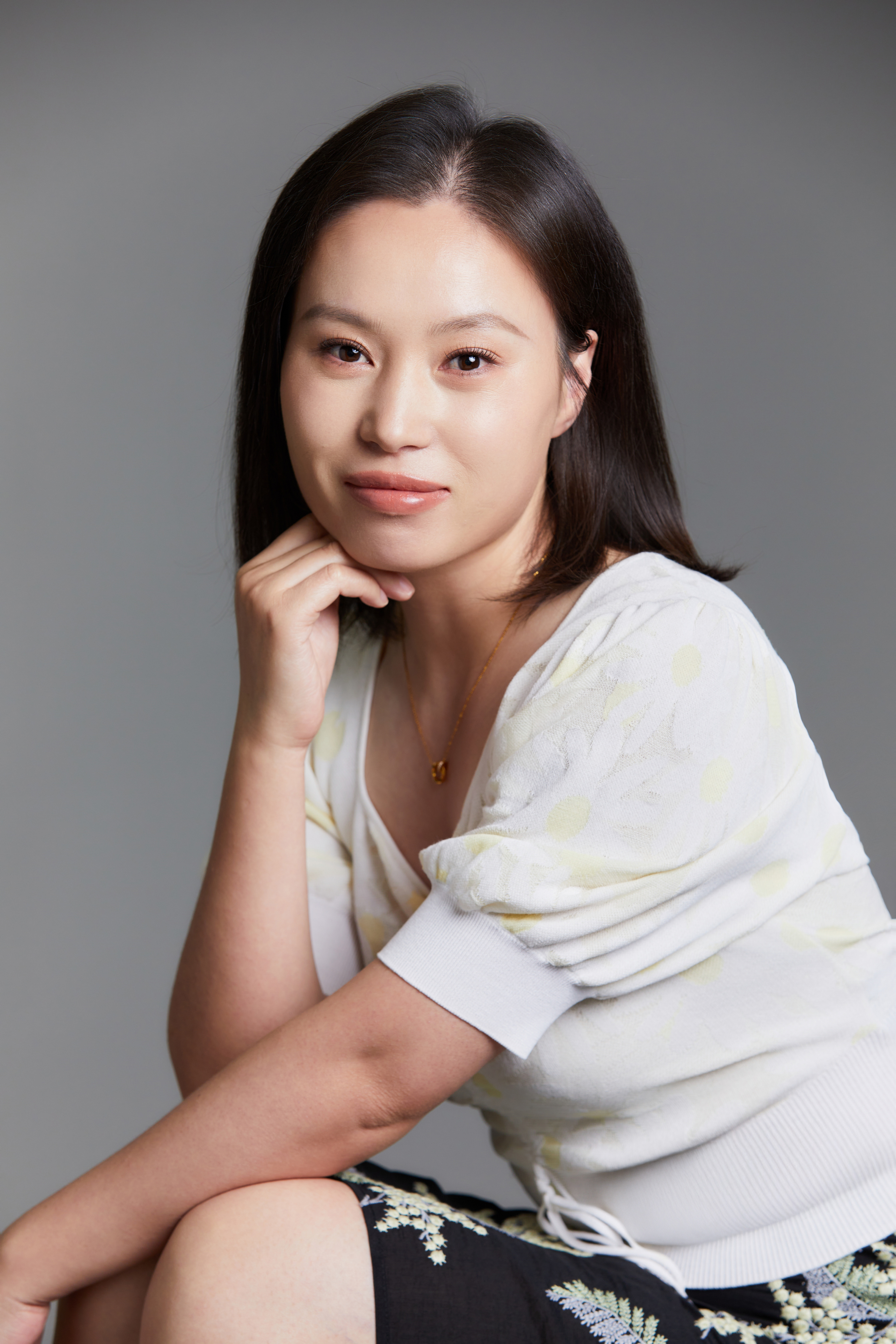}}]{Ning Yang}
is an Associate Professor at the Institute of Automation, Chinese Academy of Sciences. She received her Ph.D. degree from the University of Science and Technology Beijing, China, where she spent time as a visiting Ph.D. student at the Communications and Networking Lab, Northwestern University, USA. Her current research focuses on large language models and their applications in wireless communications.
\end{IEEEbiography}

\begin{IEEEbiography}[{\includegraphics[width=1in,height=1.25in,clip,keepaspectratio]{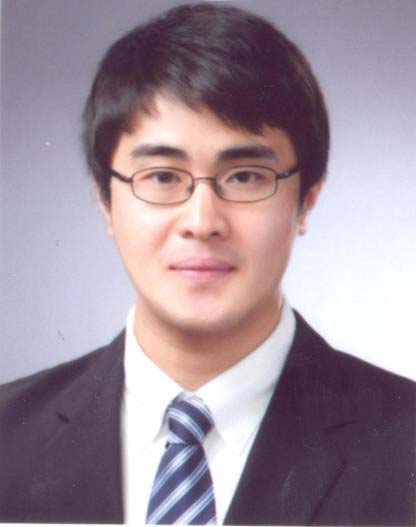}}]{Shengbin Cao}
received the Ph.D. degree in computer science from the City University of Hong Kong, Hong Kong, in 2018. His research interests include wireless networks, including medium-access control (MAC) protocol design, performance analysis, and optimization. He is currently with South China Normal University.
\end{IEEEbiography}

\begin{IEEEbiography}[{\includegraphics[width=1in,height=1.25in,clip,keepaspectratio]{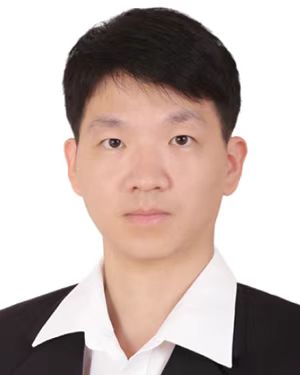}}]{Guocheng Liao}
(Member, IEEE) received the B.E. degree from Sun Yat-sen University in 2016 and the Ph.D. degree from the Department of Information Engineering, The Chinese University of Hong Kong, in 2021. He is currently an Associate Professor with the School of Software Engineering, Sun Yat-sen University. His current research interests include distributed computing and edge intelligence. He received the 2017 IEEE GLOBECOM ComSoc Young Professional Best Paper Award.
\end{IEEEbiography}

\begin{IEEEbiography}[{\includegraphics[width=1in,height=1.25in,trim=157.44bp 0 177.6bp 0,clip,keepaspectratio]{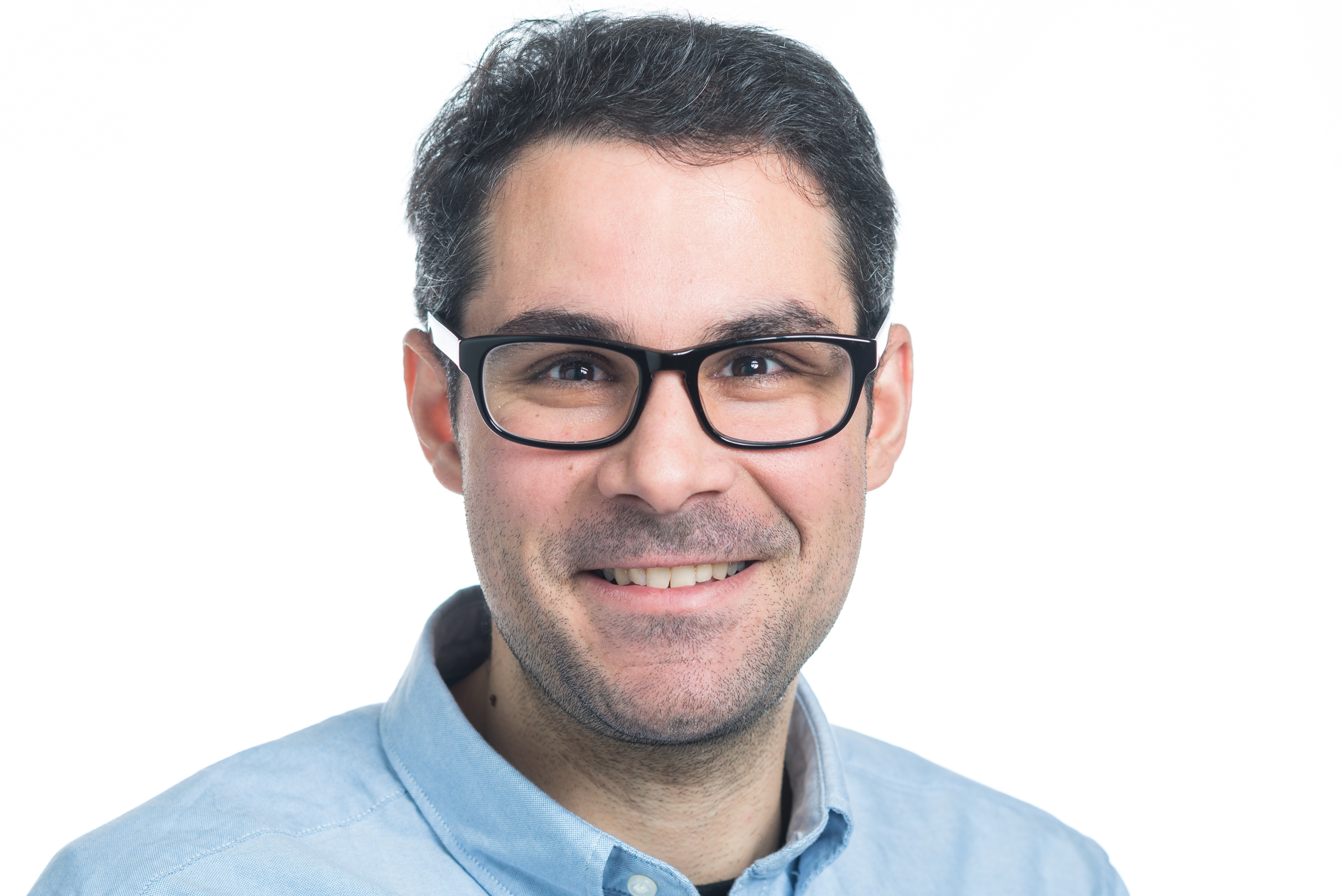}}]{Nikolaos Pappas}
(Senior Member, IEEE) received the first B.Sc. and M.Sc. degrees in computer science, the second B.Sc. degree in mathematics, and the Ph.D. degree in computer science from the University of Crete, Greece, in 2005, 2007, 2012, and 2012, respectively. From 2005 to 2012, he was a Graduate Research Assistant with the Telecommunications and Networks Laboratory, Institute of Computer Science, Foundation for Research and Technology--Hellas, Heraklion, Greece; and a Visiting Scholar with the Institute of Systems Research, University of Maryland, College Park, MD, USA. From 2012 to 2014, he was a Post-Doctoral Researcher with the Department of Telecommunications, CentraleSupelec, France. He is currently a Senior Associate Professor with the Department of Computer and Information Science, Link\"oping University, Link\"oping, Sweden. His main research interests include the field of wireless communication networks, with an emphasis on semantics-aware goal-oriented communications. He has served as the Symposium Co-Chair for the IEEE ICC 2022, IEEE WCNC 2022, IEEE GLOBECOM 2026, and the IEEE PIMRC 2027. He was the General Chair of the 23rd International Symposium on Modeling and Optimization in Mobile, Ad hoc, and Wireless Networks (WiOpt 2025). He is an Area Editor of \textsc{IEEE Open Journal of the Communications Society} and \textsc{IEEE Communications Letters}. He is an Associate Editor for \textsc{IEEE Transactions on Wireless Communications} and \textsc{IEEE Transactions on Mobile Computing}. He is Section Editor-in-Chief for the Entropy. He has served as guest editor in special issues in IEEE Internet of Things Journal, IEEE Network, IEEE Internet of Things Magazine.
\end{IEEEbiography}

\begin{IEEEbiography}[{\includegraphics[width=1in,height=1.25in,clip,keepaspectratio]{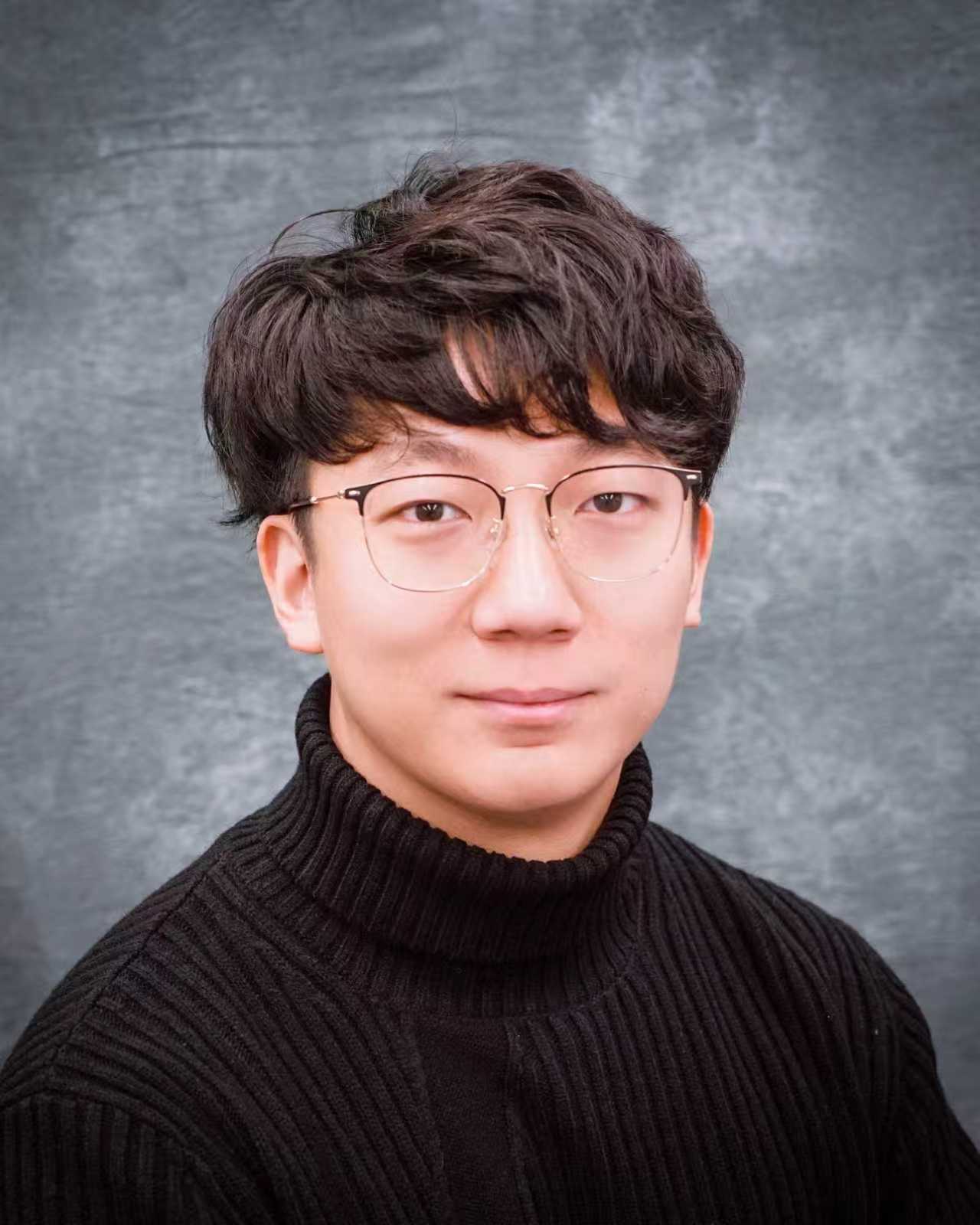}}]{Meng Zhang}
is currently an Assistant Professor with Zhejiang University/University of Illinois at Urbana-Champaign Institute, Zhejiang University. He is also an Adjunct Assistant Professor with the Electrical and Computer Engineering Department, University of Illinois at Urbana--Champaign. He received the Ph.D. degree in information engineering from The Chinese University of Hong Kong in 2019. He was a Visiting Student Research Collaborator with the Department of Electrical Engineering, Princeton University, from 2018 to 2019. He was a Post-Doctoral Fellow with the Department of Electrical and Computer Engineering, Northwestern University, from 2020 to 2021. His primary research interests are network optimization and economics, with a current emphasis on mechanism design and optimization for age of information and edge intelligence. He was a recipient of the WiOpt 2021 Best Paper Award.
\end{IEEEbiography}

\clearpage

\ifthenelse{\boolean{includeAppendix}}{
\appendix

\subsection{Simulation Setting}\label{Appen-A}

\textbf{Hardware Infomation.} We use PyGraph 2.4.0 to implement TGMS, which is trained on NVIDIA GeForce RTX 4090 and tested on AMD EPYC 7763 CPU @1.50GHz with 64 cores under Ubuntu 20.04.6 LTS.

\textbf{Model Parameters.} The parameters for the scheduler, tree generator, and Lagrangian multiplier are shown in Table \ref{Table: Scheduler Setting}, \ref{Table: Tree Generator Setting}, and \ref{Table: Lagrangian Multiplier Setting}, respectively. In practical training processes of RL algorithms, existing works commonly use a discount factor $\gamma<1$ (e.g., 0.99) to stabilize learning and reduce the variance of value estimates. In our implementation, we use $\gamma=0.99$ for both agents, as shown in the parameter tables. The undiscounted value $\gamma=1$ is used only in Lemma \ref{Lemma: equivalence between M2 and P2} to establish the finite telescoping equivalence of the tree-generation objective; it is not used as the training discount factor. To justify this $\gamma=0.99$ surrogate beyond its numerical effect on variance, we invoke a classical total-reward result: under the proper-policy assumption already adopted in Lemma \ref{Lemma: equivalence between M2 and P2}, the value function $V_\gamma(s)$ of the discounted MDP converges to the undiscounted total-reward value $V_1(s)$ as $\gamma\uparrow 1$, and the $\gamma$-discounted optimal policy is asymptotically optimal for the undiscounted total-reward problem \cite{puterman2014markov}. Consequently, training with $\gamma=0.99$ yields a policy that is close to the optimal policy of the $\gamma=1$ formulation used in Lemma \ref{Lemma: equivalence between M2 and P2}.

\begin{table}[!t]
    \centering
    \caption{Scheduler Setting}
    \begin{tabular}{ccc}
    \hline
        \textbf{Module} & \textbf{Parameter} & \textbf{Value} \\ 
        \hline
        \multirow{2}{*}{A2C} & Accumulation Steps & 32 \\
        \multirow{2}{*}{} & Discount factor ($\gamma$) & $0.99$ \\
        \multirow{4}{*}{Network} & Hidden dimensions & 8 \\
        \multirow{4}{*}{} & Attention heads & 3 \\
        \multirow{4}{*}{} & Dropout rate & $0.1$ \\
        \multirow{4}{*}{} & Activation function & PReLU \\
        \multirow{6}{*}{Actor} & Optimizer & AdamW \\
        \multirow{6}{*}{} & Learning Rate & $10^{-3}$ \\
        \multirow{6}{*}{} & Weight Decay & $10^{-4}$ \\
        \multirow{6}{*}{} & Gradient norm & $1$ \\
        \multirow{6}{*}{} & LR scheduler & CosineAnnealing \\
        \multirow{6}{*}{} & LR scheduler: $T_{\max}$ & $50$ \\
        \multirow{7}{*}{Critic} & Optimizer & AdamW \\
        \multirow{7}{*}{} & Learning Rate & $10^{-4}$ \\
        \multirow{7}{*}{} & Weight Decay & $10^{-4}$ \\
        \multirow{7}{*}{} & Gradient norm & $0.1$ \\
        \multirow{7}{*}{} & LR scheduler & ReduceLROnPlateau \\
        \multirow{7}{*}{} & LR scheduler: decay & $0.7$ \\
        \multirow{7}{*}{} & LR scheduler: patience & $20$ \\
    \hline
    \end{tabular}
    \label{Table: Scheduler Setting}
\end{table}
\begin{table}[!t]
    \centering
    \caption{Tree Generator Setting}
    \begin{tabular}{ccc}
    \hline
        \textbf{Module} & \textbf{Parameter} & \textbf{Value} \\ 
        \hline
        \multirow{2}{*}{A2C} & Accumulation Steps & 32 \\
        \multirow{2}{*}{} & Discount factor ($\gamma$) & $0.99$ \\
        \multirow{4}{*}{Network} & Hidden dimensions & 8 \\
        \multirow{4}{*}{} & Attention heads & 3 \\
        \multirow{4}{*}{} & Dropout rate & $0.5$ \\
        \multirow{4}{*}{} & Activation function & PReLU \\
        \multirow{6}{*}{Actor} & Optimizer & AdamW \\
        \multirow{6}{*}{} & Learning Rate & $10^{-2}$ \\
        \multirow{6}{*}{} & Weight Decay & $10^{-4}$ \\
        \multirow{6}{*}{} & Gradient norm & $1$ \\
        \multirow{6}{*}{} & LR scheduler & CosineAnnealing \\
        \multirow{6}{*}{} & LR scheduler: $T_{\max}$ & $50$ \\
        \multirow{7}{*}{Critic} & Optimizer & AdamW \\
        \multirow{7}{*}{} & Learning Rate & $10^{-3}$ \\
        \multirow{7}{*}{} & Weight Decay & $10^{-4}$ \\
        \multirow{7}{*}{} & Gradient norm & $0.1$ \\
        \multirow{7}{*}{} & LR scheduler & ReduceLROnPlateau \\
        \multirow{7}{*}{} & LR scheduler: decay & $0.7$ \\
        \multirow{7}{*}{} & LR scheduler: patience & $20$ \\
    \hline
    \end{tabular}
    \label{Table: Tree Generator Setting}
\end{table}
\begin{table}[!t]
    \centering
    \caption{Lagrangian Multiplier Setting}
    \begin{tabular}{cc}
    \hline
        \textbf{Parameter} & \textbf{Value} \\ 
        \hline
        Init Value & $0.05$ \\
        Training Interval & 100 \\
        Learning Rate & $10^{-5}$ \\
        \hline
    \end{tabular}
    \label{Table: Lagrangian Multiplier Setting}
\end{table}

\textbf{Examples of Graph Topology.} Some examples of the graph topology across datasets with $|\mathcal{V}|=80$ are shown in Fig. \ref{Fig: Graph Topology over Datasets}.

\begin{figure*}[pt]
    \centering
    \subfigure[An example of I080 with $|\mathcal{V}|=80$.]{
        \includegraphics[width = 0.2\textwidth]{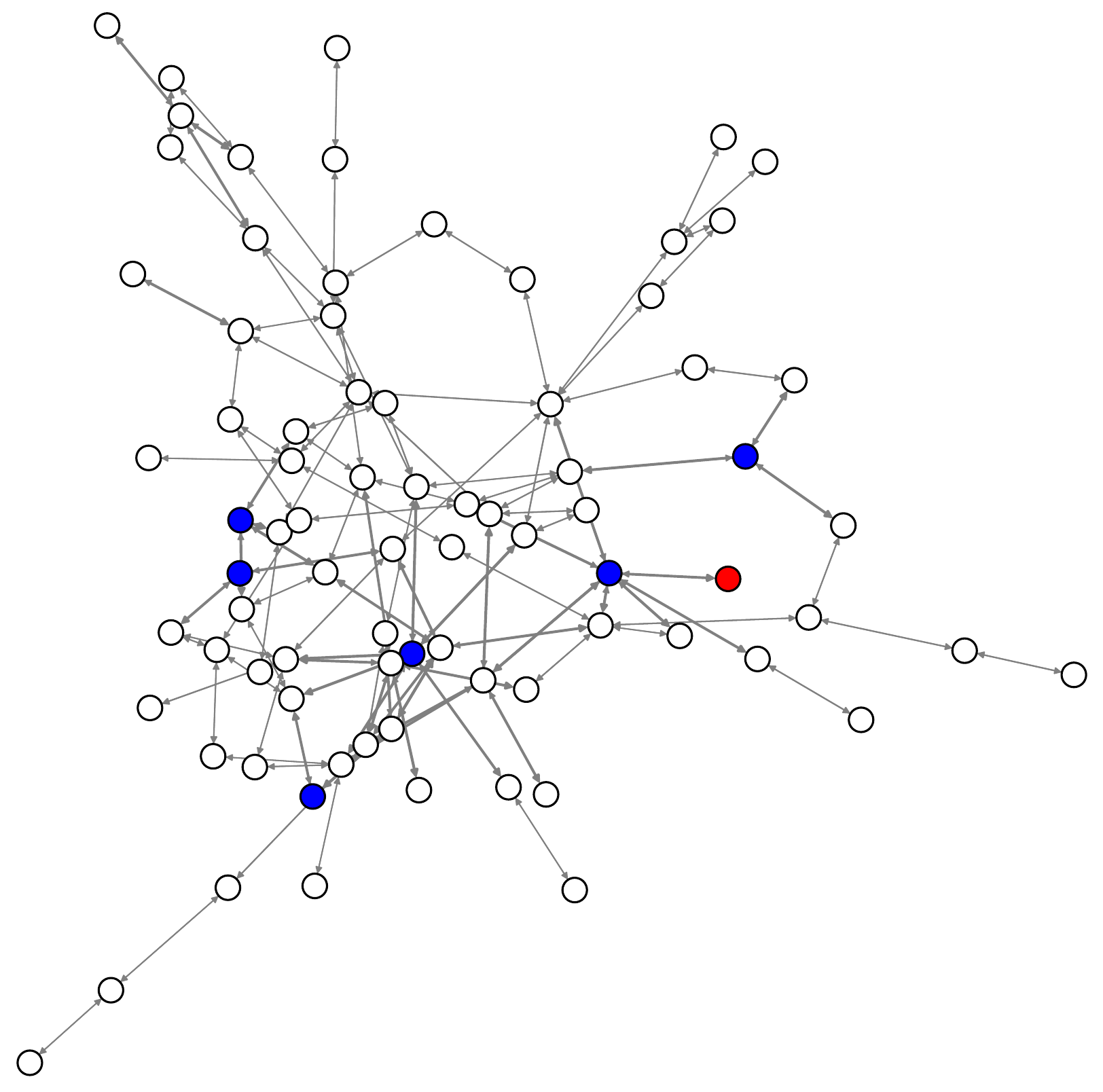}
        \label{Fig: I080}
    }\quad
    \subfigure[An example of AS-733 with $|\mathcal{V}|=80$.]{
        \includegraphics[width = 0.2\textwidth]{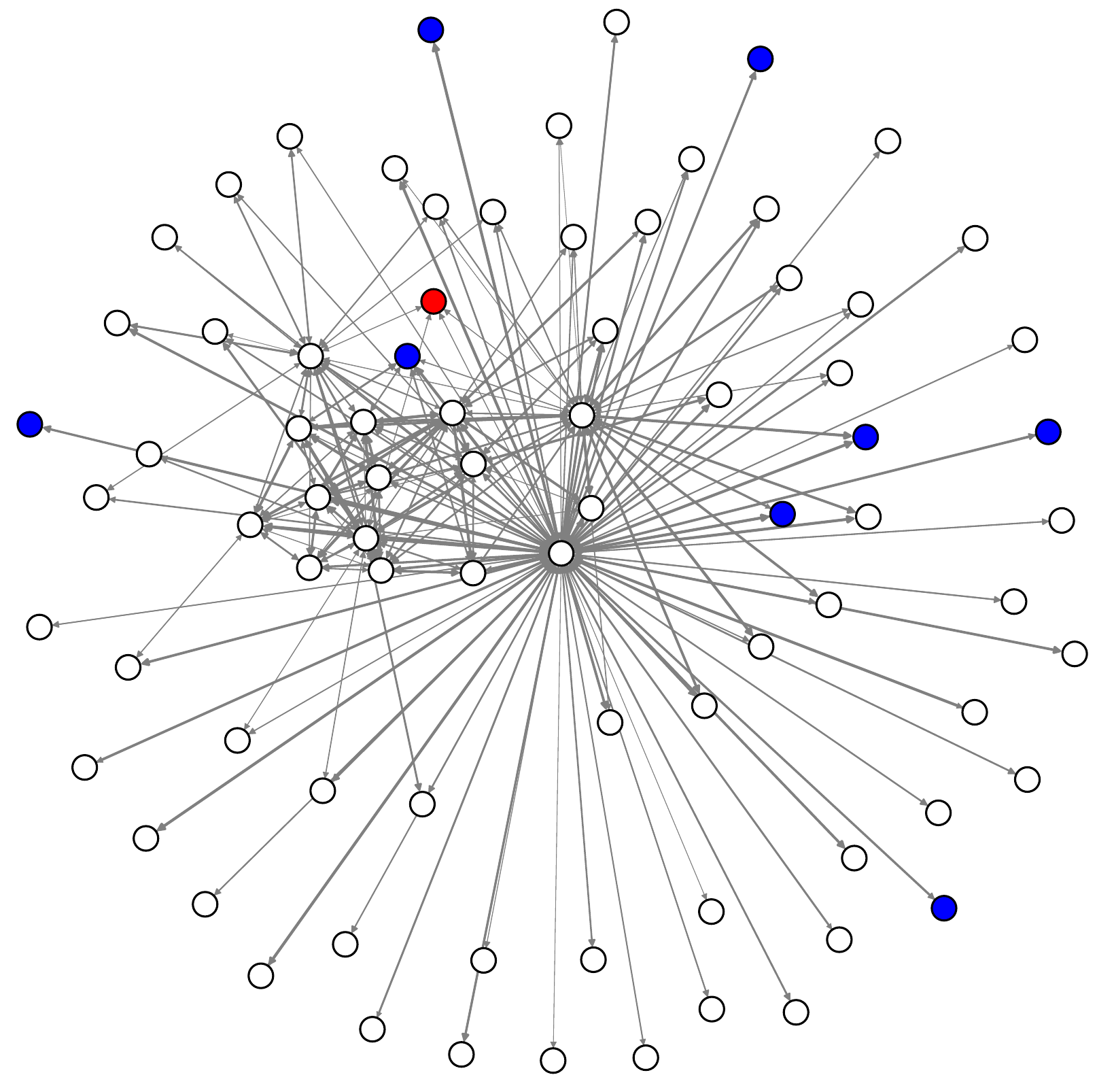}
        \label{Fig: AS-733}
    }\quad
    \subfigure[An example of BA Graphs with $|\mathcal{V}|=80$.]{
        \includegraphics[width = 0.2\textwidth]{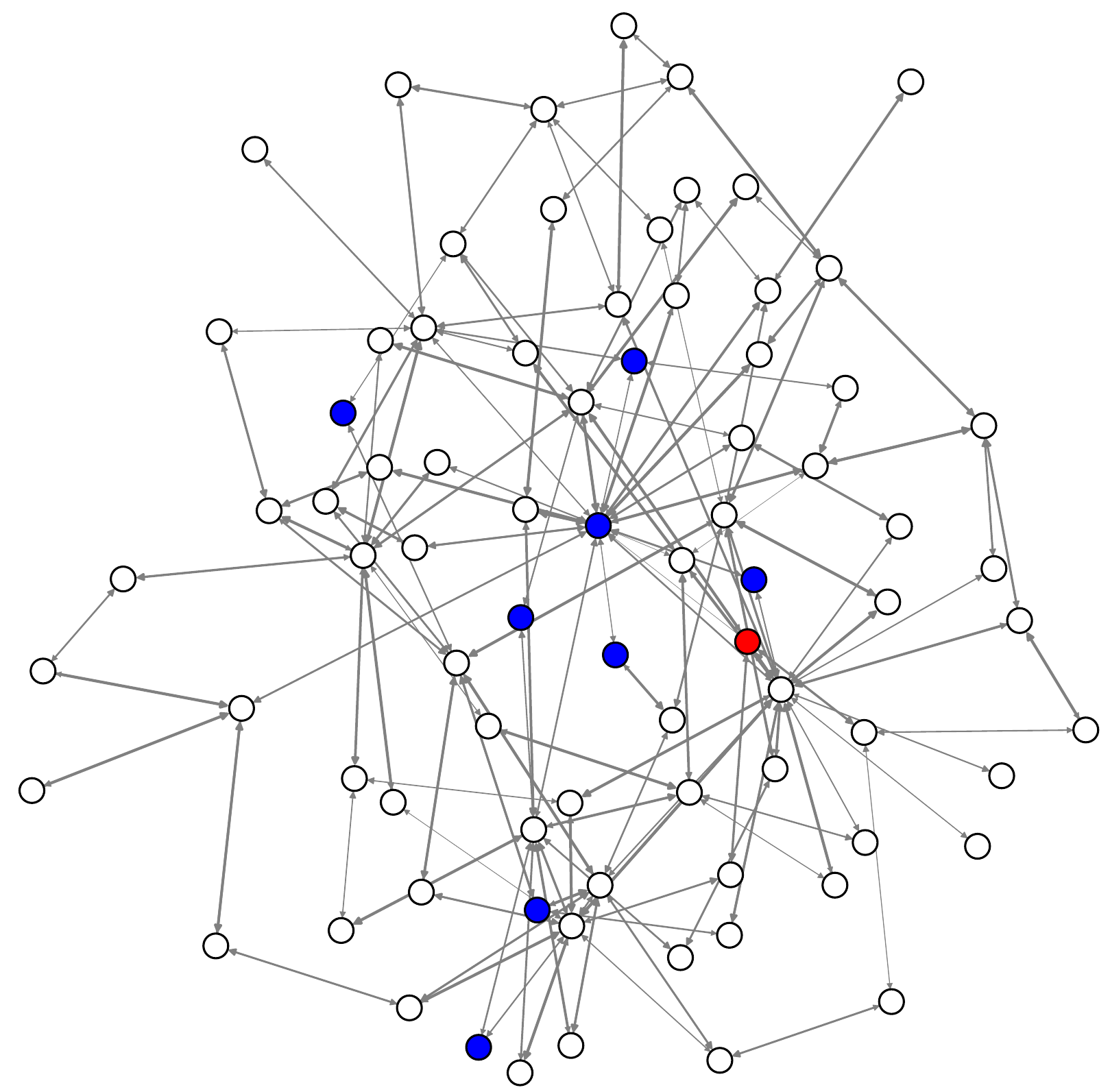}
        \label{Fig: BA}
    }\quad
    \subfigure[An example of ER Graphs with $|\mathcal{V}|=80$.]{
        \includegraphics[width = 0.2\textwidth]{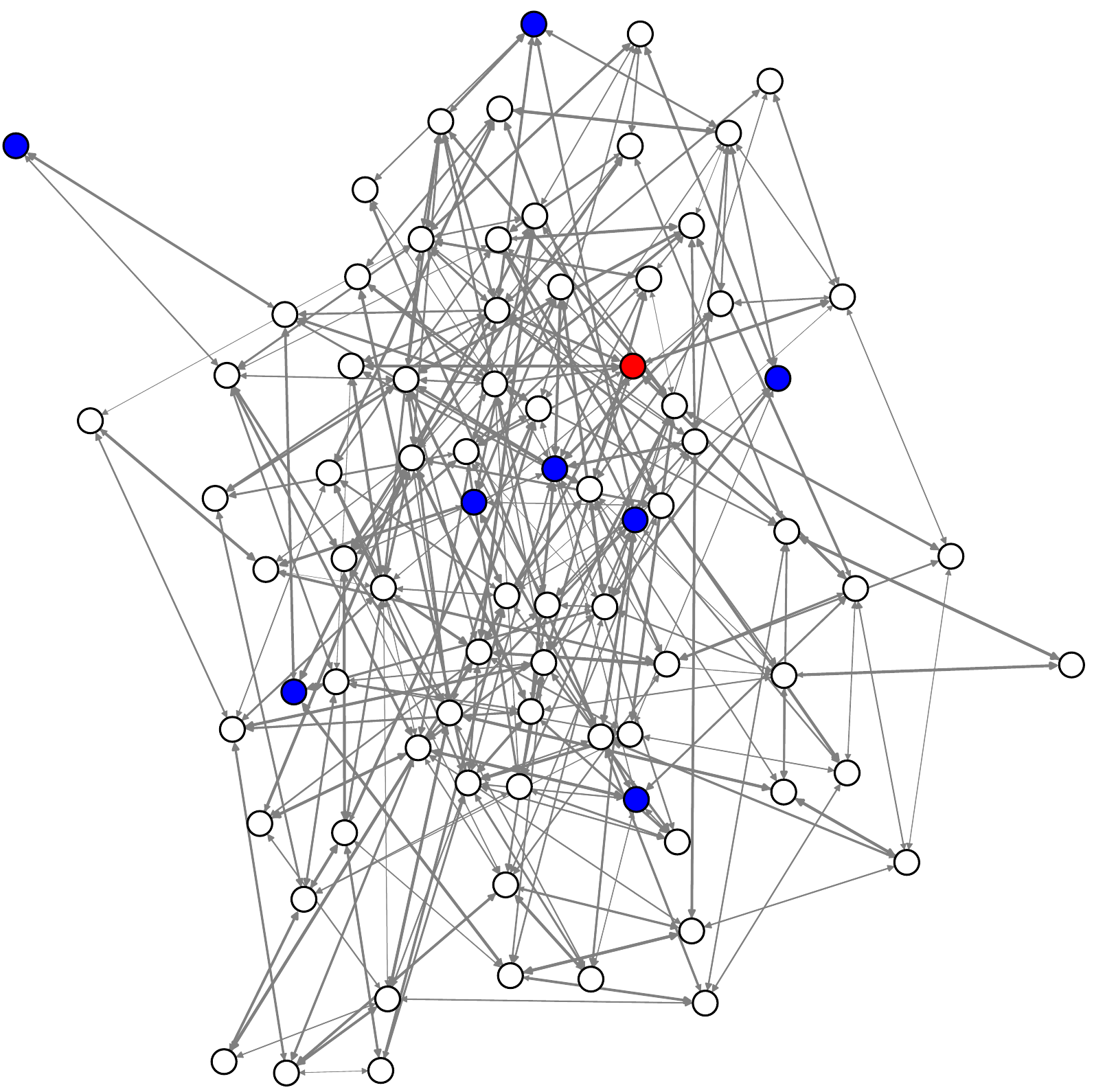}
        \label{Fig: ER}
    }
    \caption{Graph topology across datasets with $|V|=80$. The source node is indicated by the red color, the destination nodes are represented by blue, and the remaining nodes are routers. The width of the edges is proportional to the cost of the edges.}
    \label{Fig: Graph Topology over Datasets}
\end{figure*}

\subsection{MDP Formulation of STP} \label{Appe-B}
Here, we describe the MDP formulation considered in our experiments for the STP problem. This MDP formulation is similar to the tree-generating MDP. For a given graph $\mathcal{G}=(\mathcal{V}, \mathcal{E})$ with terminals $\mathcal{U}\subseteq\mathcal{V}$, the MDP formulation of STP can be defined as:
\begin{itemize}
    \item \textbf{States:} The state $s_t$ at time $t$ is defined as:
    \begin{equation}
        \label{STP: State}
        s_t=\{\mathcal{P}_t, \mathcal{U}\}, s_t\in \mathbb{R}^{\hat{\mathcal{V}}^2 + 2\hat{\mathcal{V}}}.
    \end{equation}
    The state $s_t$ includes the current partial solution $\mathcal{P}_t$ and the given terminals $\mathcal{U}$. When all terminals are covered by $\mathcal{P}_t$, $s_t$ will be a terminal state and $\mathcal{P}_t$ is the generated solution for the STP problem.
    \item \textbf{Actions:} The action space is defined as:
    \begin{equation}
        \label{STP: Action Space}
        \mathcal{A} = \{v|v\in \mathcal{V}, v\notin \mathcal{V}^{\mathcal{P}}_t, \exists u\in \mathcal{V}^{\mathcal{P}}_t, (u, v)\in \mathcal{E}\}.
    \end{equation}
    This means that the action $a_\tau\in\mathcal{A}_2$ is to choose a node $v$ that is not included in $\mathcal{P}_\tau$ and the node $v$ should be connected to at least one node $u$ in $\mathcal{P}_t$. In other words, each action $a_{t}\in\mathcal{A}_2$ is a neighbor of the current partial solution $\mathcal{P}_t$.
    \item \textbf{Transition:} The transition is deterministic here and corresponds to tagging the selected node as the last action.
    \item \textbf{Rewards:}
    The reward function is defined as:
    \begin{equation}
        \label{STP: Reward}
        r(s_t, a_t) = 
        \begin{cases}
            1 - c((v^*, a_t))/C_{\max} & \text{if } a_t \text{ is a terminal}, \\
            -c((v^*, a_t))/C_{\max} & \text{ otherwise},
        \end{cases}
    \end{equation}
    where $C_{\max}$ denote the maximum cost of all edges.
\end{itemize}

\subsection{Complete TGMS Algorithm}
The pseudo-code list of the whole TGMS algorithm architecture is described in \textbf{Algorithm \ref{PC: Whole TGMS}}.
\begin{algorithm}[!t]
    
    \caption{TGMS Framework: Complete Training and Inference Pipeline}
    \label{PC: Whole TGMS}
    \begin{algorithmic}[1]
        \Require Training dataset $\mathcal{D}$, and a network topology $\mathcal{G}_{\text{test}}$ for inference.
        \Ensure Optimized scheduling decisions $\bm{\mathcal{U}}'$ and multicast trees $\bm{\mathcal{T}}$ for $\mathcal{G}_{\text{test}}$.
        
        \Statex \textbf{// Phase 1: Initialization}
        \State Initialize Scheduler network parameters including an actor $\bm{\theta}_1$ and a critic $\bm{\omega}_1$.
        \State Initialize Tree Generator network parameters including an actor $\bm{\theta}_2$ and a critic $\bm{\omega}_2$.
        \State Initialize Lagrangian multiplier $\lambda \leftarrow 0$.
        
        \Statex \textbf{// Phase 2: Training Process}
        \State Execute \textbf{Algorithm \ref{PC: Training Pipeline} (Training Pipeline)} to train the scheduler $\{\bm{\theta}_1, \bm{\omega}_1\}$ and tree generator $\{\bm{\theta}_2, \bm{\omega}_2\}$.
        \State \quad \textbullet~Iteratively collect experiences $B_s, B_t$ and $B_\lambda$.
        \State \quad \textbullet~Update $\bm{\theta}_1, \bm{\omega}_1, \bm{\theta}_2, \bm{\omega}_2$ using \textbf{Algorithm \ref{PC: A2C} (A2C)}.
        \State \quad \textbullet~Update $\lambda$ until convergence.
        \State Freeze the trained parameters $\bm{\theta}^*_1$ and $\bm{\theta}^*_2$ for inference.

        \Statex \textbf{// Phase 3: Inference Process}
        \State Construct the initial network graph $\mathcal{G}_0$ from $\mathcal{G}_{\text{test}}$.
        \State Extract node features $\mathbf{x}_0$ and edge features $\mathbf{e}_0$ from $\mathcal{G}_0$.
        \State Execute \textbf{Algorithm 1 (TGMS)} with input $\{\mathcal{G}_0, \mathbf{x}_0, \mathbf{e}_0\}$ using the trained scheduler $\bm{\theta}^*_1$ and tree generator $\bm{\theta}^*_2$.
        \State Get scheduling decisions $\bm{\mathcal{U}}'$ and multicast trees $\bm{\mathcal{T}}$.
        \State \textbf{return} $\bm{\mathcal{U}}'$ and $\bm{\mathcal{T}}$.
    \end{algorithmic}
\end{algorithm}
The TGMS framework consists of three main phases: initialization, training, and inference. In the initialization phase, we set up the network parameters for both the scheduler and tree generator, as well as the Lagrangian multiplier. During the training phase, we iteratively collect experiences and update the network parameters using the A2C algorithm until convergence. Finally, in the inference phase, we construct the initial network graph from the test dataset, extract node and edge features, and execute the TGMS algorithm using the trained policies to obtain optimized scheduling decisions and multicast trees.

\subsection{Computational Complexity Analysis}
Here we analyze the computational complexity of TGMS. We start by analyzing the complexity of one forwarding process of the tree generator or the scheduler. Let $|\mathcal{V}|$ and $|\mathcal{E}|$ denote the number of nodes and edges in the network, respectively. For a single NGAT layer, let $d_{in}$ denote the dimension of the input node features, and $d_{out}$ denote the dimension of the output node features. We need to calculate the attention coefficients, which has a complexity of $\mathcal{O}(|\mathcal{E}| d_{in} + |\mathcal{V}| d_{in} d_{out})$. Assuming that there are $L$ NGAT layers, the total complexity of the NGAT layers is $\mathcal{O}(L (|\mathcal{E}| d_{in} + |\mathcal{V}| d_{in} d_{out}))$. The pooling layer has a complexity of $\mathcal{O}(|\mathcal{V}| d_{out})$. The policy and value heads have a complexity of $\mathcal{O}(d_{out}^2)$ each. Since $L$, $d_{in}$, and $d_{out}$ are constants, the overall complexity of one forwarding process is:
\begin{equation}
    \begin{aligned}
    &\mathcal{O}\left(L (|\mathcal{E}| d_{in} + |\mathcal{V}| d_{in} d_{out}) + |\mathcal{V}| d_{out} + 2 d_{out}^2\right) \\
    = &\mathcal{O}\left( |\mathcal{E}| + |\mathcal{V}| \right).
    \end{aligned}
\end{equation}

For the complexity of generating a multicast tree, the scheduler might select all destinations in the worst case. Thus, the tree generator may need to add all nodes to the multicast tree. Therefore, the maximum number of steps for generating a multicast tree is $|\mathcal{V}|$. For a centralized controller, the overall complexity of generating a multicast tree is:
\begin{equation}
    \begin{aligned}
    \mathcal{O}\left( |\mathcal{V}| (|\mathcal{E}| + |\mathcal{V}|) \right) &= \mathcal{O}\left( |\mathcal{V}||\mathcal{E}| + |\mathcal{V}|^2 \right) &= \mathcal{O}\left( |\mathcal{V}||\mathcal{E}| \right).
    \end{aligned}
\end{equation}
The last equality is because of $|\mathcal{E}|=\mathcal{O}(|\mathcal{V}|)$ in connected graphs. Therefore, the overall complexity of TGMS for each time slot is $\mathcal{O}(|\mathcal{V}||\mathcal{E}|)$. 

\subsection{Choices over RL algorithms}

    A2C is not a defining design element of TGMS. In principle, any RL algorithm could learn the optimal policies for our MDPs. We chose A2C primarily because it suits the unique challenges of our problem.
    Specifically, the primary advantage of A2C in our context is its ability to handle the high variance that can arise from our action space mapping. During the tree generation process, the agent's action probabilities must be constrained to a feasible action space dictated by the network topology. This remapping step can introduce significant variance into the policy gradient estimates. By using a critic to compute an advantage function, A2C subtracts a state-dependent baseline from the raw reward, which substantially reduces this variance and leads to more stable and reliable training.
    Furthermore, A2C offers practical benefits in terms of efficiency, particularly for the large-scale graphs our framework is designed for. Compared to methods like Proximal Policy Optimization \cite{schulman2017proximal} (PPO) that often rely on storing and sampling from large replay buffers, A2C is more memory-efficient as it can learn directly from sequential trajectories. This is a critical consideration, as buffer-intensive methods can lead to out-of-memory errors when applied to very large graphs, making A2C a more scalable choice for our setting.
    Finally, while more recent algorithms like PPO, Deep Deterministic Policy Gradient \cite{lillicrap2015continuous} (DDPG), or Soft Actor Critic \cite{haarnoja2018soft} (SAC) can sometimes offer further stability, they introduce additional implementation and tuning complexity, such as clipping parameters or separate target networks. A2C provided a robust and well-understood foundation that allowed us to clearly demonstrate the effectiveness of our overall graph-based multicast framework without confounding the results with an overly complex learning algorithm.
    We also compare the performance between PPO and A2C on the I080 dataset as described in Table \ref{Experiment: Different RL algorithms}. The unique parameters of PPO are set as in Table \ref{PPO parameters}. Other parameters are the same as A2C.
    \begin{table}[!t]
        \centering
        \caption{PPO Unique Parameters}
        \label{PPO parameters}
        \begin{tabular}{|c|c|}
        \hline
            \textbf{Parameter} & \textbf{Value} \\
            \hline
            Clip Ratio & 0.2 \\
            Max Buffer Size & 10 \\
            Sample Size & 3 \\
        \hline
        \end{tabular}
    \end{table}
    \begin{table}[!t]
        \centering
        \caption{Experiment Results of STP on A2C and PPO}
        \label{Experiment: Different RL algorithms}
        \begin{tabular}{|c|cccc|}
        \hline
            \multirow{2}{*}{\textbf{Algorithm}} & \multicolumn{4}{c|}{\textbf{I080}} \\
            & $\rho$ & $\rho_{worst}$ & $\rho_{best}$ & $t(s)$ \\
            \hline
            \textbf{TG-NGAT(A2C)} & 1.149 & 1.504 & 1.018 & 0.285 \\
            \textbf{TG-NGAT(PPO)} & 1.540 & 2.172 & 1.016 & 0.461 \\
        \hline
        \end{tabular}
    \end{table}
    We observe that PPO does not outperform A2C under the same hyperparameters. Due to limited academic access to GPU resources, we were unable to train PPO agents on larger graphs, further highlighting the lightweight nature of A2C for our problem.

\subsection{Robustness in Dynamic Networks} Our current model assumes a static topology for the entire duration of a packet transmission, but our formulation can naturally adapt to dynamic scenarios since the scheduling process is formulated as an MDP. To address this more rigorously, we enhance our experimental results here. We introduce the following dynamics-related parameters into our simulation environment:
\begin{itemize}
    \item \textbf{Update Interval ($\delta_{int}$):} This parameter controls how frequently the network topology is updated. A shorter interval simulates a more dynamic network.
    \item \textbf{Edge Dynamic Probability ($\delta_{prob}$):} This parameter defines the ratio of edges that may change state (specified as follows) during each network update. A higher probability indicates a more volatile network.
\end{itemize}
To measure the framework's robustness, e.g., packet losses, we consider two modes of network dynamics:
\begin{itemize}
    \item \textbf{Edge Activation/Deactivation:} In this mode, during each update interval, a fraction ($\delta_{prob}$) of existing edges are randomly deactivated (simulating link failures), and a fraction of previously inactive edges are activated (simulating link recoveries).
    \item \textbf{Random Rewriting:} In this mode, during each update interval, a fraction ($\delta_{prob}$) of edges are randomly rewired to connect different pairs of nodes, simulating more complex topology changes.
\end{itemize}
We run simulations under the above dynamic network scenarios with varying $\delta_{int}$ and $\delta_{prob}$ values to evaluate the performance of our TGMS framework. The tested dataset is I080 and the energy constraint is set to $W=4$. Other parameters are the same as in the static network scenario. An example of the dynamic network with $\delta_{int}=10$ and $\delta_{prob}=0.06$ is described in Fig. \ref{fig: little dynamic AoI}.
\begin{figure}[!t]
    \centering
    \includegraphics[width=0.48\textwidth]{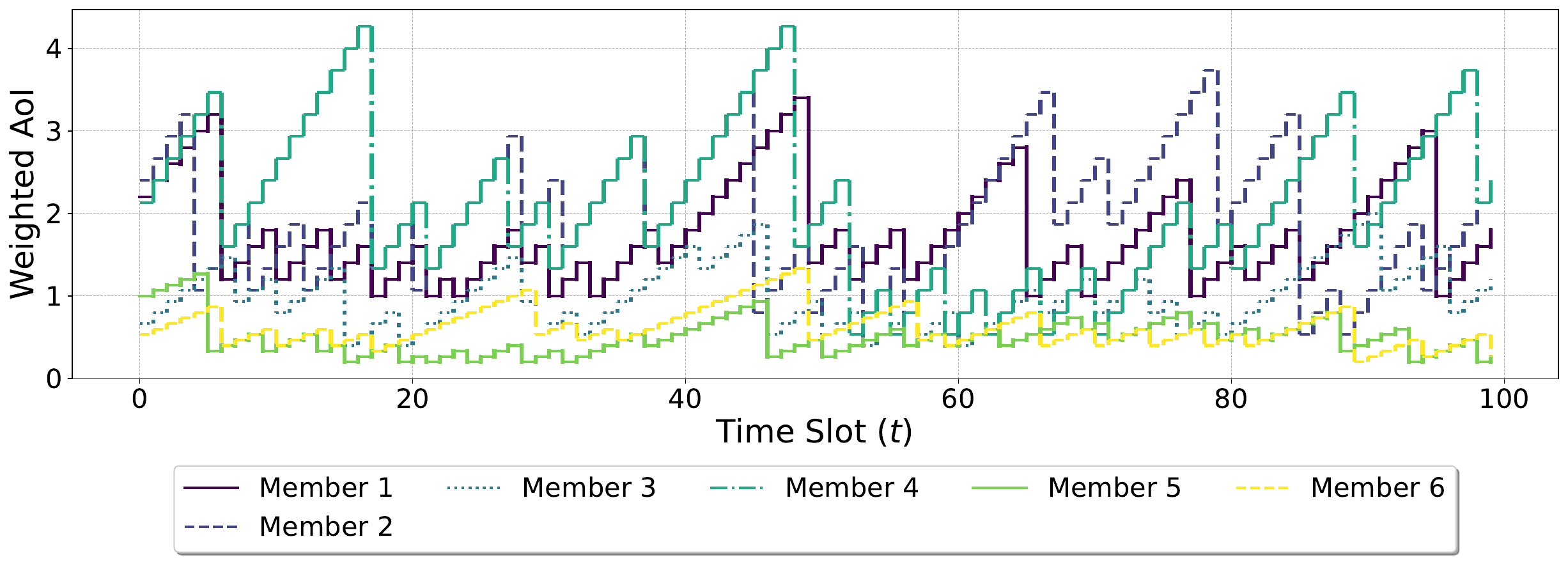} 
    \caption{An Example: Weighted AoIs under random rewriting dynamics ($\delta_{int}=10$, $\delta_{prob}=0.06$).}
    \label{fig: little dynamic AoI}
\end{figure}
We observe that the AoI of Member 1 is upgraded to different levels at different time slots due to the dynamic topology. In this scenario, the packet loss may be low since the scheduler can adaptively select other available paths to forward the packets, which shows the robustness of our framework. We also present a cumulative weighted AoI of the above example in Fig. \ref{fig: little dynamic cumulative AoI}.
\begin{figure}[!t]
    \centering
    \includegraphics[width=0.48\textwidth]{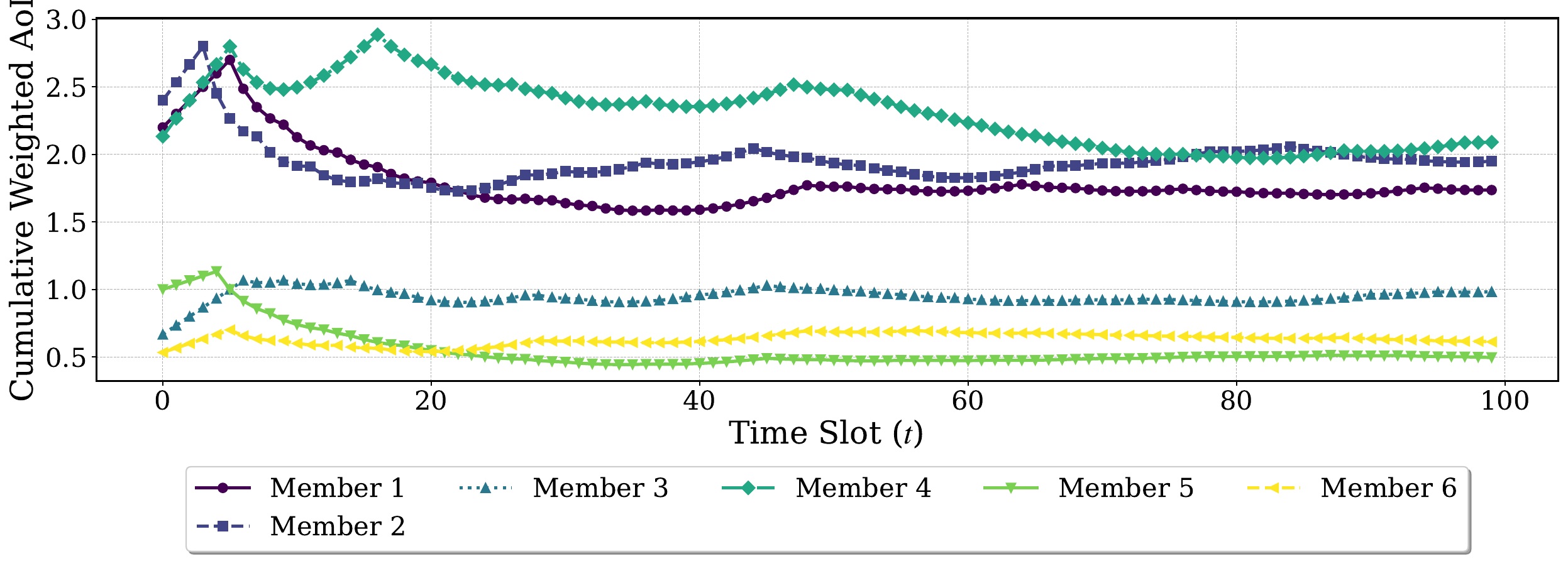}
    \caption{An Example: Cumulative weighted AoI under random rewriting dynamics ($\delta_{int}=10$, $\delta_{prob}=0.06$).}
    \label{fig: little dynamic cumulative AoI}
\end{figure}
It shows that the cumulative weighted AoI converges after about 80 time slots, which presents the effectiveness of our framework in dynamic networks. In some extreme cases, the packet loss may be high when there is no available path from the source to the destination. We present an example under $\delta_{int}=5$ and $\delta_{prob}=0.1$ in Fig. \ref{fig: huge dynamic}.
\begin{figure}[!t]
    \centering
    \includegraphics[width=0.48\textwidth]{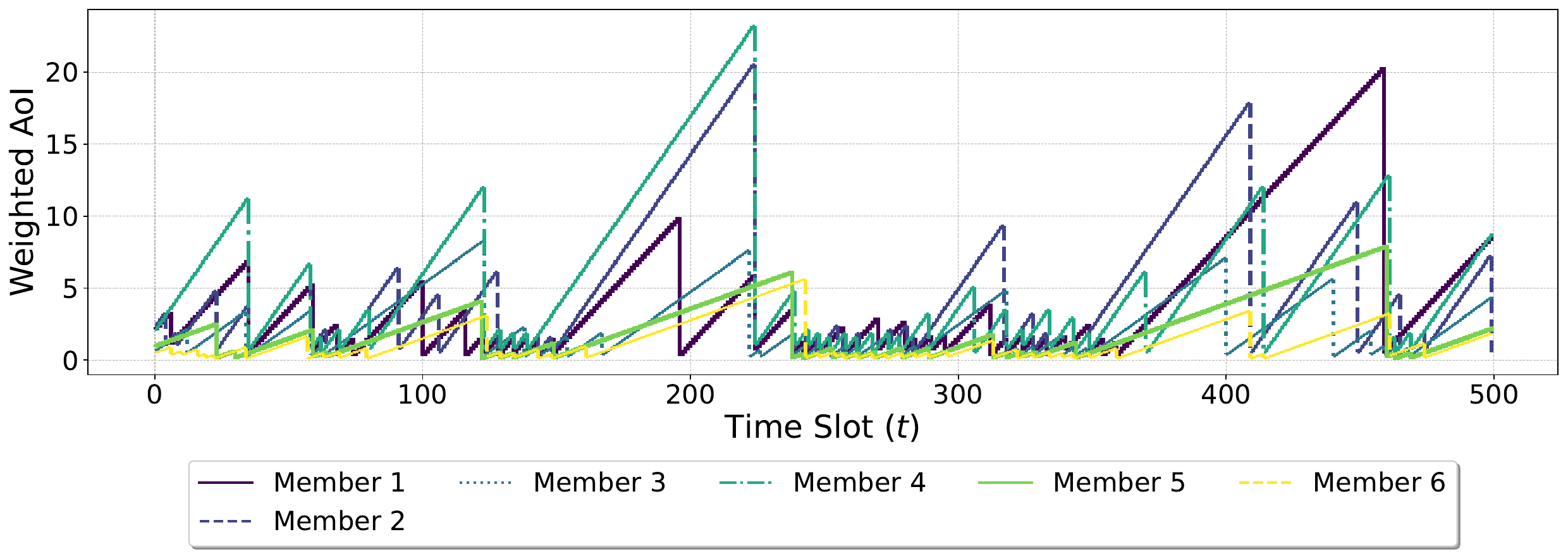}
    \caption{An Example: Extreme dynamic scenario ($\delta_{int}=5$, $\delta_{prob}=0.2$).}
    \label{fig: huge dynamic}
\end{figure}
We also observe that the AoIs of some members keep increasing over 50 time slots since there is no available path from the source to them. On the one hand, it shows that our framework also relies on feasible paths to forward packets, just like traditional routing protocols. On the other hand, it shows that our framework can recover from extreme dynamic scenarios since the AoIs of all members are finally upgraded when some paths re-appear.\par

\textbf{Robustness Evaluation.} We give the average loss rates under two dynamic modes with different network dynamics. The results are summarized in Fig. \ref{fig: activation robustness} and Fig. \ref{fig: change robustness}.
\begin{figure}[!t]
    \centering
    \includegraphics[width=0.45\textwidth]{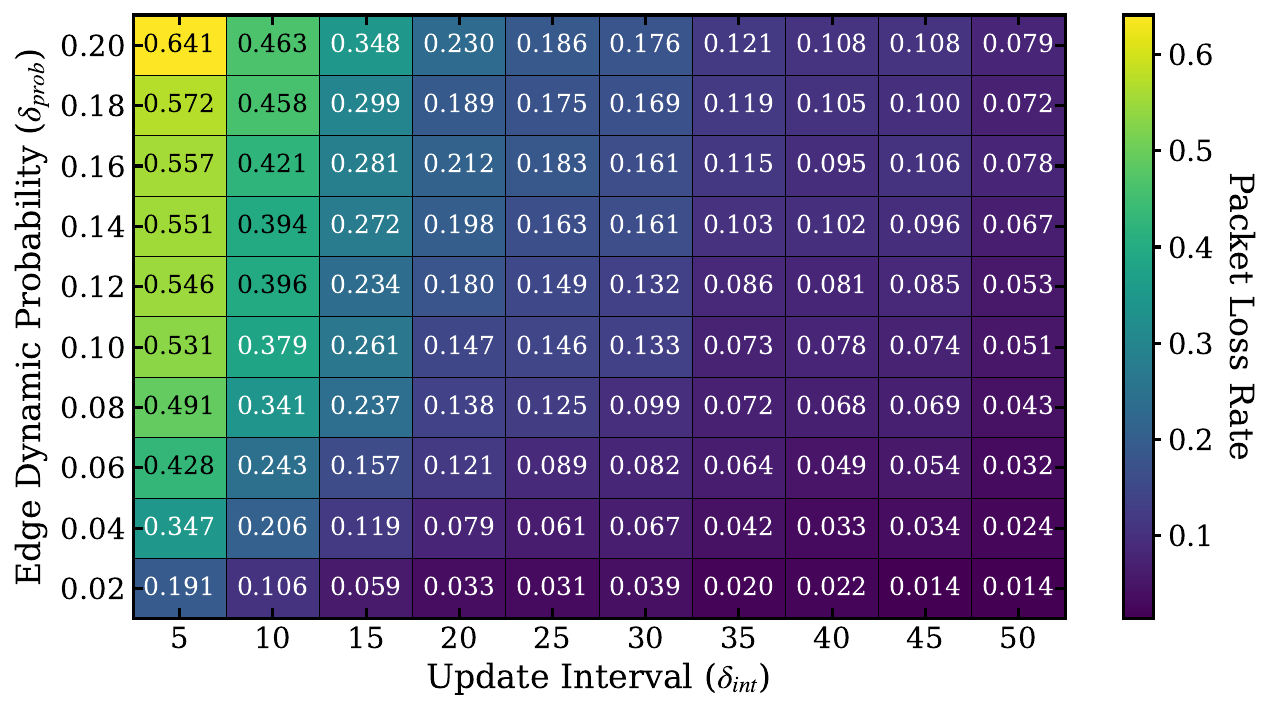}
    \caption{Packet loss rates under edge activation/deactivation dynamics.}
    \label{fig: activation robustness}
\end{figure}
\begin{figure}[!t]
    \centering
    \includegraphics[width=0.45\textwidth]{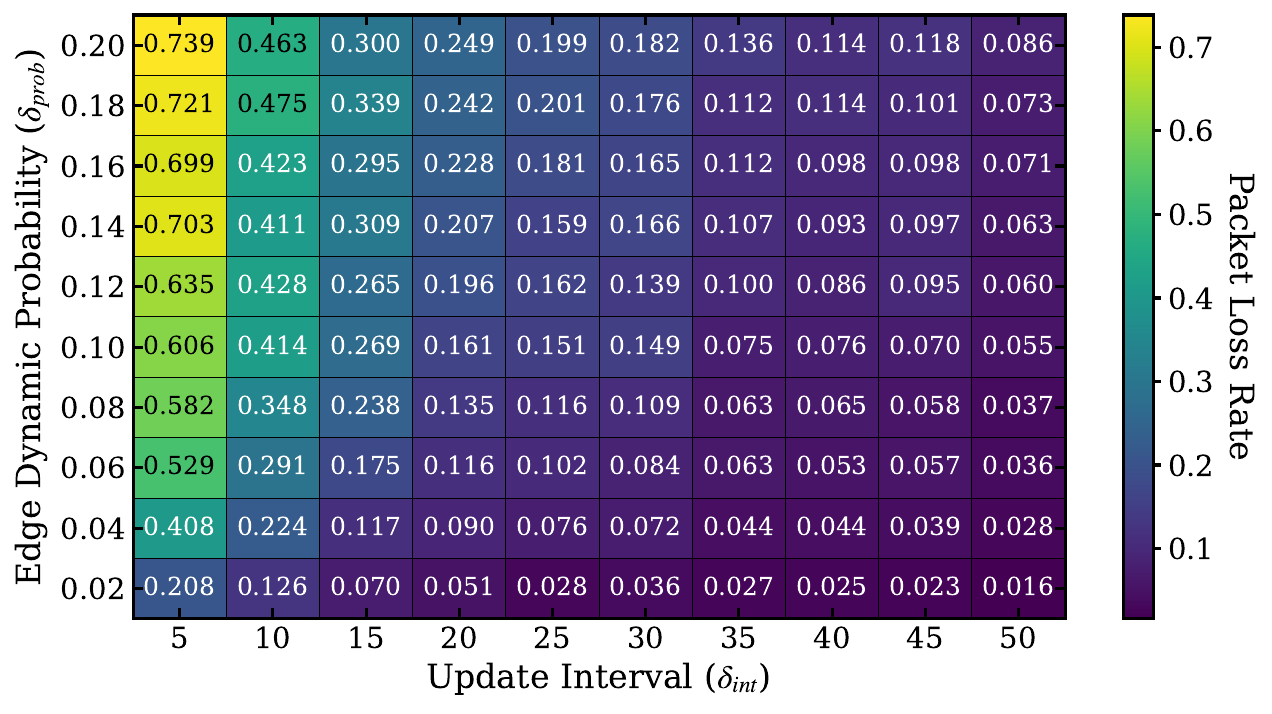}
    \caption{Packet loss rates under random rewriting dynamics.}
    \label{fig: change robustness}
\end{figure}
We observe that the packet loss rate increases with the increase of $\delta_{prob}$ and the decrease of $\delta_{int}$, which is intuitive since a more dynamic network leads to more link failures during transmissions. However, even under quite dynamic scenarios (e.g., $\delta_{int}=20$ and $\delta_{prob}=0.2$), the packet loss rate is still below 25\%. That means in a network which changes its topology about every 20 milliseconds (assume one millisecond per hop), only 25\% of packets are lost on average. This shows the robustness of our framework in dynamic networks. 
\par

\textbf{Routing Entry Cache.} To further reduce the packet loss rate in dynamic networks, we can also maintain a routing entry cache in each node to store multiple next-hop options for each destination. When a link failure occurs during transmission, the node can quickly switch to an alternative next hop from its cache, rather than waiting for the central controller to generate a new multicast tree. This can significantly improve the robustness of our framework in highly dynamic scenarios. Under a cache with 5 maximum entries, we re-evaluate the packet loss rate under the same dynamic scenarios. The results are summarized in Fig. \ref{fig: activation cache} and Fig. \ref{fig: change cache}.
\begin{figure}[!t]
    \centering
    \includegraphics[width=0.45\textwidth]{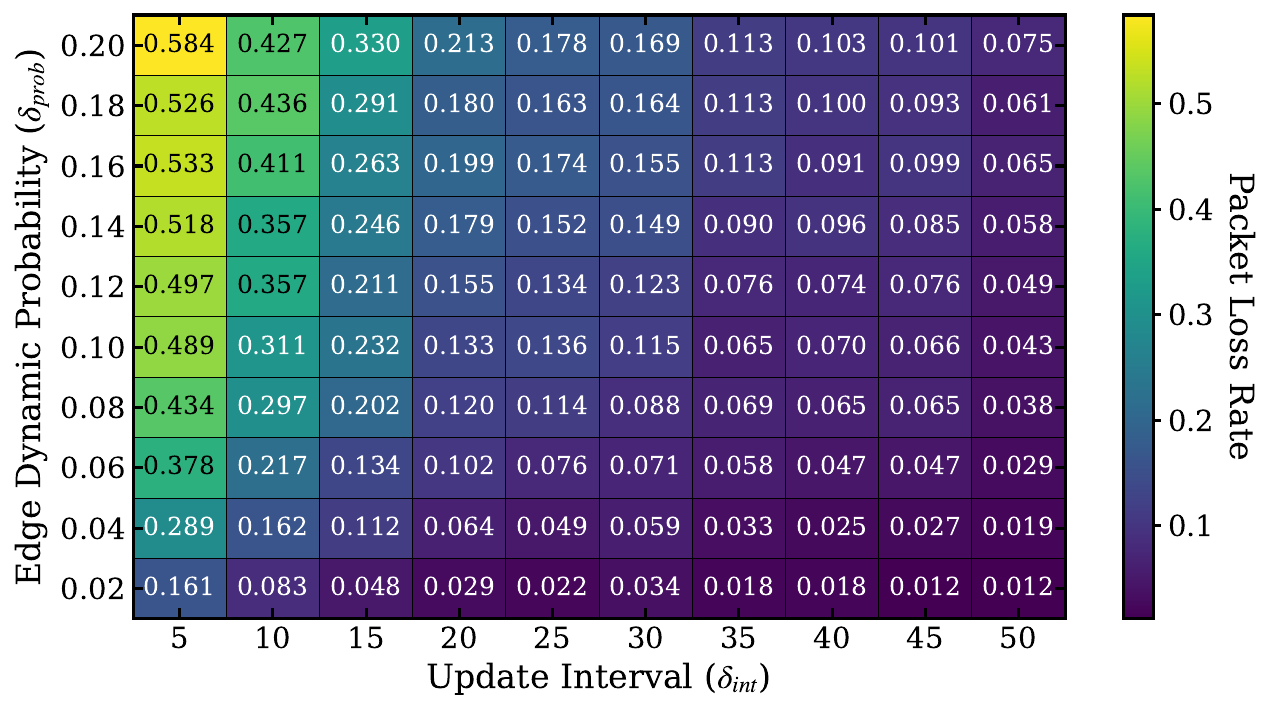}
    \caption{Packet loss rate with routing cache under edge activation/deactivation dynamics.}
    \label{fig: activation cache}
\end{figure}
\begin{figure}[!t]
    \centering
    \includegraphics[width=0.45\textwidth]{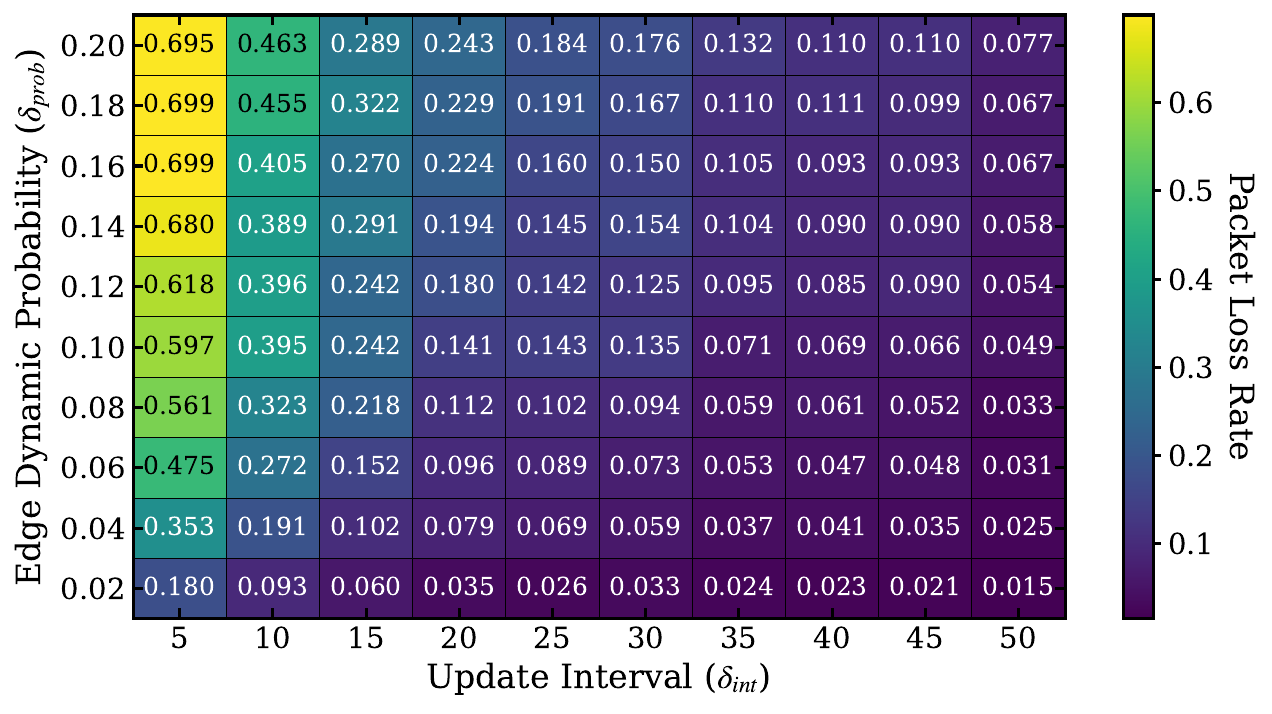}
    \caption{Packet loss rate with routing cache under random rewriting dynamics.}
    \label{fig: change cache}
\end{figure}
We can observe that the packet loss rate is reduced in most dynamic scenarios when using the routing entry cache. This shows that maintaining multiple next-hop options can effectively enhance the robustness of our framework in dynamic networks.

\subsection{Two-Timescale Update Interval}\label{App: Update Interval}
Figure~\ref{Fig: Update Interval} shows the average AoI across update intervals for the four network topologies. Over the plotted range, the average AoI varies less for AS-733 and ER-80 than for I080 and BA-80.
\begin{figure}[!t]
    \centering
    \includegraphics[width=\columnwidth]{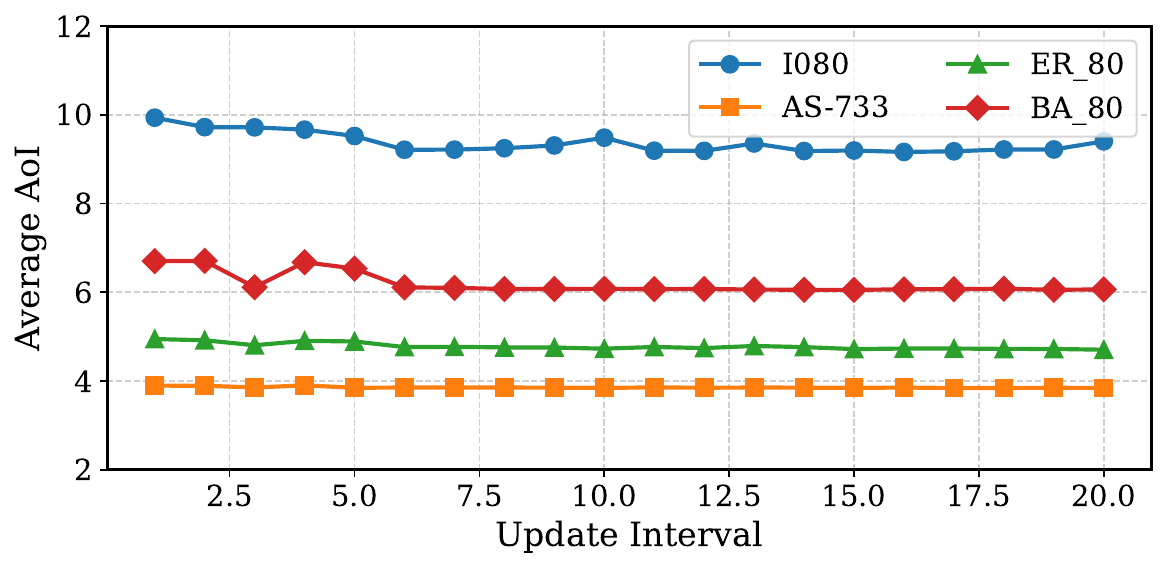}
    \caption{Average AoI versus update interval for four network topologies.}
    \label{Fig: Update Interval}
\end{figure}

\subsection{Convergence of Graph Embedding Methods.} \rev{Figure \ref{fig: GNN Convergence} reports empirical training behavior of the learned graph embedding models. While GATv2 and NGAT achieve comparable peak performance, NGAT exhibits more stable performance after approximately 100 episodes in the reported experiment, whereas the other methods degrade. These observations complement the fixed-attention analysis, but do not establish non-expansiveness of embedding-dependent attention or a general convergence guarantee for training.}

\begin{figure}[!t]
    \centering
    \includegraphics[width=0.42\textwidth]{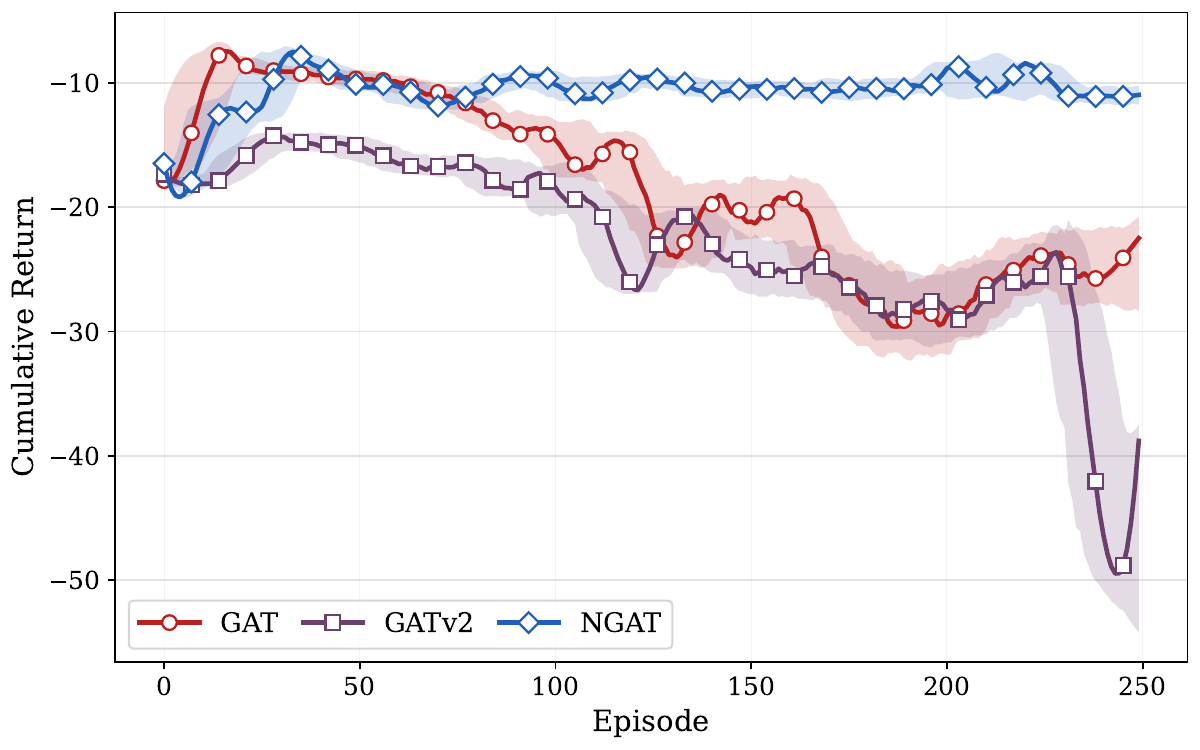}
    \caption{\rev{Empirical training behavior of GNN methods. GATv2 and NGAT achieve comparable peak performance, while NGAT exhibits more stable performance after approximately 100 episodes in this experiment. This observation does not imply a general convergence guarantee.}}
    \label{fig: GNN Convergence}
\end{figure}

\subsection{Proof of Proposition \ref{Proposition: Multicast Tree}} \label{Proof of Proposition: Multicast Tree}
From the definition of the action space $\mathcal{A}_2$, the chosen node $a_\tau\in\mathcal{A}_2$ is connected to $\mathcal{P}_{\tau'}$ and is not chosen before, which means $\mathcal{P}_{\tau'}$ is connected and acyclic. Therefore, $\mathcal{P}_{\tau'}$ is a tree. In addition, $\mathcal{P}_{\tau'}$ is initialized with a source node. From the definition of a terminal state $s_{\tau'}$, $a_t$ is included in $\mathcal{P}_{\tau'}$. Therefore, $\mathcal{P}_{\tau'}$ is a multicast tree.

\subsection{Proof of Lemma \ref{Lemma: Another form of P2}} \label{ProofLemma2}
We first analyze the relation between multicast tree $\mathcal{T}_t$ and AoI reduction:
\begin{Lemma}
    \label{Lemma: AoI Reduction}
    Given a multicast tree $\mathcal{T}_t$, if there are no other transmitting packets, the average AoI reduction of destination $u$ caused by $\mathcal{T}_t$ is $\frac{A_u(t)}{h_{\mathcal{T}_t}(u)}$ during time $[t, t+h_{\mathcal{T}_t}(u)]$.
    \begin{proof}
    Denote $\Delta A_u(t, t+h_{\mathcal{T}_t}(u))$ as the AoI reduction of destination $u$ during time $[t, t+h_{\mathcal{T}_t}(u)]$, we have:
    \begin{equation}
        \begin{aligned}
            & \Delta A_u(t, t+h_{\mathcal{T}_t}(u)) \\
            &= A^+_u(t, t+h_{\mathcal{T}_t}(u)) - A^-_u(t, t+h_{\mathcal{T}_t}(u)) \\
            &= \sum_{k=0}^{h_{\mathcal{T}_t}(u)} (A_u(t)+k) - \left(\sum_{k=0}^{h_{\mathcal{T}_t}(u)-1}(A_u(t)+k) + h_{\mathcal{T}_t}(u)\right) \\
            &= A_u(t),
        \end{aligned}
    \end{equation}
    where $A^+_u(t, t+h_{\mathcal{T}_t}(u))$ and $A^-_u(t, t+h_{\mathcal{T}_t}(u))$ denote the AoI of destination $u$ during time $[t, t+h_{\mathcal{T}_t}(u)]$ before and after the multicast tree $\mathcal{T}_t$ is generated, respectively. 
    \end{proof}
\end{Lemma}
From the multicast process, we know that the influence of any multicast tree lasts until the packet with maximum hops is received or dropped. Denote $\hat h_{\mathcal{G}_t}$ as the length of $\mathcal{G}_t$, we can rewrite the objective of problem \textbf{P2} as:
\begin{equation}
    \label{Eq: another form of P2}
    \begin{aligned}
        g(\lambda, \mathcal{U}'_t)
        &= \max_{\mathcal{T}} \frac{1}{\hat h_{\mathcal{G}_t}} \sum_{u\in\mathcal{U}'_t} \sum_{k=0}^{ h_{\mathcal{T}}(u)} -\omega_u A_u(t+k) \\
        &\quad - \lambda (C(\mathcal{T}) - \overline{C}).
    \end{aligned}
\end{equation}
From the first term of RHS of \eqref{Eq: another form of P2} in Lemma \ref{Lemma: AoI Reduction},  we have:
\begin{equation}
    \label{Eq: RHS of another form of P2}
    \begin{aligned}
        & \max_{\mathcal{T}} \frac{1}{\hat h_{\mathcal{G}_t}} \sum_{u\in\mathcal{U}'_t} \sum_{k=0}^{\hat h_{\mathcal{G}_t}} -\omega_u A_u(t+k) \\
        &= \max_{\mathcal{T}} \frac{1}{\hat h_{\mathcal{G}_t}} \sum_{u\in\mathcal{U}'_t} \omega_u \sum_{k=0}^{\hat h_{\mathcal{G}_t}}(A_u(t)+k-A_u(t+k)) \\
        &= \max_{\mathcal{T}} \frac{1}{\hat h_{\mathcal{G}_t}} \sum_{u\in\mathcal{U}'_t} \omega_u \sum_{k=h_{\mathcal{T}}(u)}^{\hat h_{\mathcal{G}_t}} (A_u(t)+k-A_u(t+k)) \\
        &= \max_{\mathcal{T}} \frac{1}{\hat h_{\mathcal{G}_t}} \sum_{u\in\mathcal{U}'_t} \omega_u (\hat h_{\mathcal{G}_t}-h_{\mathcal{T}}(u)) A_u(t) \\
        &= \max_{\mathcal{T}} \sum_{u\in\mathcal{U}'_t} \left(1-\frac{h_{\mathcal{T}}(u)}{\hat h_{\mathcal{G}_t}}\right) \omega_u A_u(t).
    \end{aligned}
\end{equation}
Substitute \eqref{Eq: RHS of another form of P2} into \eqref{Eq: another form of P2} concludes the proof. 

\subsection{Proof of Theorem \ref{Theorem: Contraction Mapping of NGAT}}
\rev{Write the embeddings and features as row-stacked matrices $\mathbf{H}$ and $\mathbf{X}$. With the same fixed $\mathbf{A}$, features, and weights for both inputs, \eqref{Eq: NGAT attention sum} becomes}
\begin{equation}
    \rev{F_{\mathbf{A}}(\mathbf{H};\mathbf{X})
    =\frac{\mathbf{A}(\mathbf{H}\mathbf{W}_1^\top+
    \mathbf{X}\mathbf{W}_3^\top)}{\|\mathbf{W}_1\|_2}.}
\end{equation}
\rev{Since $\mathbf{A}$ is nonnegative and row-stochastic, its induced infinity norm equals one and all its eigenvalues have absolute value at most one. Symmetry makes $\mathbf{A}$ orthogonally diagonalizable, hence $\|\mathbf{A}\|_2=\max_j|\lambda_j(\mathbf{A})|\leq1$. Setting $\mathbf{\Delta}=\mathbf{H}-\mathbf{H}'$, the common feature term cancels, yielding}
\begin{equation}
    \rev{\begin{aligned}
    &\|F_{\mathbf{A}}(\mathbf{H};\mathbf{X})
      -F_{\mathbf{A}}(\mathbf{H}';\mathbf{X})\|_F\\
    &=\frac{\|\mathbf{A}\mathbf{\Delta}\mathbf{W}_1^\top\|_F}
      {\|\mathbf{W}_1\|_2}\\
    &\leq\frac{\|\mathbf{A}\|_2\|\mathbf{\Delta}\|_F
      \|\mathbf{W}_1^\top\|_2}{\|\mathbf{W}_1\|_2}\\
    &\leq\|\mathbf{\Delta}\|_F=d_F(\mathbf{H},\mathbf{H}').
    \end{aligned}}
\end{equation}
\rev{The first inequality uses compatibility of the Frobenius and spectral norms and $\|\mathbf{W}_1^\top\|_2=\|\mathbf{W}_1\|_2>0$. This proves non-expansiveness of the complete embedding matrix under the fixed-attention assumptions. Applying this estimate at each layer gives the trajectory comparison in the remark when both trajectories share the same layerwise maps. No convergence or fixed-point conclusion follows from this estimate alone.} \hfill $\square$

\subsection{Proof of Theorem \ref{Theorem: Convergence Theorem}} \label{App-Proof-T2}
We consider a model consisting of NGAT layers and linear layers. \rev{Denote $\mathbf{H}^{(L)}$ as the output after a fixed finite number $L$ of NGAT layers, and use $s=\mathbf{H}^{(L)}$ as the new representation of the network state. This is a finite-depth output, not a limiting fixed point.} We can then analyze the linear part of the heads. Let $\bm{\theta}$ and $\bm{\omega}$ be the parameters of the actor and the critic, respectively. We first introduce some assumptions as follows.
\begin{Assumption}
    \label{Assumptions of Reward and Actor}
    Suppose all reward functions $r$ and the parameterized policy $\pi_{\bm{\theta}}$ satisfy the following conditions:\\
    (a) $r$ is bounded, i.e.:
    \begin{equation}
        \label{Eq: Bounded Reward}
        |r(s,a)| \leq R, \forall (s,a) \in \mathcal{S}\times\mathcal{A}.
    \end{equation}
    In our setting, this bounded-reward assumption is imposed for the convergence analysis by the finite AoI upper bound $A_u(t)\leq A_{\max}$. Since the graph is finite, $\omega_u$, $h_{\mathcal{T}}(u)/\hat h_{\mathcal{G}_t}$, and the multicast-tree cost are finite; with a bounded Lagrangian multiplier $\lambda$, the rewards $r_1$ and $r_2$ are therefore bounded.
    (b) $\nabla\log\pi_{\bm{\theta}}(a|s)$ is $L_\theta$-Lipschitz and bounded \footnote{The gradient of the policy is bounded by the gradient clipping technique.}, i.e.:
    \begin{equation}
        \|\nabla\log\pi_{\bm{\theta}_1}(a|s)-\nabla\log\pi_{\bm{\theta}_2}(a|s)\|
        \leq L_{\bm{\theta}}\|\bm{\theta}_1-\bm{\theta}_2\|, 
    \end{equation}
    \begin{equation}
        \label{Eq: Bound of Policy Gradient}
        \|\nabla\log\pi_{\bm{\theta}}(a|s)\| \leq B_{\bm{\theta}} ,\forall\theta_1,\theta_2\in\mathbb{R}^d,
    \end{equation}
    where $L_{\bm{\theta}}, B_{\bm{\theta}}$ are constants.
\end{Assumption}
\begin{Assumption}
    \label{Assumptions of Critic}
    Suppose the critic satisfies the following condition: \\
    (a) The critic can be represented by a linear function $V_{\bm\omega}(s)=\phi(s)^\top\bm\omega$, where $\phi(s)$ is the feature vector of state $s$ and $\bm\omega$ is the parameter of the critic. \\
    (b) The feature mapping $\phi(s)$ has a bounded norm $|\phi(s)| \leq 1$. \\
    (c) $V_{\bm{\omega}}(s)$ is bounded, i.e.: $|V_{\bm{\omega}}(s)| \leq B_{\bm{\omega}}$,
    where $B_{\bm{\omega}}$ is a constant.
\end{Assumption}
\rev{Assumption \ref{Assumptions of Critic}, including the unit-norm feature condition and bounded critic output, is imposed separately for the actor-critic analysis. The finite-depth NGAT bound discussed in the main text does not by itself establish these conditions.} For the objective function $J(\bm{\theta})$, we have $|J(\bm{\theta})|<\frac{R}{1-\gamma}$. As Assumption \ref{Assumptions of Critic} states, we can obtain an approximated advantage value function $\hat A_t(s_t, a_t)=r(s_t, a_t)-V_{\bm\omega}(s_t)$, where $\bm\omega$ is the parameter of the critic. Then the gradient of $J(\bm{\theta})$ can be approximated by:
\begin{equation}
    \nabla \hat J(\bm{\theta})=\mathbb{E}_{s\sim \mu_{\bm{\theta}}, a\sim\pi_{\bm{\theta}}}
    [\hat A_{\pi_{\bm{\theta}}}(s,a) \nabla\log\pi_{\bm{\theta}}(a|s) ],
\end{equation}
and $\nabla \hat J(\bm{\theta})$ is bounded as follows:
\begin{Lemma}[Bound of Approximated Policy Gradient]
    \label{Bound of Approximated Policy Gradient} Under Assumption \ref{Assumptions of Reward and Actor} and \ref{Assumptions of Critic}, the policy gradient $\nabla \hat J(\bm{\theta})$ is bounded, i.e.:
    \begin{equation}
        \| \nabla \hat J(\bm\theta) \| \leq (R+B_{\bm\omega})B_{\bm\theta}.
    \end{equation}
    \begin{proof}
        The approximated advantage value function $\hat A_t(s_t, a_t)$ can be written as:
        \begin{equation}
            \begin{aligned}
                \hat A_t(s_t, a_t) &= r(s_t, a_t) - \phi(s_t)^\top\bm\omega \\
                & \leq |r(s_t, a_t)| + |\phi(s_t)^\top\bm\omega|
                \leq R+B_{\bm\omega},
            \end{aligned}
        \end{equation}
        which implies:
        \begin{equation}
            \| \hat A_t(s_t, a_t) \nabla \log\pi_{\bm\theta}(a_t|s_t)\| \leq (R+B_{\bm\omega})B_{\bm\theta}.
        \end{equation}
        The proof is concluded.
    \end{proof}
\end{Lemma}
Now we analyze the approximation error of $\nabla \hat J(\bm{\theta})$. First, introduce the following lemma:
\begin{Lemma}[Lipschitz-Continuity of Policy Gradient \cite{zhang2020global}]
    \label{Lipschitz-Continuity of Policy Gradient}
    Under Assumption \ref{Assumptions of Reward and Actor}, the policy gradient $\nabla J(\bm{\theta})$ is Lipschitz continuous with some constant $L>0$, i.e.:
    \begin{equation}
        \|\nabla J(\bm{\theta}_1)-\nabla J(\bm{\theta}_2)\| \le L\|\bm{\theta}_1-\bm{\theta}_2\|.
    \end{equation}
\end{Lemma}
According to Lemma \ref{Lipschitz-Continuity of Policy Gradient} and (1.2.19) from \cite{nesterov2018lectures}, we have:
\begin{equation}
    \label{Policy Gradient Smoothness}
    J(\bm\theta_2) \geq J(\bm\theta_1) + \braket{\nabla J(\bm\theta_1), \bm\theta_2-\bm\theta_1}-\frac{L}{2}\|\bm\theta_1-\bm\theta_2\|^2,
\end{equation}
where $\braket{\cdot, \cdot}$ denotes the dot product of two vectors. The actor of Algorithm \ref{PC: A2C} is updated by:
\begin{equation}
    \label{Update of Actor}
    \bm\theta_{t+1}^a = \bm\theta_t^a + \alpha_t\hat{A}_t(s_t,a_t) \nabla\log\pi_{\bm\theta_t}(a_t|s_t),
\end{equation}
Substitute \eqref{Update of Actor} into \eqref{Policy Gradient Smoothness} yields:
\begin{equation}
    \label{Eq: Initial Inequality}
    \begin{aligned}
         J(\bm\theta_{t+1}) \geq J(\bm\theta_{t}) + \alpha_t\braket{\nabla J(\bm\theta_{t}), \nabla \hat J(\bm\theta_{t})} -\frac{L\alpha_t^2}{2}\|\nabla \hat J(\bm\theta_{t})\|^2.
    \end{aligned}
\end{equation}
Here, we focus on the second term on the right-hand side of the equation above, which can be decomposed as:
\begin{equation}
    \label{Eq: Error Decomposition}
    \begin{aligned}
        & \braket{\nabla J(\bm\theta_{t}), \nabla \hat J(\bm\theta_{t})} \\
        &= \braket{\nabla J(\bm\theta_{t}), \phi(s_t)^\top(\bm\omega^*-\bm\omega_t)\nabla\log\pi_{\bm\theta_t}(a_t|s_t)} \\
        &\quad + \braket{\nabla J(\bm\theta_{t}), (v(s_t)-\phi(s_t)^\top \bm\omega^*)\nabla\log\pi_{\bm\theta_t}(a_t|s_t)} \\
        &\quad + \braket{\nabla J(\bm\theta_{t}), \nabla \hat J(\bm\theta_{t})- \nabla J(\bm\theta_{t})} + \braket{\nabla J(\bm\theta_{t}), \nabla J(\bm\theta_{t})}
    \end{aligned}
\end{equation}
We can view the first term on the RHS of \eqref{Eq: Error Decomposition} as the error of the critic, the second term as the error from linear function approximation of the critic, and the third term as the Markovian noise. Therefore, the above equation can be rewritten as:
\begin{equation}
    \begin{aligned}
        \label{Eq: Error Decomposition 2}
        & \braket{\nabla J(\bm\theta_{t}), \nabla \hat J(\bm\theta_{t})} \\
        &= \Phi_{EC}(\bm\theta_{t}) + \Phi_{LFA}(\bm\theta_{t}) + \Phi_{M}(\bm\theta_{t}) + 
        \| \nabla J(\bm\theta_{t}) \|^2,
    \end{aligned}
\end{equation}
We will analyze the bounds of these terms in the following. First, $\Phi_{EC}(\bm\theta_{t})$ measures the distance between the optimal parameter $\bm\omega^*$ and the current parameter $\bm\omega_t$, which is bounded as follows:
\begin{Lemma}[Error of Critic]
    \label{Error of Critic}
    Given Assumption \ref{Assumptions of Critic}, we have:
    \begin{equation}
        \begin{aligned}
            \mathbb{E}[\Phi_{EC}(\bm\theta_{t})] \geq -B_{\bm{\theta}} \sqrt{\mathbb{E}[\|\nabla J(\bm\theta_{t})\|^2]} \sqrt{\mathbb{E}[\|\bm\omega^*-\bm\omega_t\|^2]}
        \end{aligned}
    \end{equation}
    \textit{proof:}
    From Cauchy inequality and \eqref{Eq: Bound of Policy Gradient}, we have:
    \begin{equation}
        \begin{aligned}
           & \braket{\nabla J(\bm\theta_{t}), \phi(s_t)^\top(\bm\omega^*-\bm\omega_t)\nabla\log\pi_{\bm\theta_t}(a_t|s_t)} \\
           & \geq -\| \nabla J(\bm\theta_{t}) \| \| \phi(s_t)^\top(\bm\omega^*-\bm\omega_t) \| \| \nabla\log\pi_{\bm\theta_t}(a_t|s_t) \| \\
           & \geq -\| \nabla J(\bm\theta_{t}) \| \| \bm\omega^*-\bm\omega_t \| \| \nabla\log\pi_{\bm\theta_t}(a_t|s_t) \| \\
           & \geq -B_{\bm\theta} \|\nabla J(\bm\theta_{t})\| \|\bm\omega^*-\bm\omega_t\|.
        \end{aligned}
    \end{equation}
    Taking the expectations of both sides yields the result. \hfill $\square$
\end{Lemma}
Then we analyze the approximation error between the optimal parameter $\bm\omega^*$ and the parameter $\bm\omega_t$. Suppose there is an approximation error bound $\epsilon_v>0$ that
$\|\mathbb{E}[\phi(s)^\top\bm\omega^*-v(s)]\| \leq \epsilon_v$, we have:
\begin{Lemma}[Error of Linear Function Approximation]
    \label{Error of Linear Function Approximation}
    \begin{equation}
        \begin{aligned}
            \mathbb{E}[\Phi_{LFA}(\bm\theta_{t})] \geq -\epsilon_v B_{\bm\theta}\mathbb{E}[\|\nabla J(\bm\theta_{t})\|].
        \end{aligned}
    \end{equation}
    \textit{proof:}
    From Cauchy inequality, we have:
    \begin{equation}
        \begin{aligned}
            &\braket{\nabla J(\bm\theta_{t}), (v(s_t)-\phi(s_t)^\top \bm\omega^*)\nabla\log\pi_{\bm\theta_t}(a_t|s_t)} \\
            & \geq -\|\nabla J(\bm\theta_{t})\| \|v(s_t)-\phi(s_t)^\top \bm\omega^*\| \|\nabla\log\pi_{\bm\theta_t}(a_t|s_t)\| \\
            & \geq -\epsilon_v B_{\bm\theta}\|\nabla J(\bm\theta_{t})\|. 
        \end{aligned}
    \end{equation}
    Taking the expectations of both sides yields the result. \hfill $\square$
\end{Lemma}
Suppose the Markovian noise is bounded, i.e., there exists $\epsilon_m>0$ that $-\epsilon_m \leq \Phi_{M}(\bm\theta_{t}) \leq \epsilon_m$. Substitute the inequalities above into \eqref{Eq: Error Decomposition 2} yields:
\begin{equation}
    \label{Eq: Error Decomposition 3}
    \begin{aligned}
        & \mathbb{E}[\braket{\nabla J(\bm\theta_{t}), \nabla \hat J(\bm\theta_{t})}] \\
        &\geq -B_{\bm{\theta}} \sqrt{\mathbb{E}[\|\nabla J(\bm\theta_{t})\|^2]} \sqrt{\mathbb{E}[\|\bm\omega^*-\bm\omega_t\|^2]} \\
        & -\epsilon_v B_{\bm\theta}\mathbb{E}[\|\nabla J(\bm\theta_{t})\|] -\epsilon_m + \mathbb{E}[\| \nabla J(\bm\theta_{t}) \|^2].
    \end{aligned}
\end{equation}
Substitute \eqref{Eq: Error Decomposition 3} into \eqref{Eq: Initial Inequality} and take the expectation of both sides yields:
\begin{equation}
    \begin{aligned}
        & \mathbb{E}[J(\bm\theta_{t+1}) - J(\bm\theta_{t})] \\
        & \geq -\alpha_t B_{\bm{\theta}} \sqrt{\mathbb{E}[\|\nabla J(\bm\theta_{t})\|^2]} \sqrt{\mathbb{E}[\|\bm\omega^*-\bm\omega_t\|^2]} \\
        &\quad - \alpha_t \epsilon_v \mathbb{E}[\|\nabla J(\bm\theta_{t})\|] -\alpha_t\epsilon_m \\
        &\quad + \alpha_t\mathbb{E}[\| \nabla J(\bm\theta_{t}) \|^2] -\frac{L\alpha_t^2B_j^2}{2},
    \end{aligned}
\end{equation}
where $B_j = (R+B_{\bm\omega})B_{\bm\theta}$. Let $G_t=\mathbb{E}[\| \nabla J(\bm\theta_t) \|^2]$ and $D_t=\mathbb{E}[\|\bm\omega^*-\bm\omega_t\|^2]$. By Young's inequality,
\begin{equation}
    {
    \begin{aligned}
        B_{\bm\theta}\sqrt{G_t}\sqrt{D_t}\leq \frac{1}{4}G_t+B_{\bm\theta}^2D_t,\quad
        \epsilon_v B_{\bm\theta}\sqrt{G_t}\leq \frac{1}{4}G_t+\epsilon_v^2B_{\bm\theta}^2.
    \end{aligned}}
\end{equation}
Thus,
\begin{equation}
    {
    \begin{aligned}
        \mathbb{E}[J(\bm\theta_{t+1}) - J(\bm\theta_{t})]
        &\geq \frac{\alpha_t}{2}G_t-\alpha_t B_{\bm\theta}^2D_t \\
        &\quad -\alpha_t\epsilon_v^2B_{\bm\theta}^2-\alpha_t\epsilon_m
        -\frac{L\alpha_t^2B_j^2}{2}.
    \end{aligned}}
\end{equation}
Rearranging the terms, dividing by $\alpha_t$, and summing both sides over $i=n$ to $k$, we obtain
\begin{equation}
    \label{Eq: Explicit Gradient Bound}
    {
    \begin{aligned}
        \sum_{i=n}^{k}G_i
        &\leq 2\sum_{i=n}^{k}\frac{\mathbb{E}[J(\bm\theta_{i+1})-J(\bm\theta_i)]}{\alpha_i}
        +2B_{\bm\theta}^2\sum_{i=n}^{k}D_i \\
        &\quad +(k-n+1)(2B_{\bm\theta}^2\epsilon_v^2+2\epsilon_m)
        +L B_j^2\sum_{i=n}^{k}\alpha_i.
    \end{aligned}}
\end{equation}
Since the reward is bounded, $|J(\bm\theta)|\leq R/(1-\gamma)$. Using the standard summation-by-parts argument for nonincreasing $\alpha_i$,
\begin{equation}
    {
    \begin{aligned}
        \sum_{i=n}^{k}\frac{\mathbb{E}[J(\bm\theta_{i+1})-J(\bm\theta_i)]}{\alpha_i}
        \leq \frac{2R}{\alpha_n(1-\gamma)}.
    \end{aligned}}
\end{equation}
Moreover, with $\alpha_i=c_\alpha/(1+i)^\sigma$ and $0<\sigma<1$,
\begin{equation}
    {
    \begin{aligned}
        \sum_{i=n}^{k}\alpha_i
        \leq \frac{c_\alpha}{1-\sigma}(k-n+1)^{1-\sigma}.
    \end{aligned}}
\end{equation}
Dividing \eqref{Eq: Explicit Gradient Bound} by $k-n+1$ and using $k\geq 2n-1$ gives
\begin{equation}
    {
    \begin{aligned}
        \frac{1}{k-n+1}\sum_{i=n}^{k}G_i
        &\leq 2B_{\bm\theta}^2\bar{\mathcal E}_c(n,k)
        +\frac{4R(1+n)^\sigma}{c_\alpha(1-\gamma)(k-n+1)} \\
        &\quad +\frac{L B_j^2c_\alpha}{(1-\sigma)(k-n+1)^\sigma}
        +2B_{\bm\theta}^2\epsilon_v^2+2\epsilon_m \\
        &\leq C_1\bar{\mathcal E}_c(n,k)+\frac{C_2}{n^\sigma}
        +C_3\epsilon_v^2+C_4\epsilon_m,
    \end{aligned}}
\end{equation}
where the second inequality uses $0<\sigma\leq 1/2$, so that $(1+n)^\sigma/(k-n+1)=\mathcal{O}(1/n^{1-\sigma})\leq \mathcal{O}(1/n^\sigma)$. Finally,
\begin{equation}
    {
    \begin{aligned}
        \min_{n\leq i\leq k}\mathbb{E}[\|\nabla J( \bm\theta_i)\|^2]
        \leq \frac{1}{k-n+1}\sum_{i=n}^{k}G_i \\
        \leq C_1\bar{\mathcal E}_c(n,k)+\frac{C_2}{n^\sigma}
        +C_3\epsilon_v^2+C_4\epsilon_m.
    \end{aligned}}
\end{equation}
The proof is completed. \hfill $\square$
}

\end{document}